# RESERVOIR CHARACTERIZATION: A MACHINE LEARNING APPROACH

*Thesis submitted to the*

*Indian Institute of Technology, Kharagpur*

*for award of the degree*

*of*

**Master of Science (by Research)**

*by*

**Soumi Chaki**

Under the guidance of

**Prof. Aurobinda Routray**

**Prof. William K. Mohanty**

**DEPARTMENT OF ELECTRICAL ENGINEERING**

**INDIAN INSTITUTE OF TECHNOLOGY, KHARAGPUR**

**April 2015**

# CERTIFICATE OF APPROVAL

Certified that the thesis entitled **Reservoir Characterization: A Machine Learning Approach** submitted by **Soumi Chaki**, to the Indian Institute of Technology, Kharagpur, for the award of the degree of Master of Science (by Research) has been accepted by the external examiners and that the student has successfully defended the thesis in the viva-voce examination held today.

Signature:
Name:
(Member of the DAC)

Signature:
Name:
(Member of the DAC)

Signature:
Name:
(Member of the DAC)

Signature:
Name:
(Superviser)

Signature:
Name:
(Joint Superviser)

Signature:
Name:
(External Examiner)

Signature:
Name:
(Chairman)

Department of Electrical Engineering
Indian Institute of Technology, Kharagpur
Kharagpur, India 721302.

## CERTIFICATE

This is to certify that this thesis entitled **Reservoir Characterization: A Machine Learning Approach** submitted by **Soumi Chaki**, to the Indian Institute of Technology, Kharagpur, is a record of bona fide research work carried out under our supervision and is worthy of consideration for award of the degree of Master of Science (by Research) of the Institute.

Supervisor
**Aurobinda Routray**
Department of Electrical Engineering
Indian Institute of Technology, Kharagpur
Kharagpur, West Bengal
India - 721302

Supervisor
**William K. Mohanty**
Department of Geology and Geophysics
Indian Institute of Technology, Kharagpur
Kharagpur, West Bengal
India - 721302

Place: IIT Kharagpur

Date:

# DECLARATION

I certify that

a. The work contained in the thesis is original and has been done by myself under the general supervision of my supervisor(s).

b. The work has not been submitted to any other Institute for any degree or diploma.

c. I have followed the guidelines provided by the Institute in writing the thesis.

d. I have conformed to the norms and guidelines given in the Ethical Code of Conduct of the Institute.

e. Whenever I have used materials (data, theoretical analysis, and text) from other sources, I have given due credit to them by citing them in the text of the thesis and giving their details in the references.

f. Whenever I have quoted written materials from other sources, I have put them under quotation marks and given due credit to the sources by citing them and giving required details in the references.

Soumi Chaki

# ACKNOWLEDGEMENTS

This thesis marks the end of a long and eventful journey during which there were many people whom I would like to acknowledge for their encouragement and support. I am grateful to my supervisors, Prof. Aurobinda Routray, and Prof. William K. Mohanty, for their constant support throughout the research work. Their advice and constructive criticism have helped the thesis take its present shape.

I would like to thank the members of the DAC, Prof. A. K. Deb, Prof. P. K. Dutta, Prof. M. Jenamani for their suggestions and comments during the presentations and throughout the tenure of the work. I am thankful to Head, Department of Electrical Engineering, IIT Kharagpur, for providing me the facilities to carry out my research work. I am also grateful to all other faculty members of the Department of Electrical Engineering, IIT Kharagpur for the help received during my research. I also extend my thanks to the staffs of Central Library, IIT Kharagpur, for their assistance in locating books in a library which is one of the largest of its type.

I would like to thank Geodata Processing and Interpretation Centre (GEOPIC), Dehradun, Oil and Natural Gas Corporation Limited (ONGC) for collaborating with us for the research project. My special thanks to Mrs. Puja Prakash, Mr. S. K. Das and Mr. P.K. Chaudhuri. I would like to acknowledge the help from Mr. Akhilesh Kumar Verma from Department of Geology and Geophysics, IIT Kharagpur.

I would like to appreciate the help and support of all my friends and colleagues of Real Time Embedded System Lab., Systems and Information Lab., Department of Electrical Engineering and Centre for Railway Research Lab.

Finally, I would like to thank my family who always encouraged me for higher studies and made me believe that knowledge is more powerful than wealth. My deepest gratitude goes to my parents for their unflagging love and support throughout my life.

Place: IIT Kharagpur

Date: Soumi Chaki

# Abstract

'Reservoir Characterization (RC)' can be defined as the act of building a reservoir model that incorporates all the characteristics of the reservoir that are pertinent to its ability to store hydrocarbons and also to produce them. It is a difficult problem due to non-linear and heterogeneous subsurface properties and associated with a number of complex tasks such as data fusion, data mining, formulation of the knowledge base, and handling of the uncertainty.

This present work describes the development of algorithms to obtain the functional relationships between predictor seismic attributes and target lithological properties. Seismic attributes are available over a study area with lower vertical resolution. Conversely, well logs and lithological properties are available only at specific well locations in a study area with high vertical resolution. If a functional relationship can be calibrated between seismic signals and lithological properties at available well locations, then distribution of these properties across the study area can be predicted from available seismic information. Depending on the distribution of the lithological properties, a dataset can be classified into two categories – balanced and imbalanced. Sand fraction, which represents per unit sand volume within the rock, has a balanced distribution between zero to unity. On the other hand, water saturation, oil saturation etc. has an imbalanced distribution skewed at one and zero respectively. The investigation about the sand fraction (balanced distribution) variation over the study area has been attempted as a prediction problem; whereas, the distribution of water saturation (balanced distribution) has been approached as a classification (Class low/ Class high) problem in this work.

The thesis addresses the issues of handling the information content mismatch between predictor and target variables and proposes regularization of target property prior to building a prediction model. In this thesis, two Artificial Neural Network (ANN) based frameworks are proposed to model sand fraction from multiple seismic attributes without and with well tops information respectively. The performances of the frameworks are quantified in terms of Correlation Coefficient (CC), Root Mean Square Error (RMSE), Absolute Error Mean (AEM), etc.

After successful completion of sand fraction prediction, a one-class classification framework based on Support Vector Data Description (SVDD) is proposed to classify water saturation from well logs. The designed framework is modified to include seismic variables as predictor attributes to obtain the variation of water saturation over the study area. In other words, the class labels (Class low/Class high) of water saturation belonging to a well location can be predicted from seismic attributes by the modified classification based framework. The proposed frameworks have outperformed other supervised classification algorithms in terms of g-metric means and program execution time (in seconds).

# Contents

## List of Symbols

| Symbol | Description |
|---|---|
| $X$ | A variable |
| $\mu_X$ | Mean of the $X$ |
| $\sigma_X$ | Standard deviation of the $X$ |
| $H(X)$ | Entropy |
| $p(x_i)$ | Probability of $X$ having the $i^{th}$ value $x_i$ in the dataset |
| $H(X \mid A)$ | The conditional entropy of $X$ after $A$ has been observed |
| $I(X;A)$ | Mutual information between $X$ and $A$ |
| $NMI(X;A)$ | Normalized mutual information |
| $x(t)$ | A signal |
| $X(k)$ | Fourier Transform of $x(t)$ |
| $\psi(t)$ | a mother wavelet |
| $\phi(t)$ | scaling function |
| $a_l(k)$ | the approximate coefficients at level $l$ |
| $d_l(k)$ | the detailed coefficients at level $l$ |
| $w$ | Window size of a spatial filter |
| $\text{acc}_\text{P}$ | sensitivity |
| $\text{acc}_\text{N}$ | specificity |
| g | G-metric means |
| $K(x_i, x_j)$ | Kernel function |
| $L$ | Lagrangian function |

## List of Abbreviations

| Abbreviation | Description |
|---|---|
| SF | Sand Fraction |
| MI | Mutual Information |
| NMI | Normalized Mutual Information |
| ANN | Artificial Neural Network |
| FT | Fourier Transform |
| WD | Wavelet Decomposition |
| db4 | Daubechies 4 wavelet |
| EMD | Empirical Mode Decomposition |
| IMF | Intrinsic Mode Functions |
| RMSE | Root Mean Square Error |
| AEM | Absolute Error Mean |
| CC | Correlation Coefficient |
| SI | Scatter Index |
| SVM | Support Vector Machines |
| SVDD | Support Vector Data Description |
| PSD | Power Spectral Density |
| BPNN | Back Propagation Neural Network |
| MANN | Modular Artificial Neural Network |
| SCG | Scaled Conjugate Gradient |
| GR | Gamma Ray content |
| RHOB | Bulk Density |
| RT | Deep Resistivity |
| DT | P-sonic |
| RM | Medium Resistivity |
| RS | Shallow Resistivity |
| NPHI | Neutron Porosity |
| SP | Spontaneous Potential |
| AI | Acoustic Impedance |
| API | American Petroleum Institute |
| IDWT | Inverse Discrete Wavelet Transform |
| EEG | Electroencephalography |
| SOM | Self-Organizing Map |

## List of Figures

**List of Tables**

# Chapter 1. Introduction

The act of building a reservoir model that incorporates all the characteristics of the reservoir to store hydrocarbons and also to produce them is termed as 'Reservoir Characterization (RC)' [1]. The non-linear and heterogeneous physical properties of the subsurface make reservoir characterization a difficult task. The initial step of this characterization is prediction of reservoir characteristics (such as sand fraction, shale fraction, porosity, permeability, fluid saturation, water saturation etc.) or class variations of these properties from well logs and seismic attributes. Prediction of petrophysical properties is associated with a number of complex tasks such as data fusion (i.e. integration of data from various sources), data mining (i.e. information retrieval after analysing those data), formulation of the knowledge base, and handling of the uncertainty. The applications of advanced statistical, machine learning and pattern recognition techniques to such problems have received considerable interest among the researchers in oil-gas sector [2], [3]. The objective of these types of studies is to identify potential zone for drilling a new well [4]. The fundamental characteristics of a reservoir system are typically distributed spatially in a non-uniform and non-linear manner. Extraction of lithological information from available datasets is an important step in the reservoir characterization process. Since there is no direct measurement for the lithological parameters, they are to be computed from other geophysical logs [5] or seismic attributes [6]. This process also requires repeated intervention of the experts for fine tuning the prediction results. Standard regression methods are not suitable for this problem due to the high degree of the unknown nonlinearity. The problem is further complicated because of uncertainties associated with lithological units. In this context, Artificial Neural Network (ANN) and its variants with Fuzzy Logic are considered to be useful tools to establish a mapping between lithological and well log properties [7]–[10].

Further, it is important to characterize how 3D seismic information is related to production, lithology, geology, and well log data. It is suggested that the use of 3D seismic data along with well logs can provide better insights while extrapolating reservoir properties away from the existing wells [11], [12]. ANN [13], Adaptive Neuro-Fuzzy Inference System (ANFIS) [14], Support Vector Machine (SVM) [15], type-2 Fuzzy Logic system [16], and hybrid systems [17], [18] are some the efficient machine learning tools used in the field of reservoir characterization. Now a days, one of the challenging problems for the petroleum industry is to enhance oil recovery from naturally occurring complex reservoir systems. Therefore, it is important to identify the patterns of the characteristic distributions of the pertinent reservoir parameters in the subsurface.

## 1.1 Literature Review

Hydrocarbons migrate from source rock through porous medium to reach reservoir rock for temporary preservation [19]. Finally, the mobile hydrocarbons get seized in the cap rocks. As such, the identification of hydrocarbon–enriched–formations by characterization of each layer in the borehole is of enormous necessity to the explorers. Recognition of a potential hydrocarbon–enriched zone in a prospective oil exploration field can be carried out using well logs which categorize layers into different sections such as dry, water containing, and hydrocarbon bearing layers. The lithological properties in the neighborhood of a borehole can be known from well logs, whereas these remain unknown and difficult to estimate away from the wells. In such cases, available seismic attributes can be used as a guidance to predict lithological information at all traces of the area of interest [11]. Well logs and seismic attributes are integrated at available well locations to design a reservoir model with the least uncertainty. However, mapping between lithological properties and seismic attributes is governed with nonlinear relationship and mismatch in information content. Ahmadi *et al.* [20] explored that nonlinear problems can be approached using state-of-art computer–based methods like expert systems [21], multiple regression, neural networks [22], Neuro-fuzzy Systems [23] etc.

ANN is widely used to model single or multiple target properties from predictor variables in different research domains. It has been found from literature that ANN is a natural choice of researchers and engineers because of its prediction and generalization capability. For example, it is used in climatological studies [24], ocean engineering [25], telecommunications [26], text recognition [27], financial time series [28], reservoir characterization [4], [10], [29]–[32], etc. A diverse dataset containing information assembled from multiple domains can be used for learning and validation of ANN. However, it has some inherent limitations. Firstly, performance of ANN is dependent on the selection of network structure and associated parameters. Secondly, training a complex, multilayered network is a time intensive process. Furthermore, a complex network trained with relatively smaller number of learning patterns may lead to overfitting, and thus, generalization capability is compromised. It is equally important to assess the possibility of modeling the target property from predictor variables using ANN or any other nonlinear modeling approach. Sometimes the model performance can be improved by applying suitable filtering techniques to the predictor/target variables in the pre-processing stage. Several studies have contributed on the performance analysis of ANN along with other machine learning algorithms to model a target variable from single or multiple predictors with respect to RC problem; however, the following aspects still remain unexplored such as:

- Design of appropriate pre-processing stages for effectiveness of machine learning algorithms
- Proper choice of structure and methods associated with selected machine learning algorithms (here, ANN model parameters- e.g. activation function type, number of hidden layers etc.)
- Suitable post-processing methods for the predicted output

Modelling of petrophysical characteristics from well logs and seismic data plays a crucial role in petroleum exploration. Two major challenges are faced while interpreting and integrating different kinds of datasets (mainly, well logs and seismic data); 1) nonlinear and diverse nature of reservoir variables associated with the subsurface systems, and 2) absence of any direct relationship between seismic and well log signals from a theoretical perspective. Similarly, calibration of a functional relationship between a reservoir characteristic and predictor seismic attributes is an intricate task. Linear multiple regression and neural networks are popular among statistical techniques for reservoir modelling from well logs and seismic attributes [4], [33]. Lately, several computation intensive Artificial Intelligence methods such as ANN, neuro-fuzzy, Self-Organizing Map (SOM), committee machine and Learning Vector Quantization (LVQ) have attained recognition as the potential tools to solve nonlinear and complex problems in the domain of reservoir characterization [4], [5], [11], [29], [31], [34]–[38]. Despite the difference in theory and computation, most of the modelling algorithms are applied for the same purpose. On the contrary, a single technique can serve as a potential tool for solving different problems. For instance, ANN has been applied in several areas of science and technology for different objectives such as prediction, classification, etc. Moreover, different categories of neural network architectures are capable to solve nonlinear problems; however, as complexity of the problem increases due to enhancement in the number of inputs or the complex nature of the predictor variables, the performance of the network decreases rapidly.

Several researchers have claimed that the application of modular networks and hybrid networks show better performance compared to a single algorithm [39]–[44]. In particular, due to the heterogeneous nature of the subsurface system (or reservoir variables), application of a single network for complete depth range of a well may not be sufficient to achieve adequate prediction accuracy. In this context, module based networks, so-called modular artificial neural networks (MANN), are well suited to solve complex nonlinear problems. Moreover, the concept of modularity is applied in many fields to divide a complex problem into a set of relatively easier sub-problems; then, the smaller sub-problems are solved by modules; finally, the obtained results are combined to achieve the solution of the main problem [40], [45]–[47]. The module-wise division is carried out based on different clusters and classes of the dataset. This modularity concept is implemented individually or along with another machine learning algorithm to solve different types of problem. Lithological information extraction from an integrated dataset is a major step in the reservoir modelling.

MANN has been previously used in reservoir characterization domain. Fung *et al.*, [29] used to predict petrophysical properties from a set of well logs. The smaller networks are constructed corresponding to different classes obtained from a trained LVQ network. However, seismic attributes are not considered in the study by Fung *et al.*, [29]. A similar work has been carried out to predict permeability from multiple well logs such as spectral gamma ray, electrical resistivity, water saturation, total porosity etc. recorded from four closely spaced boreholes using modular neural network [37]. The dataset is divided into three sets– 70%, 15%, and 15% for training, validation, and testing respectively. Despite improvement in the prediction results compared to a single network, this study suffers some inherent limitations. Firstly, seismic attributes are not considered as predictor variables in this study. Secondly, blind prediction of the modeled lithological property (permeability) is not carried out. Thirdly, selection of the number of networks and number of hidden layer neurons are not guided by any particular theory. The best network is finalized by trail-and-error framework. These limitations can be addressed to design a framework to model petrophysical properties from seismic attributes from a data set of diverse lithological nature along the depth.

In oil exploration, different lithological classes, clusters, zones of interest in terms of well tops and horizons are identified from the preliminary analysis of well logs and it is integrated with seismic data of the same region. In the past, many classifications and clustering techniques have been used to make different classes of data depending on their variability [29], [48]–[50]. Recently, the concept of chaotic time series data analysis namely dynamic programming [51], synchronization methods [52], [53] are applied to assess similarity between the pattern of the well log data, and which lead towards the identification of similar zones among the wells. Generally, zonation of the logs is carried out manually by experienced geoscientists. Above nonlinear approaches are aimed to provide information of similar patches in the log data or similar zones in different wells, and hence have potential application in the reservoir characterization. In the present study, well tops are identified from a combination of well logs and accordingly different zones are marked on the log data. In the literature, it has been claimed that modular (multi-nets) systems have the advantage of being easier to understand or modify as per requirement. Geoscientists' guided zone-wise division of a well log can assist in target evaluation after training several models using zone wise divided training patterns yielding improved prediction accuracy [29], [50]. Hence, a study can be carried out on module wise prediction of a reservoir property to achieve improved performance.

These modelling of reservoir characteristics from seismic data and well logs are carried out using state-of-art nonlinear approaches such as ANN, Fuzzy Logic (FL), Genetic Algorithm (GA), etc. Some applications of these methods in the field of petroleum reservoir modelling are discussed in [4], [54]–[56]. However, it has been observed that the accuracy in reservoir modelling can be improved using classification-based approaches [57]. The prediction of class labels of petrophysical properties from well logs and seismic attributes can be beneficial for reservoir studies. Depending on the distribution of the lithological properties, a dataset can be classified into two categories – balanced and imbalanced. For example, sand fraction i.e. per unit sand volume within the rock varies from zero to unity. On the other hand, water saturation, oil saturation etc. have imbalanced distributions skewed at one and zero respectively. Water saturation, which represents the fraction of formation water presents in the pore space, is an important reservoir characteristic. Now, it is a complex task whose performance depends on the available subsurface information. Supervised classifiers are generally selected over unsupervised clustering algorithms due to the complex nature of the problem. Nevertheless, the requirement of a complete and representative training dataset is must for accurate learning of these supervised classifiers. Supervised algorithms can be useful in case of classifying a balanced dataset. However, in case of an imbalanced dataset, the aforementioned constraints of the training dataset do not get satisfied. Moreover, the underrepresented training dataset may have several class distribution skews. Lately, the learning problems from imbalance datasets have received interests from researchers due to existence of such dataset in "real-world applications" [58]–[61]. Kernel-based methods have gained acceptance in classification of imbalanced dataset over other supervised classification methods, especially in remote sensing fields [62]–[64]. Support vector data description (SVDD) is a latest kernel-based algorithm which has attracted attention from researchers of different fields for its ability in learning without any a priori knowledge on distribution of dataset [65]–[67]. However, in the domain of reservoir characterization, the classification of an imbalanced dataset using one-class classification approach has not been attempted. Therefore, the study related to imbalanced dataset classification can be addressed in the scope of this thesis.

## 1.2 Research Issues

From the above cited literature review, the issues pertaining to the accurate reservoir characterization are as follows:

- **Integration of dataset**: Preparation of the master dataset combining information acquired from different sources prior to modelling and classification of lithological properties. For example, well logs and seismic attributes are collected by different methods with different sampling rates, resolutions. The problem of non-unique sampling of well log and seismic data, different scales of seismic, well logs and other reservoir data should also be adequately handled by developing generalized methodologies that may be independent of the target reservoir characteristic.
- **Thin reservoir units**: Area with thin–bedded stacked reservoirs (sand/shale units) makes it hard to identify the changes in the sand/shale fraction.
- **Poor data quality**: Data acquired from a study area with poor data quality or a limited number of well controls and seismic coverage is not helpful to carry out reservoir characterization. It is difficult to design prediction or classification based models using a poor data set. Pre-processing methodologies are to be fine-tuned to accommodate this fact. Uncertainties associated with acquired dataset also contribute to poor performance of a designed model.
- **Information content**: In case of designing a machine learning model to predict lithological properties from seismic inputs, a key challenge is the information content of the predictor variables. If the information content of the predictor variables is less than that of the petrophysical properties then a trade-off between the amount of information required and actual amount of retrieval possible, has to be carried out according to the information theory.

## 1.3 Objectives

The primary objective of this work is the development of algorithms to obtain the functional relationships between predictor seismic attributes and target lithological properties (e.g. sand fraction, water saturation etc.). This can further be divided into:

- Integration of the seismic and borehole datasets

- Addressing the issue related to the mismatch of the information content of the seismic attributes and lithological properties
- Prediction of the variation of the lithological properties from seismic attributes
- Prediction of class labels of the lithological properties from seismic attributes

## 1.4 Contributions of the Thesis

In this thesis, sand fraction and water saturation are used as target variables to develop prediction and classification based frameworks from well logs and seismic attributes.

The major contributions of this thesis are as follows:

- Development of a pre-processing scheme to improve the prediction capability of machine learning algorithms by information filtering for prediction of a lithological property (e.g. sand fraction) from seismic attributes
- Development of a complete framework to carry out well tops guided prediction of sand fraction from seismic attributes
- Development of a classification framework to classify water saturation from well logs
- Modification of the aforementioned classification framework to classify water saturation from seismic attributes

## 1.5 Organization of the Thesis

The thesis is organized as follows. Chapter 2 discusses a pre-processing scheme involving a regularization stage to improve the mapping between predictor seismic attributes and target lithological properties. Chapter 3 presents two frameworks to model sand fraction from multiple seismic attributes with and without well tops information respectively. Chapter 4 describes a one-class classification based framework to classify water saturation from well logs. In other words, the class labels variation of water saturation (Class low/ Class high) has been predicted from well logs. Then, a modification of the framework involving seismic attributes as predictor variables has been discussed. Chapter 5 concludes the thesis with the discussion and the future possibilities of research.

# Chapter 2. Pre-processing

This thesis describes the development of algorithms to obtain the functional relationships between predictor seismic attributes and target lithological properties. Seismic attributes are available over a study area with lower vertical resolution. Conversely, well logs and lithological properties are available only at specific well locations in a study area with high vertical resolution. If a functional relationship can be calibrated between seismic signals and lithological properties at available well locations, then distribution of these properties across the study area can be predicted from available seismic information.

The thesis addresses the issues of handling the information content mismatch between predictor and target variables and proposes regularization of target property prior to building a prediction model in the pre-processing stage. In case of building a machine learning model for classification, prediction related problems, the pre-processing stage plays a crucial effect on the performance of the model. This stage is also very important to interpret the nature of the working dataset.

This chapter presents a pre-processing scheme to improve the prediction of a lithological property from multiple seismic attributes using machine learning and information filtering. The pre-processing framework includes signal reconstruction, data normalization, and target signal regularization. The main contribution of this research work is attributed to the aforementioned regularization scheme. The available data of lithological properties belong to the high-resolution well logs and has far more information content than the low-resolution seismic attributes. Therefore, regularization schemes based on Fourier Transform (FT), Wavelet Decomposition (WD) and Empirical Mode Decomposition (EMD) have been proposed to shape the high-resolution target lithological property for effective machine learning. The processed dataset by the proposed scheme will be used to for prediction of a lithological property from several seismic attributes.

## 2.1 Data Description

This section describes the study area and discusses about the preparations of the dataset.

### 2.1.1 Study Area

In this study, the working dataset has been acquired from a western onshore hydrocarbon field in India. Structurally, the field is located as a broad nosing feature; thus, housing the hydrocarbons between two major synclines. The hydrocarbon is present within a series of vertically stacked sandstone reservoirs individually separated by intervening shale. The average thickness of the sand layer is in the order of 5-6 m. The imaging of the seismic data and mapping between seismic attributes and well logs are difficult due to discrete sand deposition and lesser thickness of sand (only 5-6 m) in the larger depth of the subsurface (around 3000 m). The basin that contains the hydrocarbon field is an intra-cratonic basin, the basin by lava flow during the period of cretaceous age above the formerly deposited Mesozoic sediments. Deccan Trap acted as the basement for deposition of a huge thickness of Tertiary-Quaternary sediments.

### 2.1.2 Preparation of Data

A spatial database containing seismic attributes and well logs has been acquired from the study area in SEG-Y and Log ASCII Standard (LAS) format respectively. The SEG-Y file format is one of numerous standards developed by the Society of Exploration Geophysicists (SEG) for storing geophysical data [68]. On the other hand, the Canadian Well Logging Society presented LAS format to standardize the organization of digital log curves information in 1989 [69]. As the workflow is developed on MATLAB platform, the .sgy data files are converted in .mat format (MATLAB software compatible format) for MATLAB compatibility.

The database contains seismic attributes over the study area and borehole dataset at four well locations in the study area. These four wells will be hereafter referred as A, B, C, and D in terms of inlines and crosslines (xlines). The depth of each well is around 3000 meter from the ground, whereas the zone of interest varies from around 2720 meter to 2975 meter under surface. The borehole dataset contains basic logs such as gamma ray, resistivity, density along with derived geo-scientific logs such as sand fraction, permeability, porosity, water saturation, etc. These well logs are treated as one-dimensional signals for further processing in this study. On the other hand, the seismic dataset contains impedance, instantaneous frequency and seismic amplitude across the volume. The difference between the maximum displacement of a seismic wave and from the null point (point of no displacement) is defined as the seismic amplitude. It is recorded over a study area by converting the mechanical energy due to the motion of seismic wave through the depositional layers in subsurface into electrical energy by a geophone. The product of density and seismic velocity through different types of rock layers represents the seismic impedance. The third predictor variable i.e. the instantaneous frequency is defined as the rate of change of the phase of seismic amplitude signal [1], [70]. Sand fraction (SF) represents per unit sand volume within the rock [1]. In this study, the sand fraction is used as target lithological property.

Other geophysical techniques such as Ground Penetrating Radar (GPR) [71] is a near-surface technique mainly used in archaeological researches. Similarly, remote sensing imaginary dataset is used in detection and classification of objects on the surface or in the oceans, atmosphere. However, the range of penetration depths for these techniques are very small compared to 3000 meter i.e. depth of the wells. Therefore, these datasets cannot be used instead of seismic dataset for prediction of lithological parameters in deep subsurface.

## 2.2 Pre-Processing

Pre-processing plays a crucial role on the performance tuning of a machine learning algorithm. In this chapter, an efficient pre-processing approach is proposed as part of the adopted methodology to obtain a functional relationship between seismic attributes and sand fraction.

### 2.2.1 Signal Reconstruction

The borehole data are recorded at specific well locations along the depth with a high vertical resolution. The seismic data are acquired in the time domain where depth is measured in milliseconds two-way travel-time instead of meters. The time required for the sound wave to reach the reflector from the source and return to the receiver after hitting the reflector is termed as two-way travel time. In case of shallow reflectors, high frequencies are reflected, whereas, the lower frequency content of the sound signal penetrates the ground further down. The velocity and wavelength increase with the depth unlike the frequency. Thus, the seismic resolution reduces with increasing depth under subsurface [72]. For this particular dataset, the seismic data are collected spatially in the time domain with a sampling interval of two milliseconds. First, the well logs are converted from the depth domain to the time domain at 0.15 milliseconds sampling interval using the given velocity profile resulting from well-seismic-tie. The sampling intervals of both kinds of data are different. The seismic attributes and the sand fraction can be integrated either by upsampling the seismic signal or downsampling the latter. As a band-limited signal can be reconstructed from its samples based on Nyquist-Shannon theorem, and downsampling reduces the size of the dataset, the first option i.e. upsampling the seismic signal is opted. Hence, the band-limited seismic attributes are reconstructed at each time instant corresponding to the well logs by a sinc interpolator while adhering to the Nyquist–Shannon sampling theorem [73].

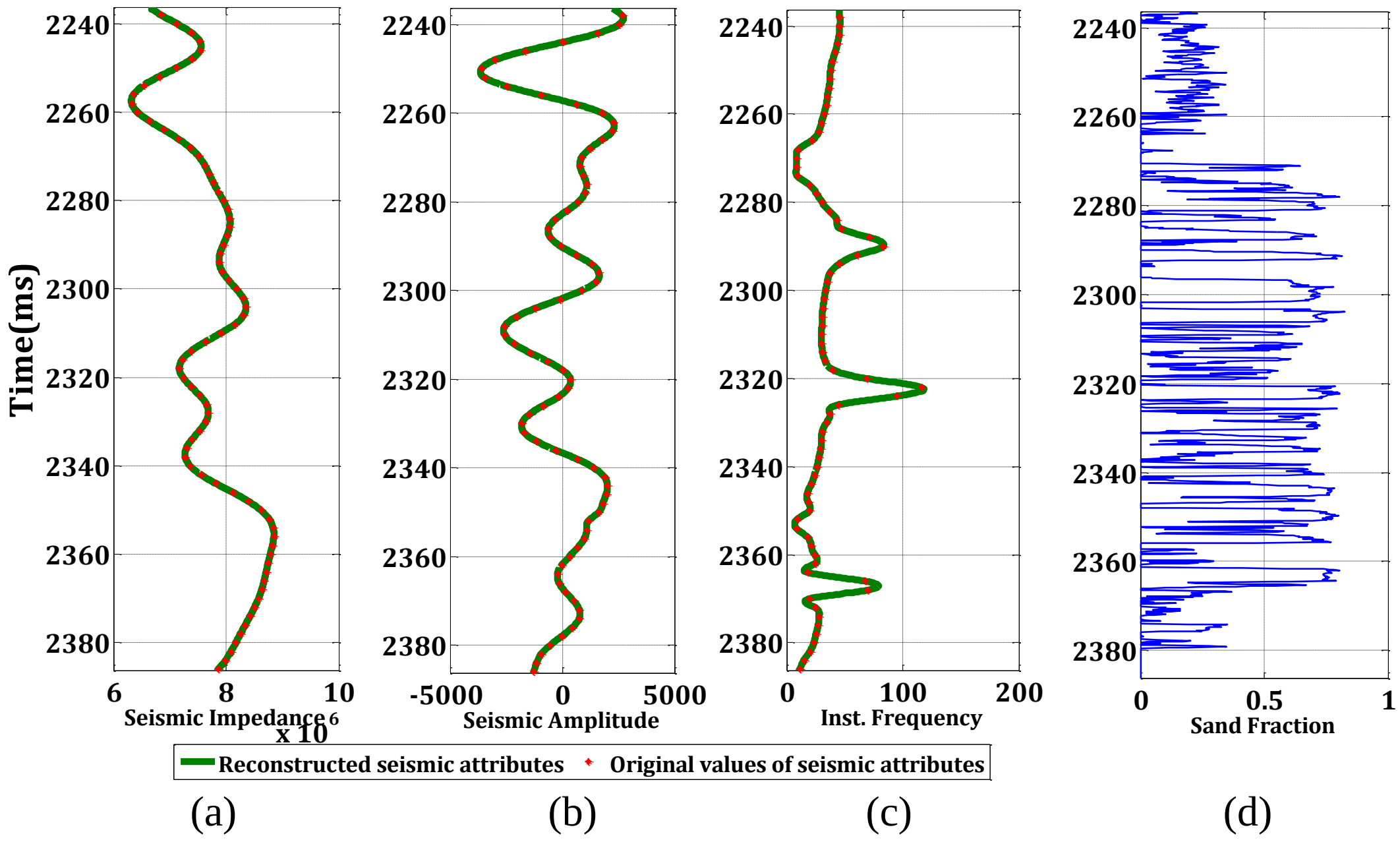

Fig. 2-1: Seismic and sand fraction signals along the well A

Fig. 2-1 represents three seismic attributes- (a) seismic impedance, (b) amplitude, (c) instantaneous frequency and the target lithological property- (d) sand fraction along the Well A. The red dots on the seismic attributes represent original values at time interval of two milliseconds and the green curves represent reconstructed signals at the time instants marked on the well logs. Fig. 2-1 (d) demonstrates a blue high-frequency curve representing sand fraction along the same well (Well A).

During the experimentation with the current dataset, we propose different normalization schemes for predictor and target variables. The predictor variables are normalized using the Z-score normalization. The values of attribute *X* are normalized using the mean and standard deviation of the *X*. The normalized value is obtained following the equation:

$$normalized_val = \frac{val - \mu_X}{\sigma_X} \qquad (2\text{-}1)$$

where $\mu_X$ and $\sigma_X$ represent the mean and standard deviation of the attribute *X*.

Then, the target variable is normalized using the min-max normalization that performs a linear transformation on original data. The relationships among the original data are preserved in this normalization.

For this particular dataset, the range of normalized target variables is selected as [0.1, 0.9]. The normalized dataset is used in the regularization stage.

### 2.2.2 Data Regularization and Re-sampling

It can be observed from Fig. 2-1 that the frequencies present in seismic signals are much lower compared to that of the sand fraction. In other words, the sand fraction carries much higher information as compared to the seismic attributes. According to laws of information theory, a higher information-carrying signal cannot be modelled using single or multiple lower information-carrying predictor signals. Only a part of the target variable that is dependent on the predictor variables can be modelled. Thus, the necessity of information filtering through regularization is established. In this chapter, three different signal processing approaches are selected and implemented in order to filter the target signal. The parameters belonging to this stage are tuned following the changes in entropy before and after filtering along with visual inspection of the output signal with respect to that of the original target signal. The entropy has been computed from the Power Spectral Density (PSD) of the signal. The average amount of information gained from a measurement that specifies $X$ is defined to be the entropy $H(X)$ of a system. It can be formally defined as

$$H(X) = -\sum_i p(x_i)\log_2 p(x_i) \qquad (2\text{-}2)$$

where $p(x_i)$ is the probability of $X$ having the $i^{th}$ value $x_i$ in the dataset.

This is known as Shannon entropy [74], [75]. If $A$ is another random variable described on the same dataset then the mutual information between the two can be expressed as

$$I(X;A) = H(X) - H(X \mid A) \qquad (2\text{-}3)$$

where, $H(X \mid A)$ is the conditional entropy of $X$ after $A$ has been observed. A reservoir property (here sand fraction) can be represented by $X$ and seismic attribute e.g. seismic impedance can be represented by $A$. The statistical property of $I(X;A)$ can be interpreted as the reduction in the uncertainty of the reservoir property, due to observing the attribute $A$. The statistical property of $I(X;A)$ can be interpreted as the reduction in the uncertainty of the reservoir property, due to observing the attribute $A$. In [76], Normalized Mutual Information (NMI) is defined as the mutual information normalized by minimum entropy of both the variables.

$$NMI(X;A) = I(X;A) / \min(H(X), H(A)) \qquad (2\text{-}4)$$

In this study, the *NMI* computed between predictor and target signal has been used to adjust the parameters of the information filtering algorithms.

**Fourier Transform (FT) based Regularization**

The first regularization approach is based on FT (Algorithm 2-1). Here, the spectrums of target and predictor variables are compared, and higher frequency components of the target signal are truncated. Then, the target signal is reconstructed using Inverse Fourier Transform (IFT). Comparing the FT of the sand fraction (Fig. 2-2(a)) with that of the seismic impedance (Fig. 2-2 (b)), the presence of higher order frequencies is evident in the former. It can be observed from Fig. 2-2 (b) that the spectrum of band-limited seismic impedance diminishes beyond frequency range (-0.2: +0.2 hertz). Then, the part of the sand fraction spectrum belonging to slightly a wider frequency range (green curve, Fig. 2-2 (a)) is reconstructed to obtain regularized target. The wider range of frequencies is chosen with the assumption that Neural Networks as nonlinear predictors are capable of mapping input signals of lower frequencies to output signals with higher frequencies. Of course it needs an entirely different research to find the prediction capability of a given nonlinear mapping process. The original and regularized target signals are presented in Fig. 2-2(c) by the blue and red curves respectively. As shown in Table 2-1, the information content of the original sand fraction is higher as compared to that of the seismic predictor variables which makes it difficult to model the target (sand fraction) from predictor attributes. The regularization process decreases the information content in the sand fraction as seen in Table 2-1. The dependency between predictor and target variables in terms of NMI (Table 2-2) also improves as a result of regularization.

Algorithm 2-1 : SF Regularization based on FT

**Task** : Regularizing target sand fraction based on FT

**Input** : Predictor signal $x(t)$ and target signal $y(t)$

a) The target $y(t)$ and predictor signal $x(t)$ are extracted from raw dataset.

b) Compute Fourier Transform of $x(t)$: $X(k) = \sum_{j=1}^{N} x(j)\omega_N^{(j-1)(k-1)}$

where, $\omega_N = e^{-2\pi i/N}$ is the $N^{th}$ root of unity

Similarly, FT of target $y(t)$ is computed as:

$Y(k) = \sum_{j=1}^{N} y(j)\omega_N^{(j-1)(k-1)}$, where, $\omega_N = e^{-2\pi i/N}$ is the $N^{th}$ root of unity

c) Compare the spectrums of target and predictor signals

d) Select the bandwidth parameter $\xi_{\max}$ Hz

e) The part of the target spectrum exceeding $\xi_{\max}$ Hz is truncated to zero.

Modified Target : $Y_{\text{mod}}(k)$

f) Construct regularized target signal $y_r(t)$ by carrying out IFT of the truncated spectrum:

$$y_r(t) = \frac{1}{N}\sum_{k=1}^{N} Y_{\text{mod}}(k)\, w_N^{-(j-1)(k-1)}$$

g) Calculate entropies of predictors (seismic attributes here) as well as original and regularized target signals (sand fraction here).

h) If entropy of regularized target is comparable with that of the predictor signal and the regularization result is satisfactory, then regularization is completed else go to step d).

**Output :** Regularized target signal $y_r(t)$

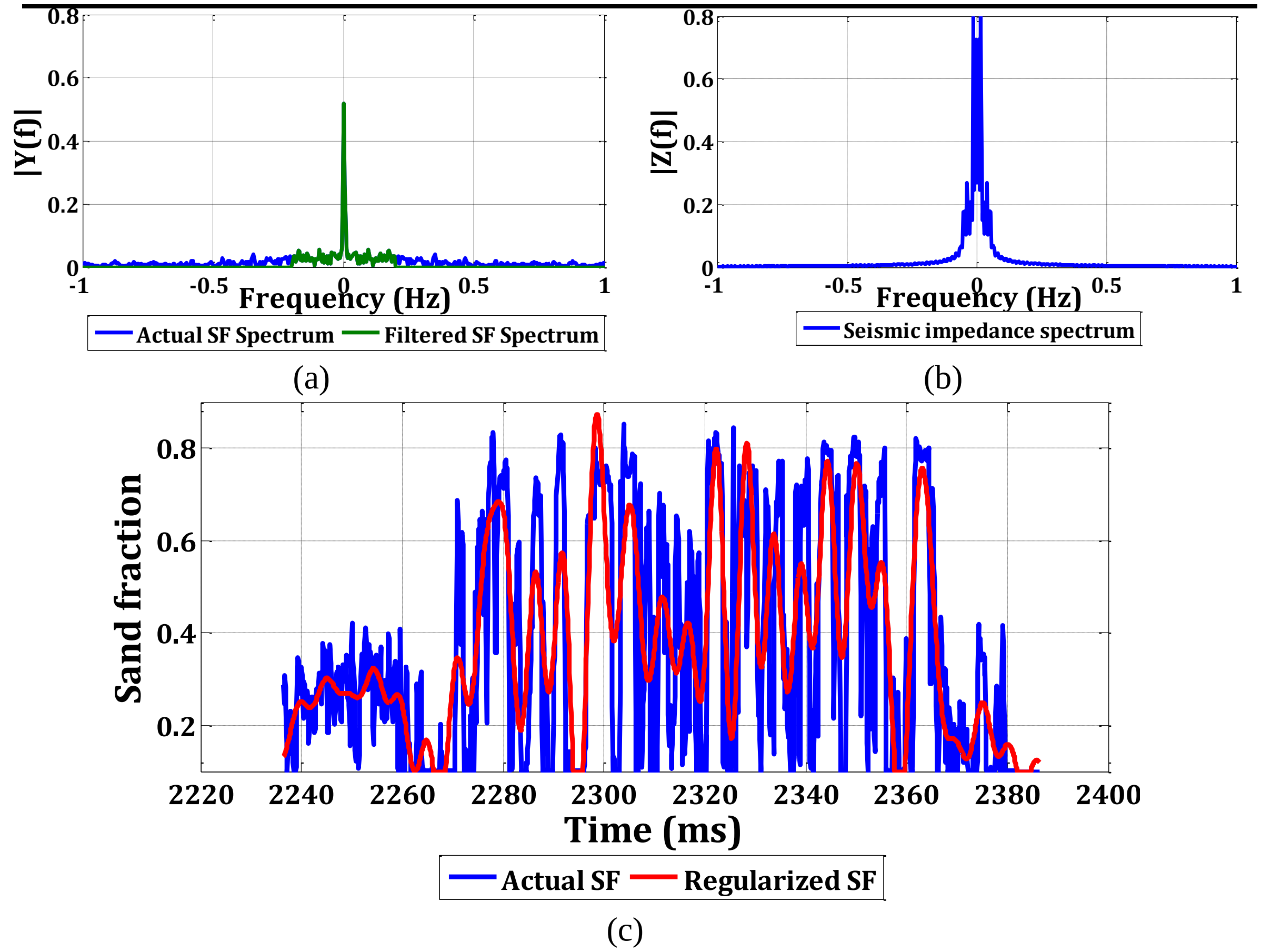

Fig. 2-2: Regularization based on FT to reconstruct sand fraction signal along Well A

Table 2-1: Entropy (in bit) of PSD of signals for Well A

| Variables | | Entropy Value |
|---|---|---|
| Seismic Impedance | | 0.15 |
| Inst. Amplitude | | 0.16 |
| Inst. frequency | | 0.12 |
| Original Target Signal | | 0.28 |
| Regularized SF | FT | 0.17 |
| | WD | 0.20 |
| | EMD | 0.24 |

Table 2-2: NMI among predictors and target sand fraction for Well A

| Predictor Variable | | Seismic Impedance | Inst. Amplitude | Inst. Frequency |
|---|---|---|---|---|
| Original Signal | | 0.12 | 0.11 | 0.09 |
| Regularized Signal | FT | 0.16 | 0.16 | 0.14 |
| | WD | 0.15 | 0.14 | 0.11 |
| | EMD | 0.14 | 0.14 | 0.12 |

**Wavelet Decomposition (WD) based Regularization**

A time-frequency representation of a one-dimensional non-stationary signal is obtained using wavelet analysis. Recent literatures reveal that the wavelet decomposition is used in different fields of research. For example, it is applied in Electroencephalography (EEG) signal analysis for artefact removal to detect the effect of sleep deprivation [77], study of geomagnetic signals [78]–[80] etc. *Wavelets* are small wave like oscillating functions that are localized in time and frequency [77], [81]–[83]. A finite energy time domain signal can be decomposed and expressed in terms of scaled and shifted versions of a mother wavelet $\psi(t)$ and a corresponding scaling function $\phi(t)$ in the discrete domain. The scaled and shifted form of the mother wavelet $\psi_{l,k}(t)$ and the corresponding scaling function $\phi_{l,k}(t)$ are arithmetically represented as

$$\psi_{l,k}(t) = 2^{l/2}\psi(2^l t - k), l, k \in R \tag{2-5}$$

$$\phi_{l,k}(t) = 2^{l/2}\phi(2^l t - k), l, k \in R \tag{2-6}$$

The original signal $X(t)$ is first decomposed into high-frequency and low-frequency components using high pass and low pass filters. After each filtering step, the output time series is down-sampled by two. The low-frequency part approximates the signal while the high-frequency part denotes residuals between original and approximate signal. At successive levels, the approximate component is further decomposed using the same set of high-pass and low-pass filters. A signal $X(t)$ can be expressed mathematically in terms of the above wavelet $\psi_{l,k}(t)$ and corresponding scaling function $\phi_{l,k}(t)$ at level $l$ as

$$X(t) = \sum_k a_l(k)\phi_{l,k}(t) + \sum_k d_l(k)\psi_{l,k}(t) \qquad (2\text{-}7)$$

where $a_l(k)$ and $d_l(k)$ are the approximate and detailed coefficients at level $l$. These coefficients are computed using filter bank approach as in [84].

Fig. 2-3 describes the steps of WD for three levels. Here, a signal is decomposed into approximate and detailed coefficients using low pass $H(k)$ and high pass $G(k)$ filters respectively. After decomposition, the coefficients can be modified. In case of signal reconstruction, the modified approximate and detailed coefficients are up-sampled by two and then convolved with respective synthesis filters and then the resulting pair is summed. Finally, modified signal is acquired following $l$ level synthesis.

The performance of wavelet analysis is dependent on the mother wavelet selection and decomposition level. The Daubechies family of wavelets has a compact support with relatively more number of vanishing moments [81]. Therefore, in most of the cases different variants of Daubechies family wavelets are used for signal analysis. The initial wavelet selections can be modified if necessary. Algorithm 2-2 describes the steps associated with WD based regularization.

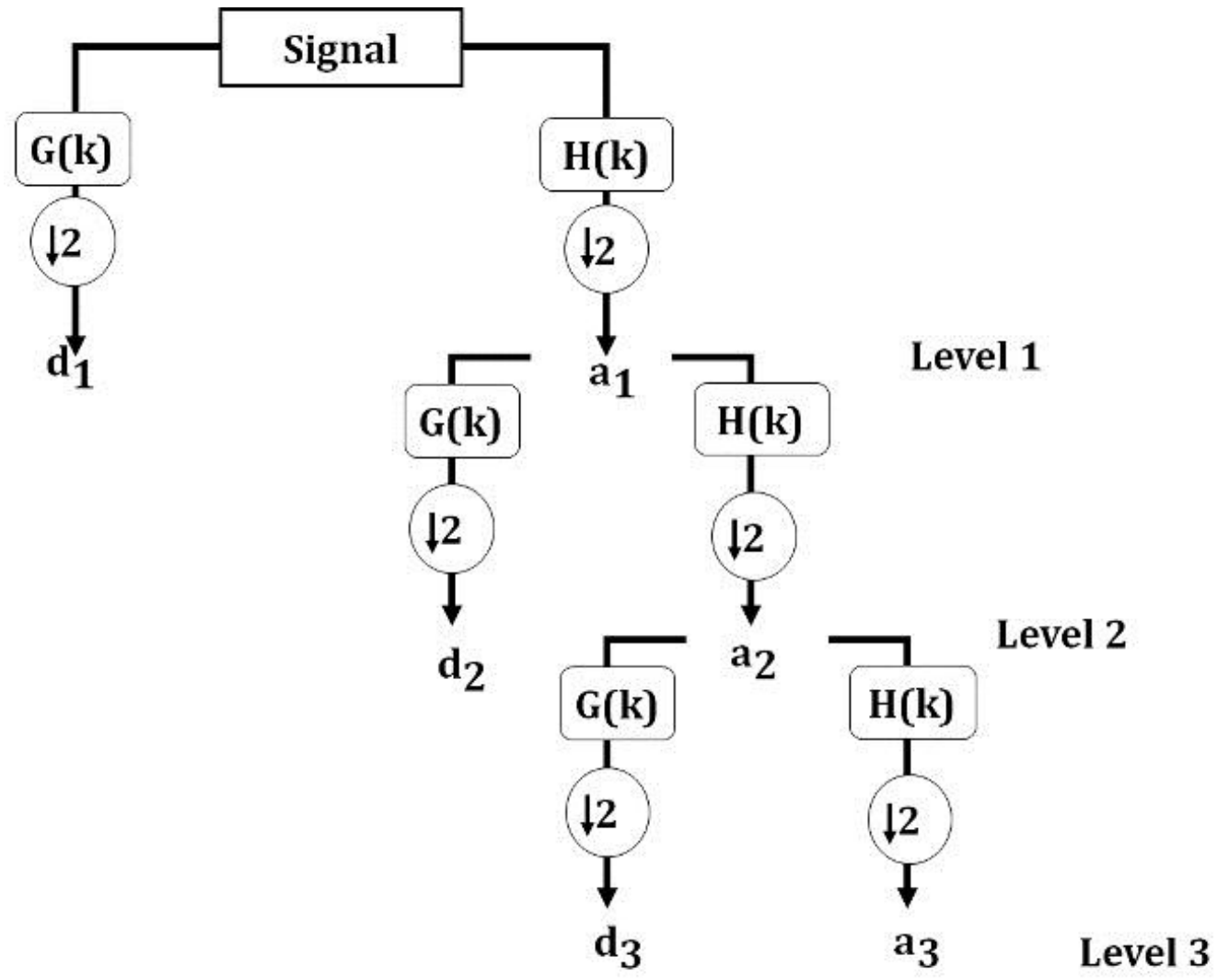

Fig. 2-3: Demonstration of wavelet decomposition of a signal for level 3

| Algorithm 2-2: SF Regularization based on WD |
|---|
| **Task** : Regularizing target sand fraction based on WD<br>**Input** : Predictor signal $x(t)$ and target signal $y(t)$<br>a) Same as Algorithm 2-1<br>b) Select the wavelet type and number of decomposition levels. |

c) Apply the procedure as in Fig.5 to the target signal.
d) Decide: detailed coefficients to be truncated for regularization by looking at the seismic amplitude and its decomposition result
e) The selected detailed coefficients are made zero.
f) The regularized target signal is reconstructed from the modified coefficients.
g) Calculate entropies of predictors as well as original and regularized target signals.
h) If entropy of regularized target is comparable with that of the predictor signal and the regularization result is satisfactory, then regularization is completed, else go to step d).

**Output :** Regularized target signal $y_r(t)$

For this study, fourth order Daubechies wavelet (db4) with six levels of decomposition has been chosen. Fig. 2-4 (a)-(b) represent the results of WD–based regularization of target sand fraction with predefined wavelet type, and decomposition level. The first three detailed coefficients of the original sand fraction signal are demonstrated in Fig. 2-4 (a).

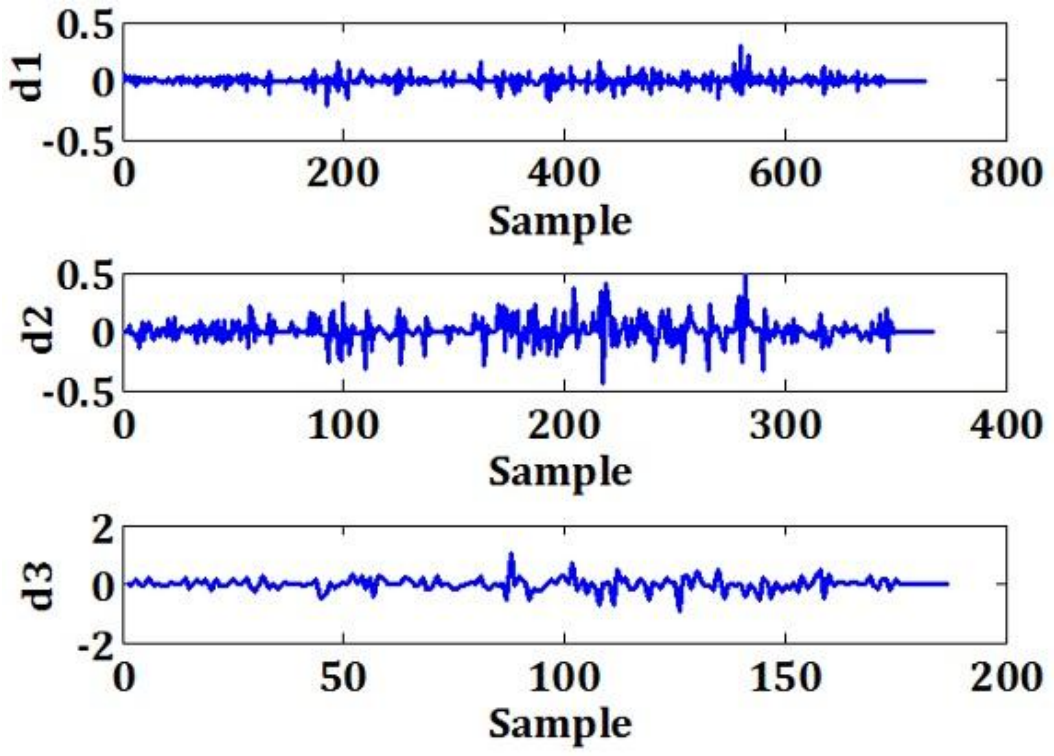

(a)

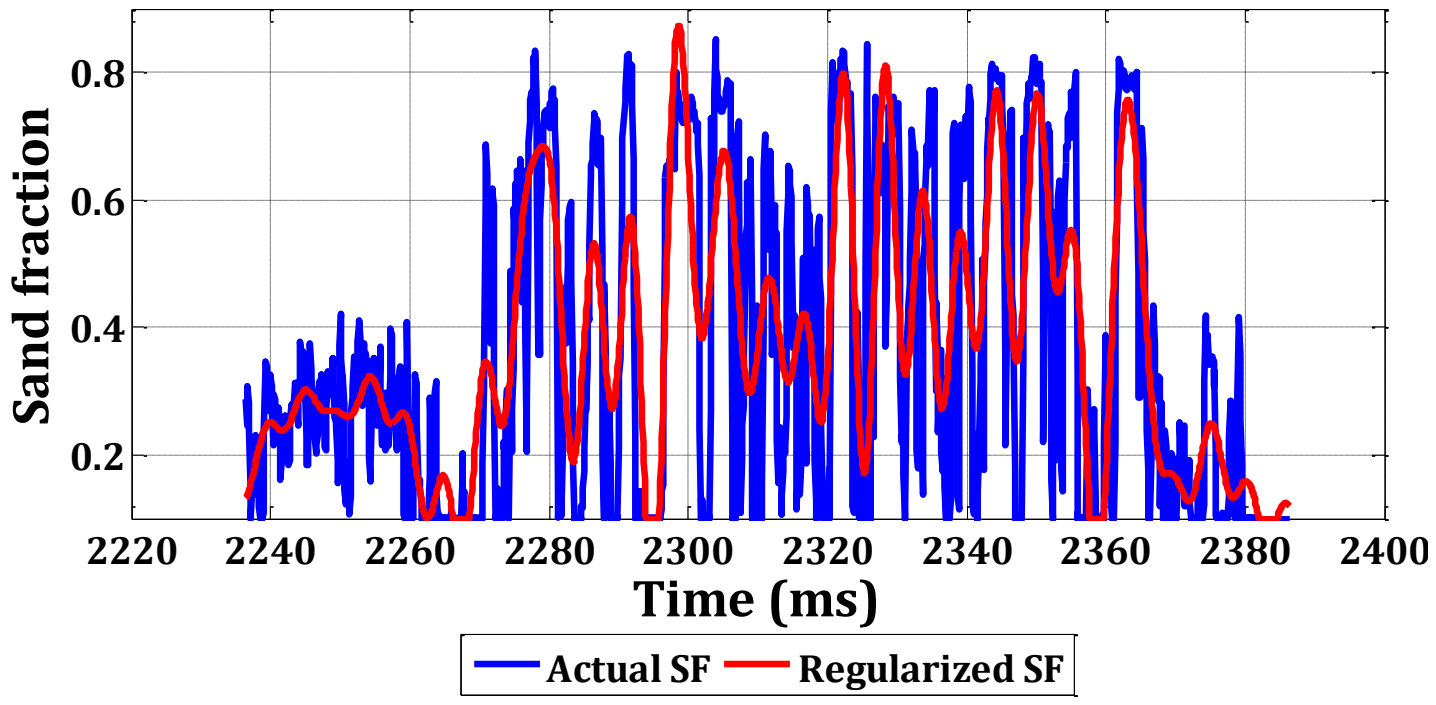

(b)

Fig. 2-4: Regularization based on WD to reconstruct the sand fraction along Well A

After the decomposition, the detailed coefficients of initial levels of the original target signal are made zero and the regularized signal is constructed by performing Inverse Discrete Wavelet Transform (IDWT) from the modified coefficients. For Well A, the first five detailed coefficients are truncated and the regularized target is reconstructed from the approximate and detailed coefficients of the sixth level by IDWT. The regularization result for Well A is presented in Fig. 2-4 (b), where the blue and red curves represent the original and regularized target sand fraction signals respectively.

Table 2-1 and Table 2-2 reveal the changes in information content of the original and regularized sand fraction by WD and increase of the dependency between target and predictor variables as a result of regularization.

### Empirical Mode Decomposition (EMD) based Regularization

Seismic and well log signals are non-stationary signals. Reports suggest that in most of the cases the frequency analysis of signals are carried out in selected windows with respect to a given orthogonal basis [85]–[87]. The disadvantage of basis decomposition techniques is the mismatch between signal trend and constant basis functions. These necessitate a new decomposition method, namely EMD.

Algorithm 2-3: SF Regularization based on EMD

**Task** : Regularizing target sand fraction based on EMD
**Input** : Predictor signal $x(t)$ and target signal $y(t)$

a) Same as Algorithm 2-1
b) Initialize: $r_0(t) = x(t)$, $i = 1$
c) Extract the $i^{th}$ IMF:
   i. Initialize: $h_0(t) = r_i(t)$, $j = 1$
   ii. Extract the local minima and maxima of $h_{j-1}(t)$
   iii. Create upper envelope $e_{\max}(t)$ and lower envelope $e_{\min}(t)$ of $h_{j-1}(t)$ by interpolating local maxima and minima
   iv. Calculate mean envelope: $m_{j-1}(t) = \dfrac{e_{\max}(t) + e_{\min}(t)}{2}$
   v. $h_j(t) = h_{j-1}(t) - m_{j-1}(t)$
   vi. If $h_j(t)$ is an IMF,
   then, $imf_i(t) = h_j(t)$
   else go to the step-(ii). with $j = j+1$
d) $r_i(t) = r_{i-1}(t) - imf_i(t)$
e) If $r_i(t)$ has at least two extrema,
   Then, go to c) with $i = i+1$
   Else, $x(t) = \sum_{i=1}^{n} imf_i(t) + r_n(t)$ is decomposed into $n$ numbers of IMFs and residue signal.
f) EMD of target signal $y(t)$ is carried out following steps b)-e)

$$y(t) = \sum_{i=1}^{p} imf_i(t) + r_p(t)$$

g) The number of intrinsic mode functions (IMF) and distribution of IMFs are observed for target $y(t)$ and predictor $x(t)$ : $p > n$

h) Decide $p_1$: number of IMF truncated from the EMD of $y(t)$ for regularization where, $p_1 < p$

i) Construct regularized target signal $y_r(t)$: $y_r(t) = \sum_{i=1}^{p_1} imf_i(t) + r_p(t)$

j) Calculate entropies of predictors as well as original and regularized target signals.

k) If entropy of regularized target is comparable with that of the predictor signal and the regularization result is satisfactory, then regularization is completed else go to step h).

**Output :** Regularized target signal $y_r(t)$

EMD is an algorithmic decomposition method which decomposes the input signal into a set of Intrinsic Mode Functions (IMFs) and a residue signal [88]. There are two properties associated with IMFs such that (1) the numbers of zero–crossings and extrema present in IMFs are same, and (2) IMFs are symmetric with respect to the local mean [88]. In other words, EMD detects and extracts the highest frequency component in the signal [81], [89]–[95]; such that, in step ($k$+1*)*, the extracted IMF contains lower frequency component compared to that extracted in step $k$. Moreover, being an adaptive data-driven method, EMD decomposes an input signal into a variable number of components. Thus, EMD overcomes the inherent limitation of deciding *a priori* the number of decomposition levels as in WD. Algorithm 2-3 describes the detailed steps associated with EMD based regularization of target SF.

In Fig. 2-5 (a), the first three IMFs of the sand fraction log along the Well A are plotted. The comparison between EMD results of the sand fraction and seismic impedance reveals that the number of IMFs obtained is higher in case of target signal (sand fraction) than its predictor counterpart (seismic-impedance). The first IMF component is suppressed and the other IMFs are used to reconstruct the regularized sand fraction. The superimposed plots of actual (blue curve) and regularized sand fraction (red curve) signals are presented in Fig. 2-5 (b).

The user decides the regularization result is satisfactory or not based on visual inspection of original and regularized target variables. The regularized target is smoother compared to the original signal; nevertheless, the trend of the original signal is preserved even after information filtering based on either of the three proposed regularization methods.

Table 2-1 represents the entropies of the predictor and target attributes for Well A. The improvement in mutual dependency between the predictor and target variables in terms of NMI is evident from Table 2-2.

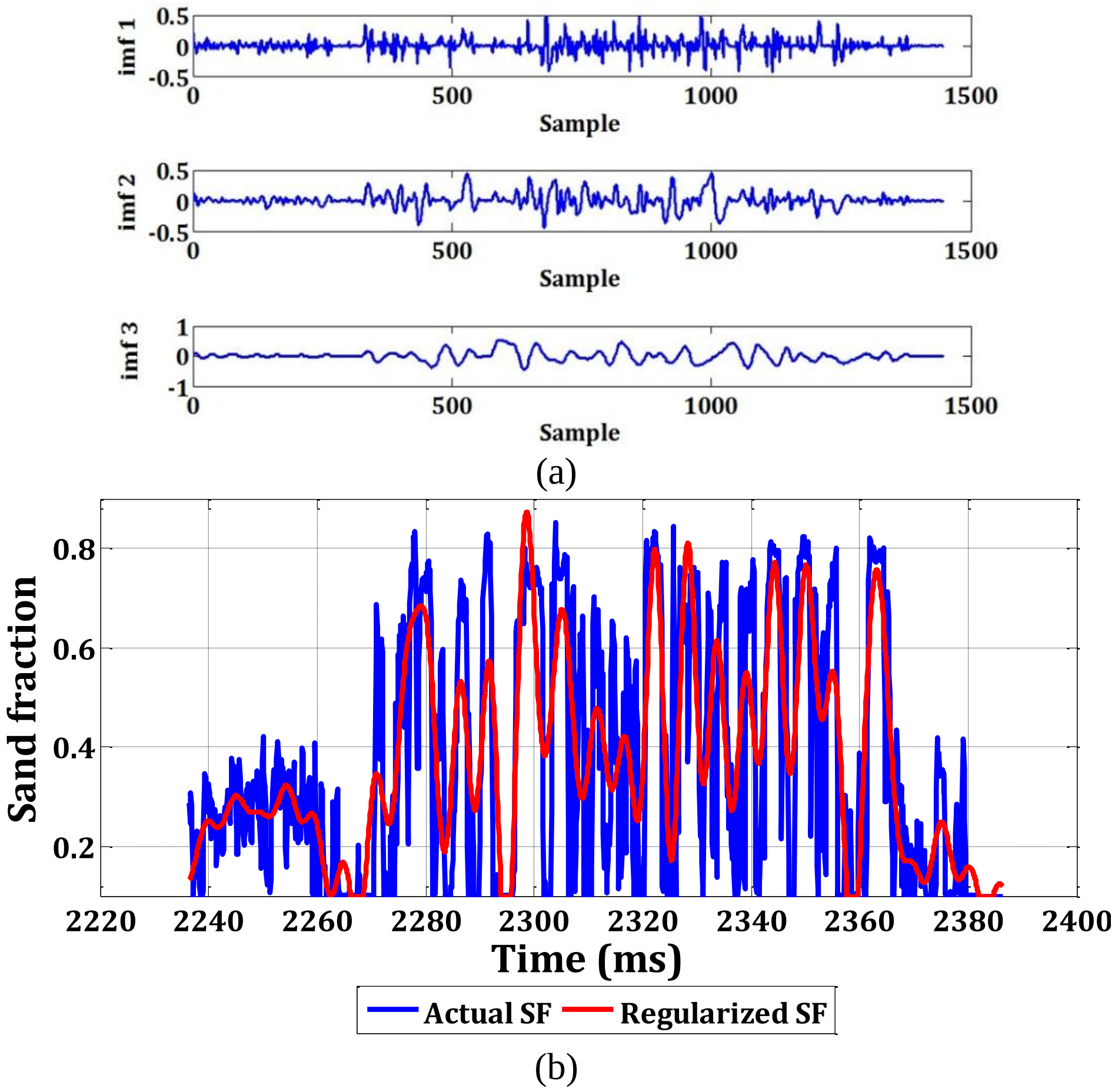

Fig. 2-5: Regularization based on EMD to reconstruct sand fraction signal along Well A

## 2.3 Conclusion

Three regularization approaches based on FT, WD, EMD are proposed and implemented to shape the target sand fraction in order to improve the predictability of the target from predictor seismic attributes. The improvement of information content and dependency between target and predictor variables are apparent from the entropy and NMI respectively. The advantages and disadvantages of the three approaches are now briefly discussed. To start with, FT is a linear transform where the frequency spectrum of a time-domain signal is obtained. Comparing the frequency spectrum of the predictor seismic impedance and target SF, the regularization parameter is selected. Depending on the selected regularization parameter the result of FT based regularization changes. However, seismic signals are non-stationary signals. Therefore, the next two approaches based on WD and EMD are opted to cross-validate the performance of FT based regularization. The selection of the wavelet and the decomposition level is important for the decomposition of the original target signal. Then, the detailed coefficients to be truncated are decided. The regularized target signal is constructed from the modified coefficients. The variable parameters i.e. wavelet type, decomposition level, and detailed coefficients to be truncated are finalized empirically based on the regularization result. However, in case of EMD, number of selection parameters are less. Only, the number of IMFs to be used for reconstruction of the target is decided. In all the three cases, it is maintained that the regularized target signal should preserve the trend of the original signal. In case of FT based approach, the regularization parameter is decided based on the FT results of both the predictor as well as the target signal. However, in the case WD, the predictor signal is regularized to observe that the frequency content of the individual coefficients are less compared to the target signal case with the selected decomposition level and wavelet type. Similarly, the number of decomposed IMFs of the seismic attributes are less compared to the SF counterpart in case of EMD. However, in both cases (WD and EMD), the number of detailed coefficients and IMFs are decided irrespective of the decomposition results of the predictor seismic signals.

In the next chapter, the regularized sand fraction is modelled using ANN based prediction frameworks. Therefore, the three regularization approaches would cross-validate each other in terms of evaluators over the network performance using original SF as target signal in Section 3.2.

# Chapter 3. Prediction of Sand Fraction

Sand fraction (SF), which represents per unit sand volume within the rock, is an important lithological property. Higher values of the sand fraction represent increased probability of porosity in a layer, and also indicate a higher probability of presence of hydrocarbon in case of a porous medium. Hence, the sand fraction is an important attribute to be modelled in reservoir characterization.

This chapter describes two different frameworks for prediction of a lithological property (sand fraction) from three predictor seismic attributes using two different datasets. The first section of this chapter describes the theory associated with ANN in brief. The next section of this chapter deals with the first set of data combining sand fraction logs at four well locations and seismic attributes logs at the corresponding well locations. The next section discusses the works with another set of data including seismic attributes from a western onshore hydrocarbon field of India and borehole data along with borehole information. In case of the second dataset, the borehole data are acquired from eight wells in the study area. Additionally, well tops at the eight well locations and horizon information over the area of interest are provided with seismic attributes and well logs.

To start with, a framework is designed to benchmark the proposed pre-processing stage as in Chapter 2 through a complete framework consisting of three stages (pre-processing, model building and validation, and finally, post-processing). The same dataset as in Chapter 2 is used in this work. An ANN with conjugate-gradient learning algorithm is used to model the sand fraction. The input data sets are segregated into training, testing and validation sets. The test results are primarily used to change the network structure and activation functions. Once the network passes the testing phase with an acceptable performance in terms of standard performance indicators, the validation phase follows. In the validation stage, the prediction is tested against unseen data. The network that yielded satisfactory performance in the validation stage is used to predict the target lithological property using seismic attributes as predictor variables throughout the given volume. Finally, a post-processing scheme using 3-D spatial filtering is carried out for smoothing the sand fraction in the volume.

Section 3.3 describes the prediction framework to predict sand fraction from the same seismic predictor attributes using the concept of MANN. At first, the acquired dataset is integrated and normalized. Then, well log analysis and segmentation of the total depth range into three different units (zones) separated by well tops are carried out. Then, three different networks are trained corresponding to three different zones using combined dataset of seven wells and then the trained networks are validated using the remaining test well data. The target property of the test well is predicted using three different tuned networks corresponding to three zones; and then the estimated values obtained from three different networks are concatenated to represent the predicted log along the complete depth range of the test well. It has been observed that the application of multiple simpler networks instead of a single one improves the prediction accuracy in terms of performance evaluators– correlation coefficient, root mean square error, absolute error mean and program execution time. Then, volumetric prediction of reservoir properties is carried out using calibrated network parameters. This stage is followed by post-processing to improve visualization. Thus, a complete framework, that includes pre-processing, model building and validation, volumetric prediction, and post-processing, is designed for successful mapping between seismic attributes and a reservoir characteristic. The proposed framework has performed better than a single ANN with reduced prediction error, program execution time and improved correlation coefficient as a result of the application of the MANN concept.

### 3.1 Artificial Neural Network (ANN)

Computers have become immensely powerful over the recent years; therefore, it has become easier to emulate a simple task carried out by a human being. For example, recognition of a particular letter can be carried out with the guidance of a teacher (*supervised learning*). On the other hand, it can be carried out without the help from a teacher; a child can learn a particular letter by correcting his mistakes several times (*unsupervised learning*). These types of simple tasks such as character recognition, differentiation between different objects are carried out by humans very easily. Similarly, these tasks can be addressed with the help of a high performance computer [96].

An ANN is an information–processing system that emulates certain performance characteristics of a biological neuron. An ANN can be characterized by three properties such as:

- ANN architecture (i.e. connections between neurons)
- Deciding the weights (by training/ learning)
- Activation functions

ANN is used in different research domains such as signal processing, control, pattern recognition, speech recognition etc. to solve different classification and prediction problems.

Multilayer ANN is a network, which processes one or several hidden layers joining input and hidden layers. Fig. 3-1 represents a typical example of a *n*-input single output multilayer ANN with a single hidden layer in it. The nodes corresponding to the hidden layer are connected to input and output nodes through weights and biases are connected to each node in the hidden and output layers. As shown in the figure, hyperbolic tangent sigmoid transfer function is used in the hidden layer nodes to facilitate the learning process [97]. In contrast, log-sigmoid transfer function is used in the output node. The hyperbolic tangent sigmoid transfer function and log sigmoid transfer function (used in output layer node) have the following forms respectively:

$$\tan sig(x) = \frac{1-e^{-2x}}{1+e^{-2x}} \tag{3-1}$$

$$\log sig(x) = \frac{1}{1+e^{-x}} \tag{3-2}$$

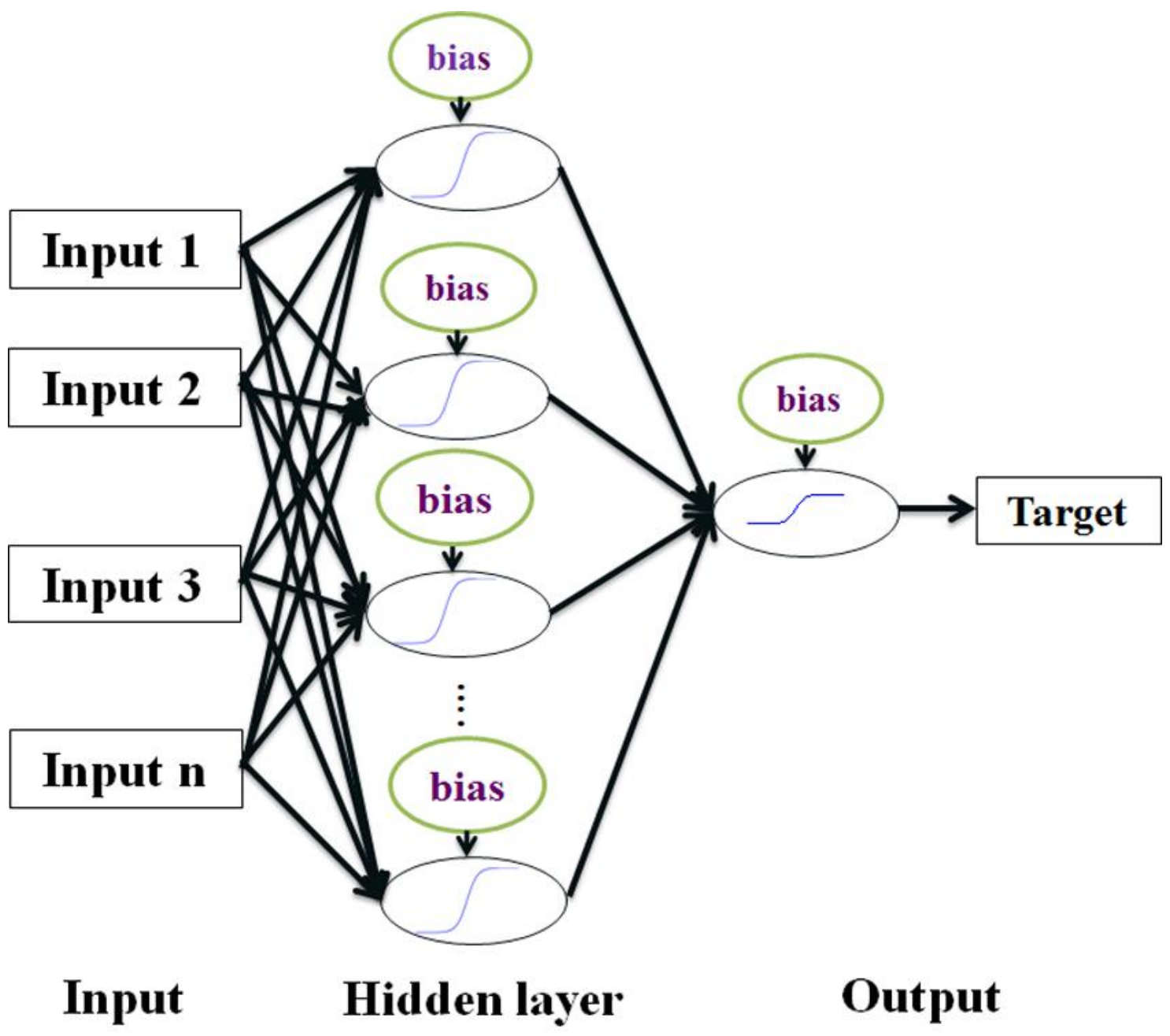

Fig. 3-1: Structure of an ANN with single hidden layer n-input single output

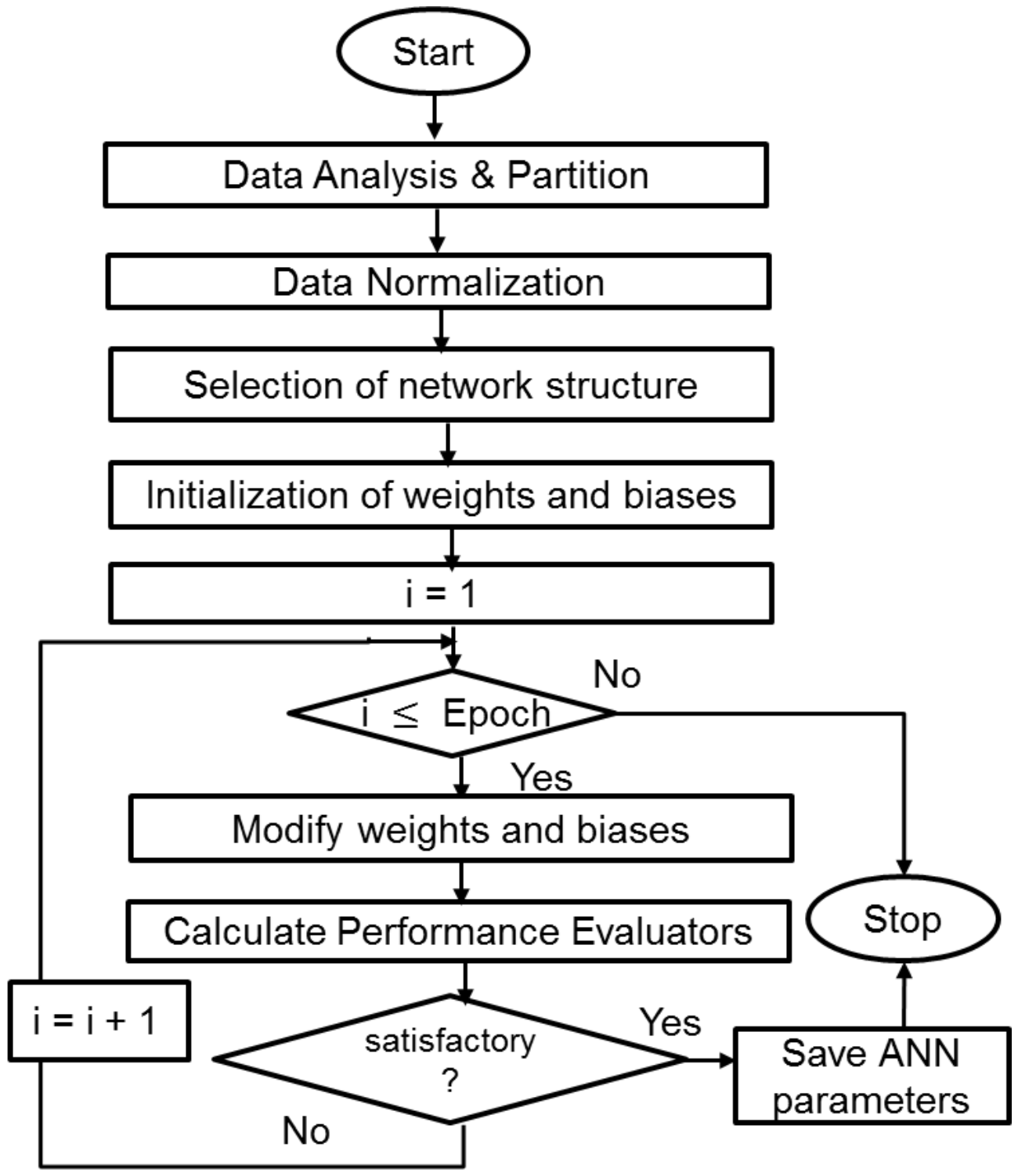

Fig. 3-2: Training framework of an ANN

The training procedure of an ANN is demonstrated in Fig. 3-2. At first, the dataset is analysed and partitioned into training and testing sets followed by dataset normalization. Then, the network structure (number of hidden layers, nodes, activation functions) are selected and weights and biases are initialized. The training starts with selected learning algorithm. The weights and biases are modified and performance evaluators (such as RMSE, CC, and AEM etc.) are calculated. The training process continues until the satisfactory values of the performance evaluators are achieved. The back-propagation learning of a multilayer perceptron is carried out in two phases. The synoptic weights of the network are constant and the input signal is propagated through the hidden layers to the output layer. The changes only take place in the activation potentials and output of the neurons [13]. In contrast, an error signal is computed as the difference between network output and actual (desired) target (response). The error signal is propagated through layers in backward direction from output layer to input layer in the backward phase. The training procedure is described in details as follows.

Assume,

$$\tau = \{x(n), d(n)\}_{n=1}^{N} \qquad (3\text{-}3)$$

to be the training samples to be used for training of a back propagation multilayer perceptron. The input vector $x(n)$ is applied to the input layer nodes and the desired output vector $d(n)$ is present at the output node to compute the error between desired and actual output. The stages associated with the back-propagation algorithm can be summarized as follows:

1. Initialization: The synaptic weights and thresholds are selected from a uniform distribution with zero mean. The variance of the distribution is selected such that the standard deviation of the induced local fields of the neurons lie at the transition between the linear and standard sections of the sigmoid activation function.
2. Presentations of training patterns: The forward (step 3) and backward propagation (step 4) are carried out for each sample of the training datasets.
3. Forward propagation: The forward signal is propagated from input layer to output layer through one or multiple hidden layers. The induced local field for neuron $j$ in layer $l$ can be computed from the output of neuron $i$ in the previous layer $(l-1)$ at iteration $n$ i.e. $y_i^{l-1}(n)$ and the synaptic weight $w_{ji}^{l}(n)$ that is connected from neuron $i$ in $(l-1)$ layer and is expressed as

$$v_j^{(l)}(n) = \sum_i w_{ji}^{(l)}(n) y_i^{(l-1)}(n) \qquad (3\text{-}4)$$

In case of output layer (here, $l = L$ and $L$ is the network depth), the output of neuron $j$ is written as

$$y_j^{(L)}(n) = o_j(n) \qquad (3\text{-}5)$$

Therefore, the error is computed as

$$e_j(n) = d_j(n) - o_j(n) \qquad (3\text{-}6)$$

where, $d_j(n)$ is the $j$ th element of the desired response vector $d(n)$.

4. Backward computation: In this step, the local gradients of the network are computed as:

$$\partial_j^{(l)}(n) = e_j^{(L)}(n)\varphi_j^{'}(v_j^{(L)}(n)) \quad (3\text{-}7)$$

for neuron $j$ in output layer $L$ and

$$\partial_j^{(l)}(n) = \varphi_j^{'}(v_j^{(l)}(n))\sum_k \partial_k^{(l+1)}(n) w_{kj}^{(l+1)}(n) \quad (3\text{-}8)$$

for neuron $j$ in output layer $l$

where, the prime in $\varphi_j^{'}(\bullet)$ represents the differentiation with respect to the argument.

The synaptic weights of the network in layer $l$ according to the generalized delta rule

$$w_{ji}^{(l)}(n+1) = w_{ji}^{(l)}(n) + \alpha[w_{ji}^{(l)}(n-1)] + \eta\partial_j^{(l)}(n) y_i^{(l-1)}(n) \quad (3\text{-}9)$$

where, $\eta$ and $\alpha$ are the learning-rate parameter and the momentum constant respectively.

5. Iterations: The forward and backward propagations are carried out until the selected stopping criterion is reached. The training examples are randomized in each epoch and the momentum and learning-rate parameter are modified.

The supervised learning of a multilayer ANN can be viewed as a problem of *numerical optimization*. The error surface of a multilayer ANN is a nonlinear function of weight vector $w$. Assume that, *the error energy averaged over the training samples* or the *empirical risk* be $\xi_{avg}$. Using (3-6), $\xi_{avg}$ can be computed as

$$\xi_{avg} = \frac{1}{2N}\sum_{n=1}^{N}\sum_{j\in C} e_j^2(n) \quad (3\text{-}10)$$

where, the set $C$ contains all the neurons in the output layer. The second derivative of the cost function $\xi_{avg}$ with respect to the weight vector $w$ is called the Hessian matrix and denoted by $H$ so that,

$$H = \frac{\partial^2 \xi_{avg}}{\partial w^2} \quad (3\text{-}11)$$

The Hessian matrix is considered as positive definite unless mentioned. There are several algorithms to train an ANN. In case of conjugate gradient methods, the computational complexity and memory usage are large because of calculation and storage of the Hessian matrix at each stage. The indefiniteness of $H$ by a scalar parameter $\lambda_k$ in case of the scaled conjugate gradient (SCG). The other parameters $\tilde{r}_k$ and $p_k$ represent the search direction and the steepest descent direction respectively. In this work, SCG algorithm is selected over other supervised algorithms to train the ANN. The steps associated with SCG can be presented as follows [98]:

1. Selection of parameters: Weight vector $w_1$ and $0<\sigma\le 10^{-4}$, $0<\lambda_1\le 10^{-4}$, $\bar{\lambda}_1=0$.

   Assume, $p_1=r_1=-E'(w_1), k=1$ and success = true
2. If success = true, then compute the second-order information:

$$\sigma_k=\sigma/\left|p_k\right|$$

$$\tilde{s}_k=(E'(w_k+\sigma_k p_k)-E'(w_k))/\sigma_k \tag{3-12}$$

$$\partial_k=p_k^T\tilde{s}_k$$

3. Modify $\partial_k$:

$$\partial_k=\partial_k+(\lambda_k-\bar{\lambda}_k)\left|p_k\right|^2 \tag{3-13}$$

4. If $\partial_k\le 0$, then make the Hessian matrix positive definite

$$\bar{\lambda}_k=2(\lambda_k-\partial_k/\left|p_k\right|^2)$$

$$\partial_k=-\partial_k+\lambda_k\left|p_k\right|^2 \tag{3-14}$$

$$\lambda_k=\bar{\lambda}_k$$

5. Evaluate step size:

$$\mu_k=p_k^T\tilde{r}_k \tag{3-15}$$

$$\alpha_k=\mu_k/\partial_k$$

6. Compute the comparison parameter:

$$\Delta_k=2\partial_k[E(w_k)-E(w_k+\alpha_k p_k)]/\mu_k^2 \tag{3-16}$$

7. If $\Delta_k\ge 0$, then error can be reduced:

$$w_{k+1}=w_k+\alpha_k p_k$$

$$\tilde{r}_{k+1}=-E'(w_{k+1}) \tag{3-17}$$

$$\bar{\lambda}_k=0,\ \text{success = true}$$

   If $k$ mod $N=0$ then

$$p_{k+1}=\tilde{r}_{k+1} \tag{3-18}$$

   else:

$$\beta_k=(\left|\tilde{r}_{k+1}\right|^2-\tilde{r}_{k+1}^T\tilde{r}_k)/\mu_k \tag{3-19}$$

$$p_{k+1}=\tilde{r}_{k+1}+\beta_k p_k$$

   If $\Delta_k\ge 0.75$, modify the scale parameter,

$$\lambda_k=0.25\lambda_k \tag{3-20}$$

   Else

$$\lambda_k = \lambda_k, \text{ success} = \text{false} \tag{3-21}$$

8. If $\Delta_k < 0.25$, increase the scale parameter

$$\lambda_k = \lambda_k + (\partial_k (1 - \Delta_k) / |p_k|^2) \tag{3-22}$$

9. Check: If the steepest descent direction $\tilde{r}_k = 0$, then $w_{k+1}$ as the desired minimum.

   else $k = k + 1$ and goto Step 2.

The value of the parameter $\sigma$ is kept small indicating that it is not critical for the performance of SCG. The fact that SCG does not involve any user dependent parameter that is critical for its performance is an advantage of this algorithm.

### 3.2 Prediction of Sand Fraction from Seismic Attributes without Well Tops Information

The integrated dataset of seismic attributes and sand fraction at four well locations, which is prepared following the pre-processing stage as in Chapter 2, is used in this study. The predictor attributes used in this study are three seismic variables such as seismic amplitude, seismic impedance, and instantaneous frequency and the target variable is sand fraction. The available borehole data along with the 3-D seismic attributes have been used to benchmark the proposed pre-processing stage using a methodology (Fig. 3-3) which consists of three stages, i.e., pre-processing, training and post-processing. The available sand fraction belongs to the high-resolution borehole data and has far more information content than the low-resolution seismic attributes. Therefore, three alternative regularization schemes based on FT, WD and EMD as described in Chapter 2 have been used in the pre-processing stage to shape the high-resolution sand fraction data for effective machine learning.

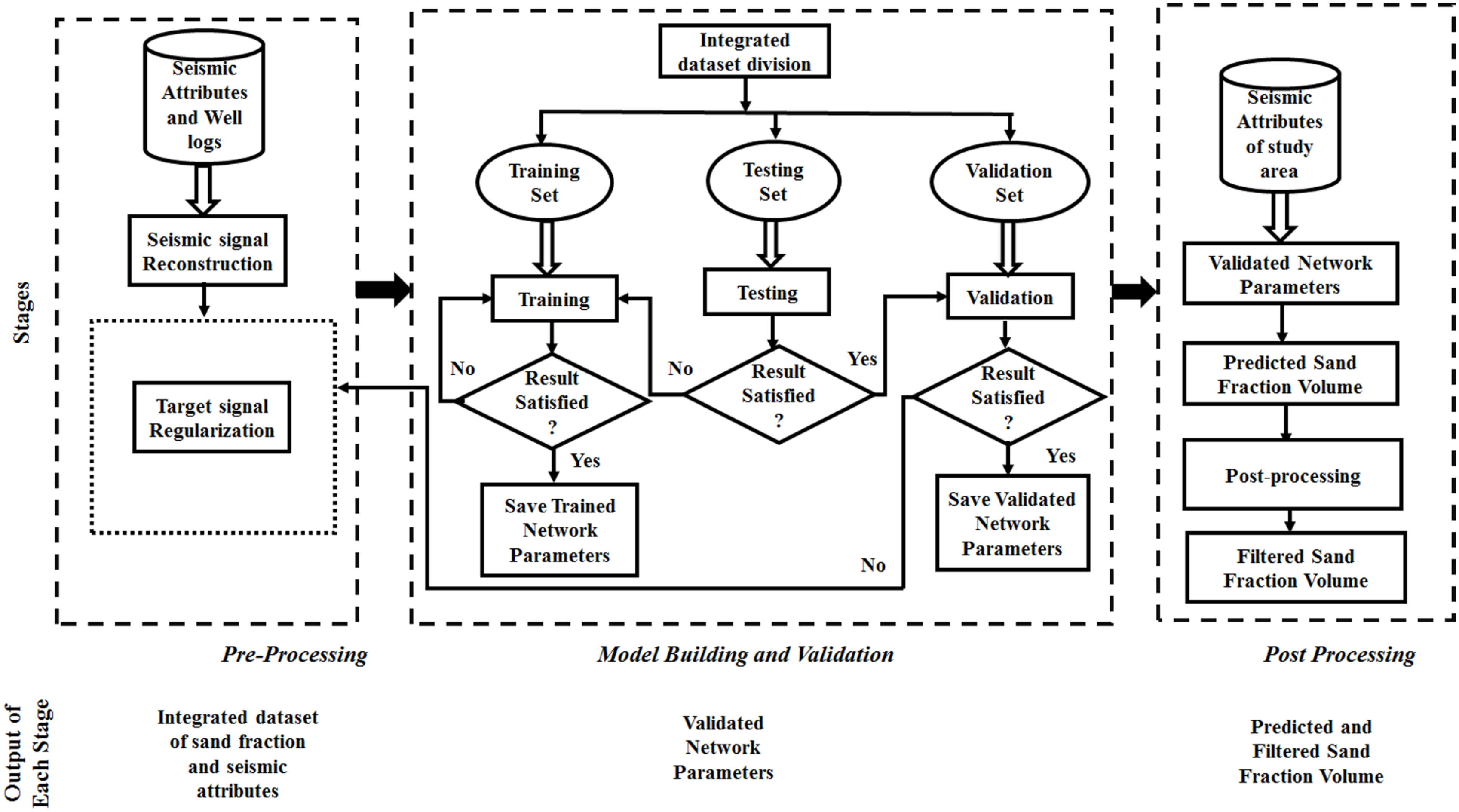

Fig. 3-3: The prediction framework to benchmark regularization stage

### 3.2.1 Pre-processing

Pre-processing plays a pivotal role in the performance fine-tuning of a machine learning algorithm. The detailed pre-processing scheme consisting of seismic signal reconstruction and target lithological properties regularization by three alternative signal processing approaches based on FT, WD, and EMD respectively is described in details in Chapter 2. The pre-processed dataset is used here for model building and validation.

### 3.2.2 Model Building and Validation

The task of model building and validation stage is implemented to evaluate the network performances using original and regularized sand fraction as target variable, respectively. In each case, the seismic inputs and sand fraction values are normalized using the data from all four wells taken together. The target variable is normalized within the range of output activation function with some offset from the limiting value of the activation function. Otherwise, the back propagation algorithm tends to drive the free network parameters to infinity. As a result, the learning process will slow down [13]. Hence, the target variables are normalized between 0.1 and 0.9 to avoid any overlap with the saturation region of the log-sigmoid function. Then, the training dataset is created by aggregating 70% sample patterns from each of the wells. The training patterns are scrambled to remove any trend along the depths. The remaining 30% samples from each of the four wells are combined and scrambled again. Then, the samples are divided into two parts to create the testing and the validation datasets. First, the network is trained using training patterns with initial parameter values. Then, the network structure and activation functions are tuned using testing patterns.

The testing phase is important for evaluating the generalization capability of the trained network [28]. The network that performs satisfactorily in terms of performance evaluators is then chosen to enter the validation stage. The performance of the trained networks is evaluated using four parameters –CC, RMSE, AEM, and SI. The scatter index represents the ratio of RMSE to mean of in situ observations [24]. These statistical characteristics are defined as

Correlation coefficient (CC): $CC = \sum_{i=1}^{N}(X_i - \overline{X_i})(Y_i - \overline{Y_i}) / \sqrt{\sum_{i=1}^{N} X_i - \overline{X_i})^2 (Y_i - \overline{Y_i})^2}$

Root mean square error: $RMSE = \sqrt{\sum_{i=1}^{N}(X_i - Y_i)^2 / N}$

Absolute error mean: $AEM = \frac{1}{N}\sum_{i=1}^{N}|X_i - Y_i| = \frac{1}{N}\sum_{i=1}^{N}|e_i|$

Scatter index: $SI = RMSE / \overline{Y}$

where, $X_i$ and $Y_i$ ( i = 1, 2, 3,…,N) represent modelled and observed values, respectively, $\overline{X}$ and $\overline{Y}$ are their corresponding mean values. Total number of data points are $N$ and $e_i$ denotes absolute error.

The statistical analysis of the errors involved in the model is important for proper understanding of the performance. The initial network structure is decided intuitively depending on the nature of the problem and the amount of available training patterns. In this study several runs of training, testing and validation of neural network structures with varying number of neurons, layers, activation functions and learning methods have been carried out to decide the best structure as well as the most effective learning algorithm. Different network structures for each instance are experimented and systematically changed keeping the improvement direction in view. Finally in the hidden layer, hyperbolic tangent sigmoid transfer function has been used. The tangent sigmoid transfer function is an automatic choice for researchers to use in the hidden layer to achieve the bi-directional swing [97], [99]. The activation function used in the output layer is log-sigmoid which is non-symmetric. It is reported that the learning rate of the network is faster when the network is anti-symmetric [13].

Finally, a network with a single hidden layer is trained using the Scaled Conjugate Gradient (SCG) Backpropagation Algorithm [98]. The advantage of this algorithm is that it does not contain any user-dependent parameters. Moreover, it is faster than other second order algorithms as it avoids the time-consuming line search per learning iteration by using a step size scaling mechanism. The number of nodes in the input layer is same as the number of predictor attributes to be used to model the ANN. For example, in case of predicting sand fraction from three predictor attributes– namely, seismic impendence, instantaneous frequency and seismic amplitude, the number of input and output nodes will be three and one respectively.

Table 2-2 reveals that the NMIs between the instantaneous frequency and SF (original/regularized) are relatively lower compared to other cases. In the first attempt, two input attributes (seismic impedance and amplitude) have been used to build a prediction model and the corresponding results are documented in Table 3-1 in terms of the performance evaluators. Though, the variation of instantaneous frequency is comparatively lower, it is an important attribute. Therefore, all three seismic attributes have been used as inputs to the prediction model (Table 3-2) in the second attempt. The results reported in Table 3-1 and Table 3-2 lead to two important observations.

- The performance of the trained networks is improved while using regularized target signals. This is quantified in terms of higher CCs and lower error values.
- In all cases, inclusion of the instantaneous frequency as the third predictor improved the prediction.

It can be observed from Table 3-1 and Table 3-2 that the network performance is superior in case of regularization based on WD with two predictor variables. On the other hand, with three predictors FT based regularization outperformed the other two regularization approaches in terms of performance evaluators. However, for all cases, the performance is improved while using the regularized sand fraction as target instead of the original log. It can be envisaged that a user can select any of the three proposed regularization techniques that best suit with the working dataset.

Table 3-1: Statistics of validation performance (Two predictors)

| Well Name | Original Target Signal | | | | Regularized Target Signal by FT | | | | Regularized Target Signal by EMD | | | | Regularized Target Signal by WD | | | |
|---|---|---|---|---|---|---|---|---|---|---|---|---|---|---|---|---|
| | CC | RMSE | AEM | SI | CC | RMSE | AEM | SI | CC | RMSE | AEM | SI | CC | RMSE | AEM | SI |
| A | 0.63 | 0.21 | 0.17 | 0.7 | 0.69 | 0.15 | 0.12 | 0.52 | 0.71 | 0.17 | 0.13 | 0.64 | 0.74 | 0.15 | 0.12 | 0.53 |
| B | 0.57 | 0.19 | 0.15 | 0.6 | 0.63 | 0.15 | 0.12 | 0.5 | 0.62 | 0.16 | 0.12 | 0.49 | 0.65 | 0.15 | 0.12 | 0.5 |
| C | 0.68 | 0.16 | 0.12 | 0.53 | 0.78 | 0.12 | 0.09 | 0.37 | 0.76 | 0.12 | 0.08 | 0.37 | 0.77 | 0.11 | 0.08 | 0.36 |
| D | 0.54 | 0.18 | 0.14 | 0.58 | 0.63 | 0.14 | 0.11 | 0.46 | 0.64 | 0.12 | 0.09 | 0.41 | 0.66 | 0.12 | 0.09 | 0.39 |

Table 3-2: Statistics of validation performance (Three predictors)

| Well Name | Original Target Signal | | | | Regularized Target Signal by FT | | | | Regularized Target Signal by EMD | | | | Regularized Target Signal by WD | | | |
|---|---|---|---|---|---|---|---|---|---|---|---|---|---|---|---|---|
| | CC | RMSE | AEM | SI | CC | RMSE | AEM | SI | CC | RMSE | AEM | SI | CC | RMSE | AEM | SI |
| A | 0.76 | 0.17 | 0.13 | 0.62 | 0.94 | 0.07 | 0.05 | 0.23 | 0.92 | 0.09 | 0.06 | 0.31 | 0.93 | 0.08 | 0.06 | 0.28 |
| B | 0.65 | 0.17 | 0.13 | 0.56 | 0.86 | 0.09 | 0.07 | 0.30 | 0.88 | 0.09 | 0.06 | 0.28 | 0.87 | 0.10 | 0.08 | 0.34 |
| C | 0.76 | 0.14 | 0.11 | 0.45 | 0.91 | 0.07 | 0.05 | 0.24 | 0.87 | 0.08 | 0.06 | 0.25 | 0.89 | 0.08 | 0.05 | 0.25 |
| D | 0.68 | 0.15 | 0.12 | 0.49 | 0.83 | 0.1 | 0.07 | 0.33 | 0.80 | 0.09 | 0.06 | 0.32 | 0.84 | 0.08 | 0.06 | 0.28 |

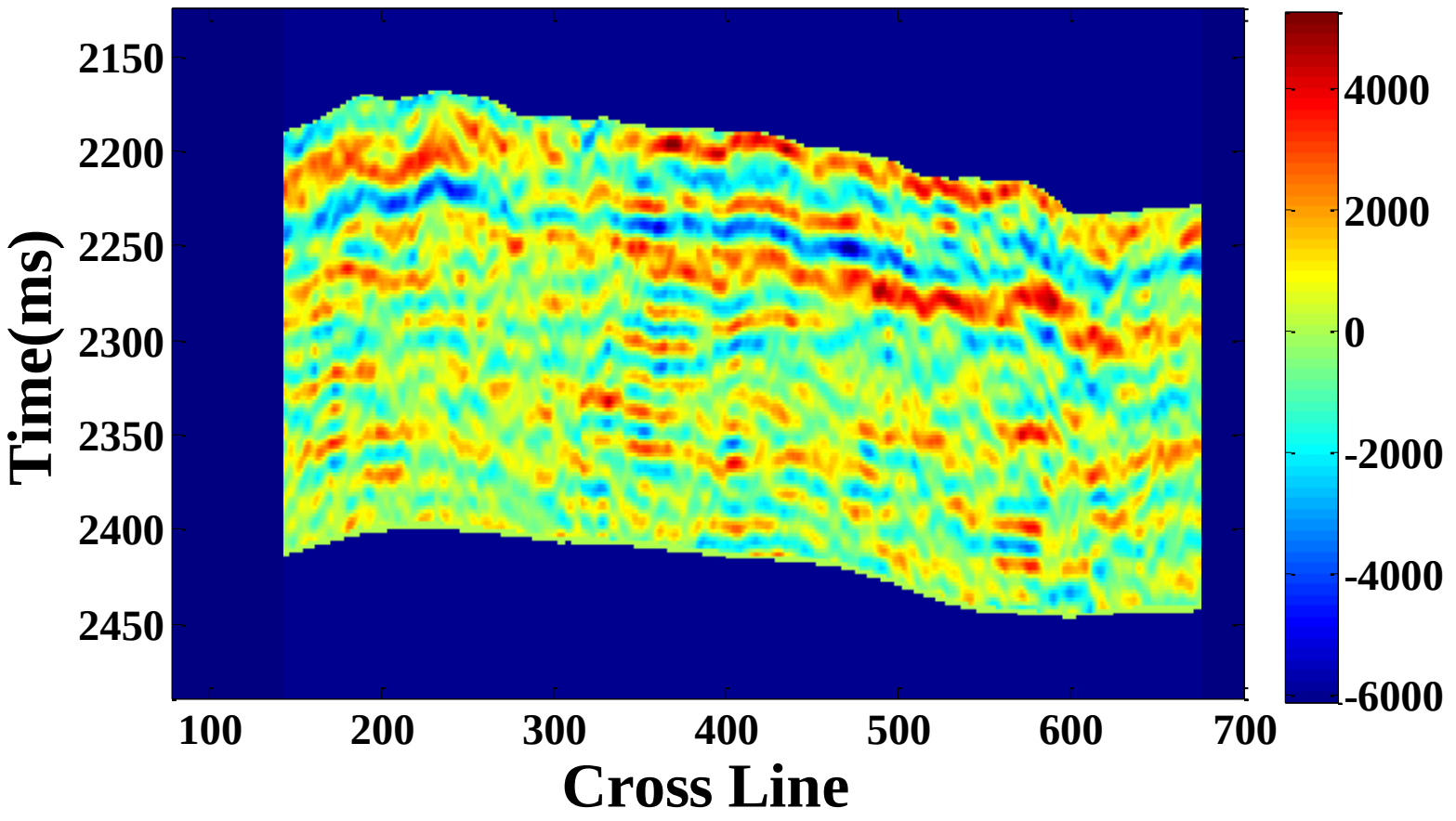

Fig. 3-4: Visualization of amplitude at inline 136

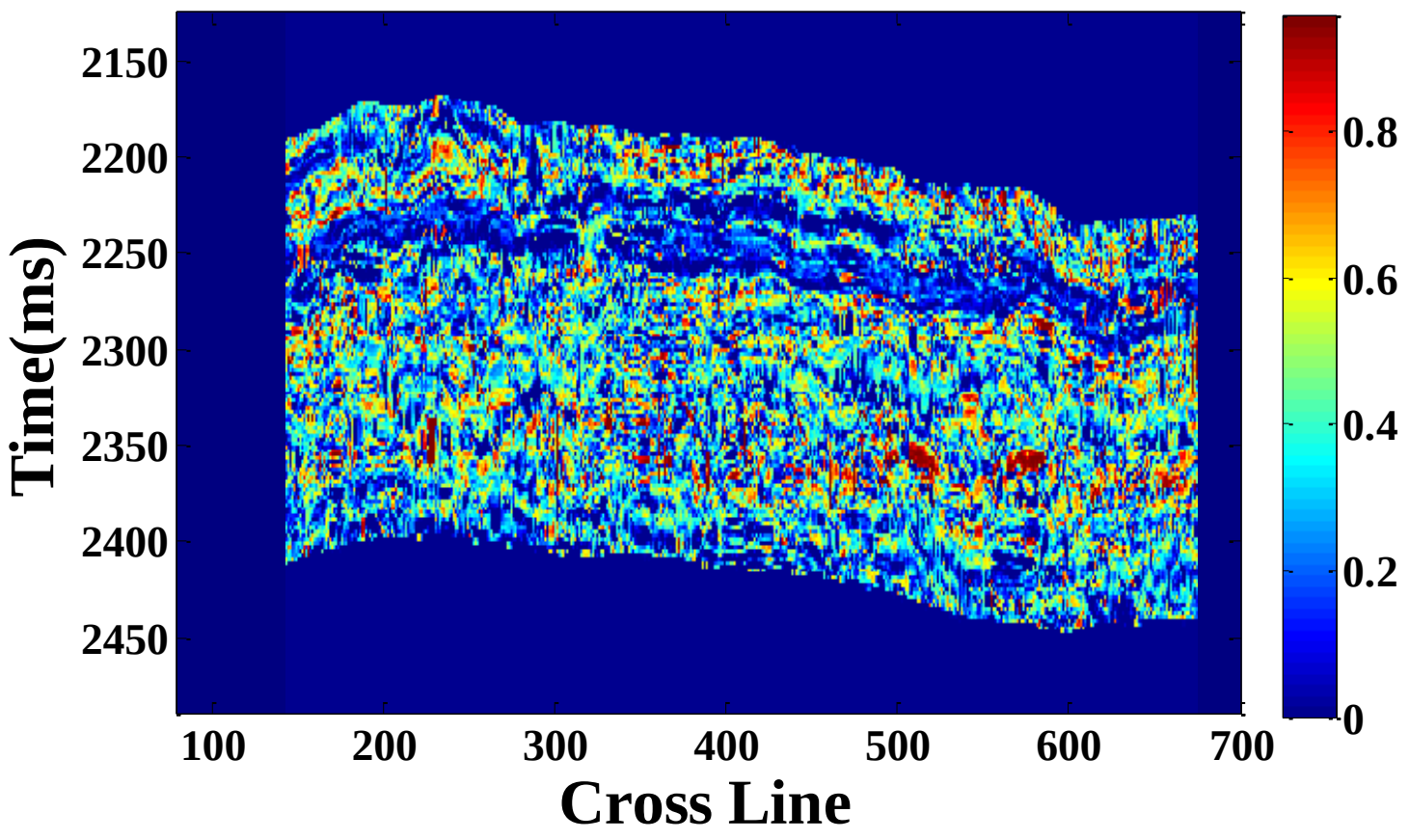

Fig. 3-5: Result of prediction of sand fraction at inline 136

The reported results reveal that the prediction accuracy increases with the inclusion of all available seismic attributes as predictors and regularized sand fraction value as target. The prediction in the entire volume along each inline and cross line is done using the trained network finalized during the validation step. Fig. 3-4 represents the variation of the seismic amplitude at a specific inline (inline 136 containing Well D). The sand fraction variation at the same inline (inline 136) as in Fig. 3-5 is obtained by prediction of the sand fraction over the study area from available seismic attributes using the validated network parameters. The network used for prediction over the study area has been calibrated using EMD–regularized–sand fraction as target. The reported results in Table 3-1and Table 3-2 are better than the performances of multilayer perceptron, SVM, and Co-Active Neuro-Fuzzy Inference System (CANFIS) for permeability modelling in terms of CC as reported in [100]. Thus, it can be said that improved performance is achieved for testing with unseen data using the framework described in Section 3.2.

It can be envisaged from the predictor seismic attributes as in Fig. 3-4 that the variation of the predicted sand fraction over the study area is smooth. However, the sand fraction across the study area as shown in Fig. 3-5 changes abruptly. Therefore, need of the post–processing stage is established to obtain a smooth sand fraction variation across the volume.

### 3.2.3 Post Processing

In reality, the sand fraction across the volume cannot change abruptly. The transition should be smoother and more or less agree to the patterns of seismic data. To incorporate this rationale, the predicted values are filtered through a 3-D median filter.

Median filter [101], [102] is a popular order-statistics filter where the value of a pixel is replaced by the median of a neighbourhood centred at that particular pixel. Selection of window value is crucial for the degree of smoothing.

| | | |
|---|---|---|
| $x_{j-1,k-1}$ | $x_{j,k-1}$ | $x_{j+1,k-1}$ |
| $x_{j-1,k}$ | $x_{j,k}$ | $x_{j+1,k}$ |
| $x_{j-1,k+1}$ | $x_{j,k+1}$ | $x_{j+1,k+1}$ |

Fig. 3-6: Representation of a pixel in 2-dimensional space with eight adjacent points

Fig. 3-6 represents a pixel $x_{j,k}$ and eight points surrounding it at location $(\mathrm{j},\mathrm{k})$. For a two dimensional with window size $(2M+1)\times(2N+1)$, the complete pixel vector around the centre pixel $x_{j,k}$ is $\{x_{j-M,k-N}, x_{j-M+1,k-N}, \ldots, x_{j,k}, \ldots, x_{j+M,k+N}\}$. The pixel at centre location $(\mathrm{j},\mathrm{k})$ is replaced by median value of the pixel vector. If $M=1, N=1$, the matrix will be filtered using a $3\times3$ window. In case of 3-D median filter having a, the median of the pixel vector within $3\times3\times3$ window size can be evaluate after carrying out six $3\times3$ 'partial-sort operation' [103].

The predicted sand fraction in the volume is used as input to the post-processing operation. Every element in the volume is considered as a pixel and is smoothened using 3-D median filter with respect to its neighbourhood within a 3x3x3 window size. The missing values along the boundaries are ignored.

The sand fraction value along inline 136 is extracted from the smoothened sand fraction 3-D volume. Fig. 3-7 represents the result of median filtering along inline 136. The effect of localizing different levels of sand fraction values can be observed by comparing Fig. 3-5 and Fig. 3-7.

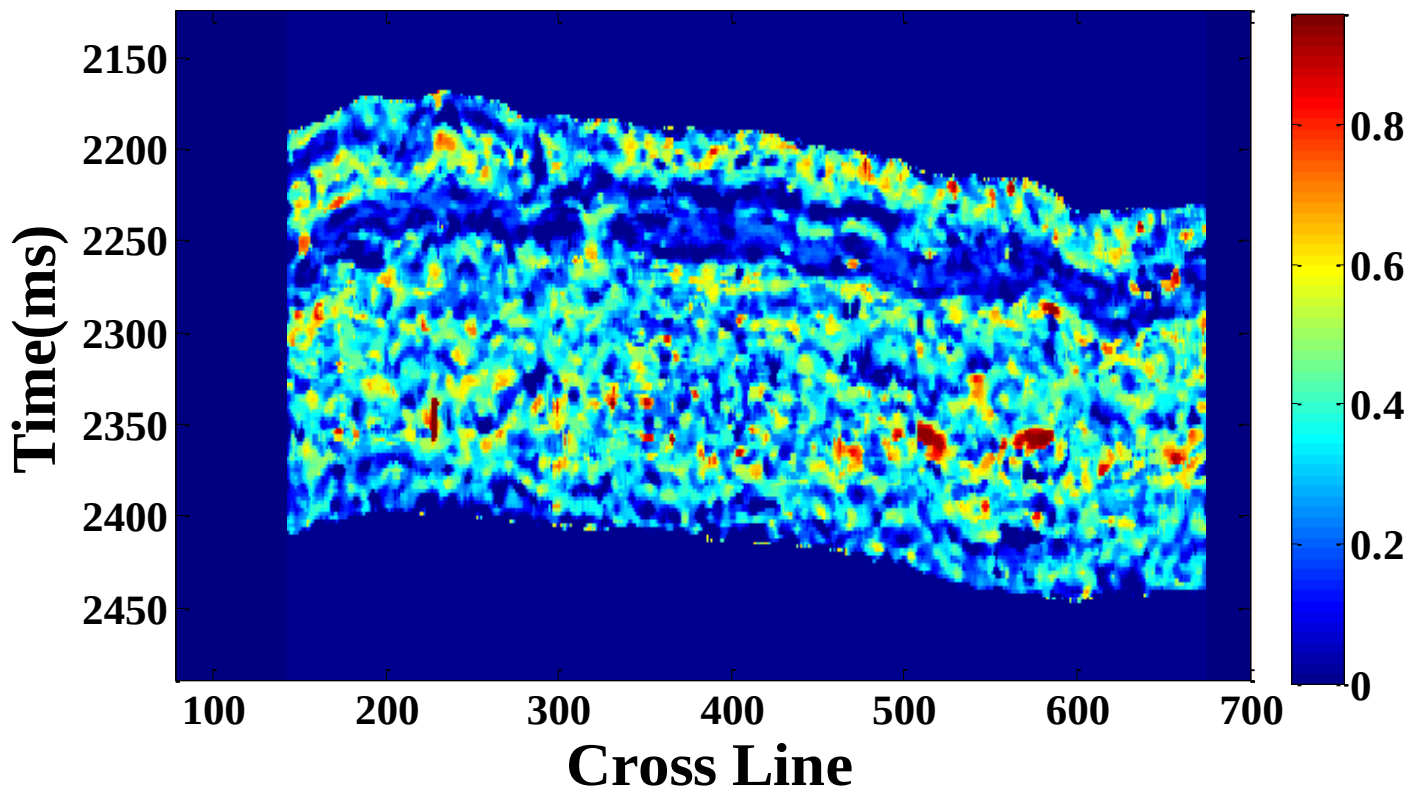

Fig. 3-7: Result of median filtering on sand fraction at inline 136

Thus, the complete framework including pre-processing, learning and validation, and finally post-processing, successfully carries out mapping between seismic attributes and the sand fraction.

### 3.2.4 Discussion

Chapter 2 brings out an elegant regularization step in pre-processing to enhance the learning capability of ANN to carry out mapping between seismic and lithological property (sand fraction) successfully. Then, the improvement in mapping between seismic attributes and target sand fraction with regularization step is established from the performance analysis Section 3.2. Another contribution of this study is the improvement in the predicted sand fraction over the volume using 3-D spatial filtering.

## 3.3 Prediction of Sand Fraction from Seismic Attributes with Well Tops Guidance

Section 3.3 proposes a complete framework consisting pre-processing, modelling, and post-processing stages to carry out well tops guided prediction of a reservoir property (sand fraction) from three seismic attributes (seismic impedance, amplitude, and instantaneous frequency) using MANN algorithm.

We can encapsulate the work done in this section as a motivated outcome of the concepts of modularity and synchronization together. The idea of well tops guided division of the dataset is evolved from synchronization or similarity. The similarity between well logs sections belonging to a certain horizon is more compared to that of the similarity between multiple complete length logs. Then, the modularity concept enables to divide a complex problem into a set of relatively simple sub-problems. The borehole data are available at specific well locations; whereas seismic attributes are acquired over the area of interest. If a functional relationship can be established between seismic and well log signals (petrophysical properties), then, the variation of the reservoir characteristic over the area can be predicted from the seismic attributes itself. As the predictor and target signals are from two different domains, therefore, a single ANN structure may not be able to successfully represent the mapping function between these two types of signals, which is the current research problem in this domain [37]. In this section, we have attempted to devise a complete framework, with the objective of overcoming the limitations of the previous studies.

In this study, two well tops (namely Top1 and Top2) are identified after analysing well logs and seismic data. In the process of mapping sand fraction from seismic attributes, first, seismic attributes are extracted from 3D seismic cube at eight well locations. Next, integration of seismic and borehole data are carried out using time-depth relationship information at the available well locations. The pre-processed master dataset is then divided into three zones based on the two well tops such as 1$^{st}$ available patterns to Top 1, Top 1 to Top 2, and Top 2 to last available data pattern. In the model building and validation stage, first, three networks have been designed for three different zones separately. Sand fraction and three seismic attributes corresponding to seven wells are used for training and testing, and patterns corresponding to the remaining well are used for blind prediction. The satisfactory performance in blind testing encourages to carry out volumetric prediction of sand fraction using the three trained models. Then, results evaluated by three different networks (zone-wise) are merged to form a volumetric cube containing the estimated sand fraction values across the study area along the entire depth range. After model building and validation, a post-processing is carried out to improve the visualization quality.

### 3.3.1 Methodology based on MANN Concept

In recent years, ANN is widely used to solve nonlinear modeling problems in the fields of science and technology such as computer science, electronics, mathematics, geosciences, medicine, physics, etc., [104]. ANN and its hybrid approaches have also proven to be useful in the nonlinear mapping of reservoir properties from well logs and seismic data [104], [105]. ANN performs satisfactorily in non-linear data mapping, pattern recognition and classification problems; however, the execution speed is slow in cases involving large dataset. Therefore, there is a significant scope of improvement in order to accelerate the training process by modifying the basic algorithm while not compromising with the prediction accuracy. Hence, MANN, which is a special category of ANN based on data categorization, is introduced as a potential tool for machine learning with efficient estimation capability and high speed [29], [50], [105]–[107]. The concept of modularity is derived from the principle of divide and conquers. Here, a complex computational task is subdivided into smaller and simpler subtasks. Each local computational model performs an explicit, interpretable and relevant job according to the mechanics of the problem involved. Finally, the output of the model will be the combination of individual results of dedicated local computational systems. In this approach, module wise networks are trained and tested, and the outputs of all modules are integrated to achieve complete sequence of the target variable.

*Description of Dataset*

The well logs and seismic data used in this section are acquired from a hydrocarbon field located at the western onshore of India. The borehole dataset includes basic logs such as gamma ray, resistivity, density and other derived logs such as sand fraction value, permeability, porosity, water saturation, etc. Conversely, the seismic dataset includes different attributes, i.e., seismic impedance, instantaneous frequency, seismic envelope, seismic sweetness, etc. This section involves seismic impedance, amplitude and instantaneous frequency to model sand fraction using an integrated dataset of seismic and sand fraction signals available at eight well locations.

The lithological properties along the depth range of a well vary in a non-linear and heterogeneous fashion. The variations of lithological properties along depth for well 4 and well 5 can be observed in Fig. 3-8. Two well tops namely Top 1 (red line) and Top 2 (green line) are shown in the figure. Integrated dataset corresponding to seven wells are used for learning of the three zone-wise prediction models and data from the eighth well is used for validation purpose.

Present study discusses the application of MANN concept for prediction of a reservoir property from seismic attributes. Well tops represent abrupt changes in the log data that corresponds to the changes in lithology denoting the corresponding zone boundaries. In this case, two well tops (Top 1 and Top2) are marked on the logs of petrophysical properties by expert geologists which in turn segments a log into three zones: Zone1: starting of log to Top 1, Zone 2: between Top 1 and Top 2, Zone 3: Top 2 to end of the log. Previous studies [53] reported that the similar zones in a well log reveals similar characteristic. Based on this hypothesis, the number of modules is decided as three same as the number of zones. Therefore, the master dataset combining seismic and borehole data is first divided into three zones (Z-1, Z-2 and Z-3) based on well tops guidance. Fig. 3-9 represents a schematic diagram depicting application of MANN concept for the present study.

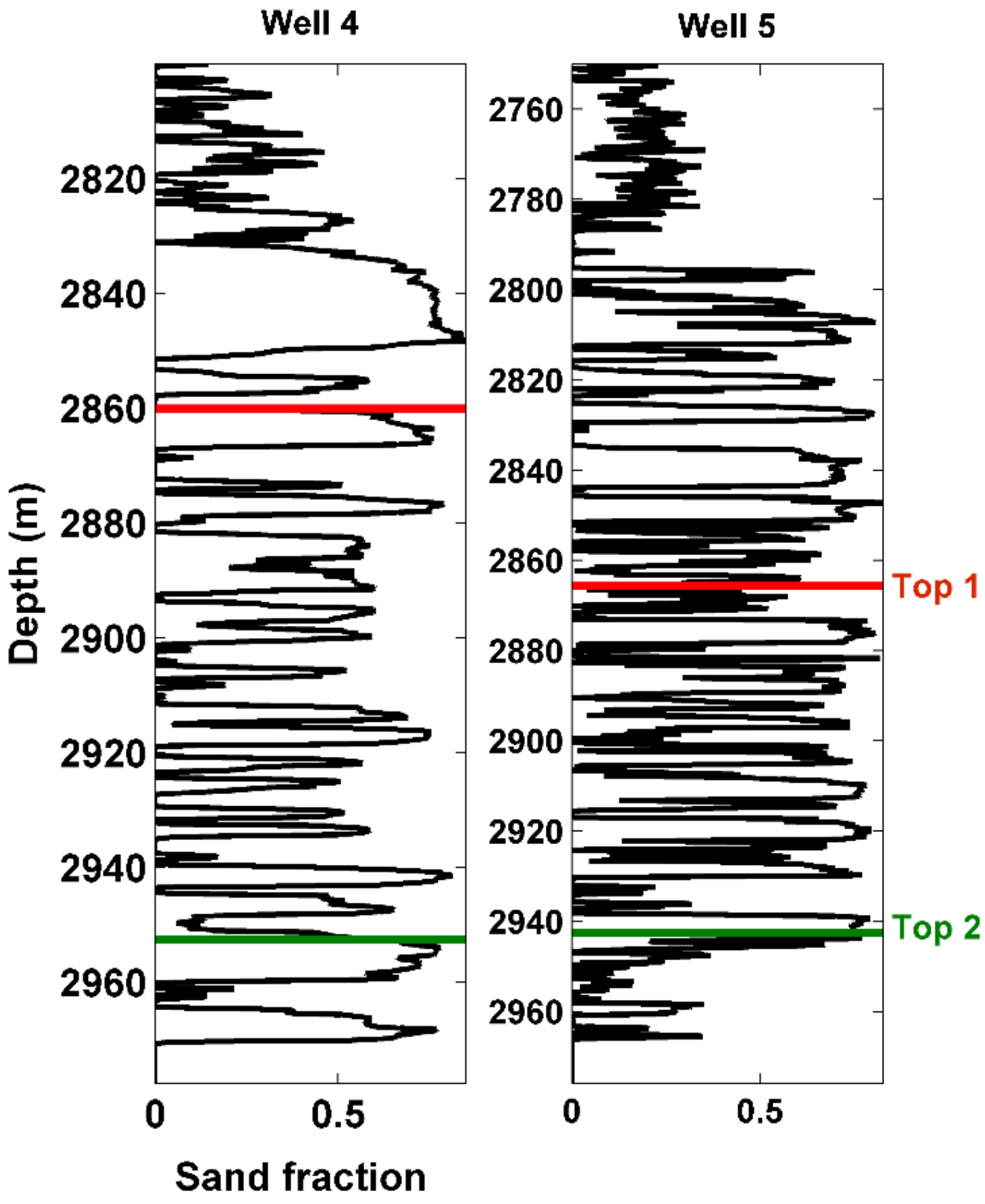

Fig. 3-8: Well log data (sand fraction) along with well tops and zoning; Z1: above Top 1; Z2: between Top 1 and Top 2; Z 3: below Top 2.

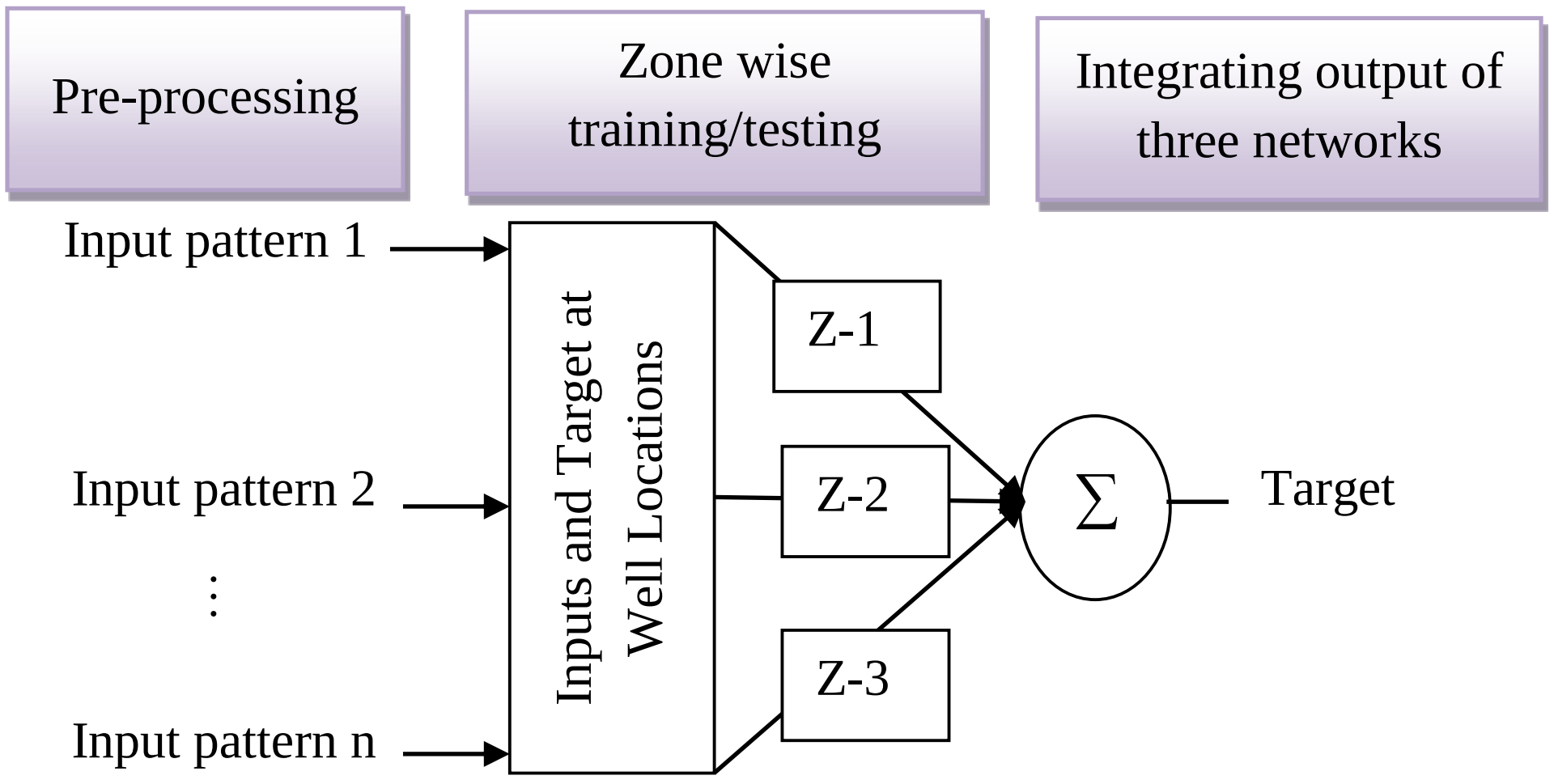

Fig. 3-9: Schematic diagram of application of MANN concept with *n* inputs and a single target

Fig. 3-10 depicts the preparation steps of the zone-wise database starting from data compilation to zone-wise division. Each of the input patterns contains three inputs (seismic impedance, amplitude, and instantaneous frequency) pertaining to three input layer nodes for all models along with single output layer neuron denoting the target sand fraction.

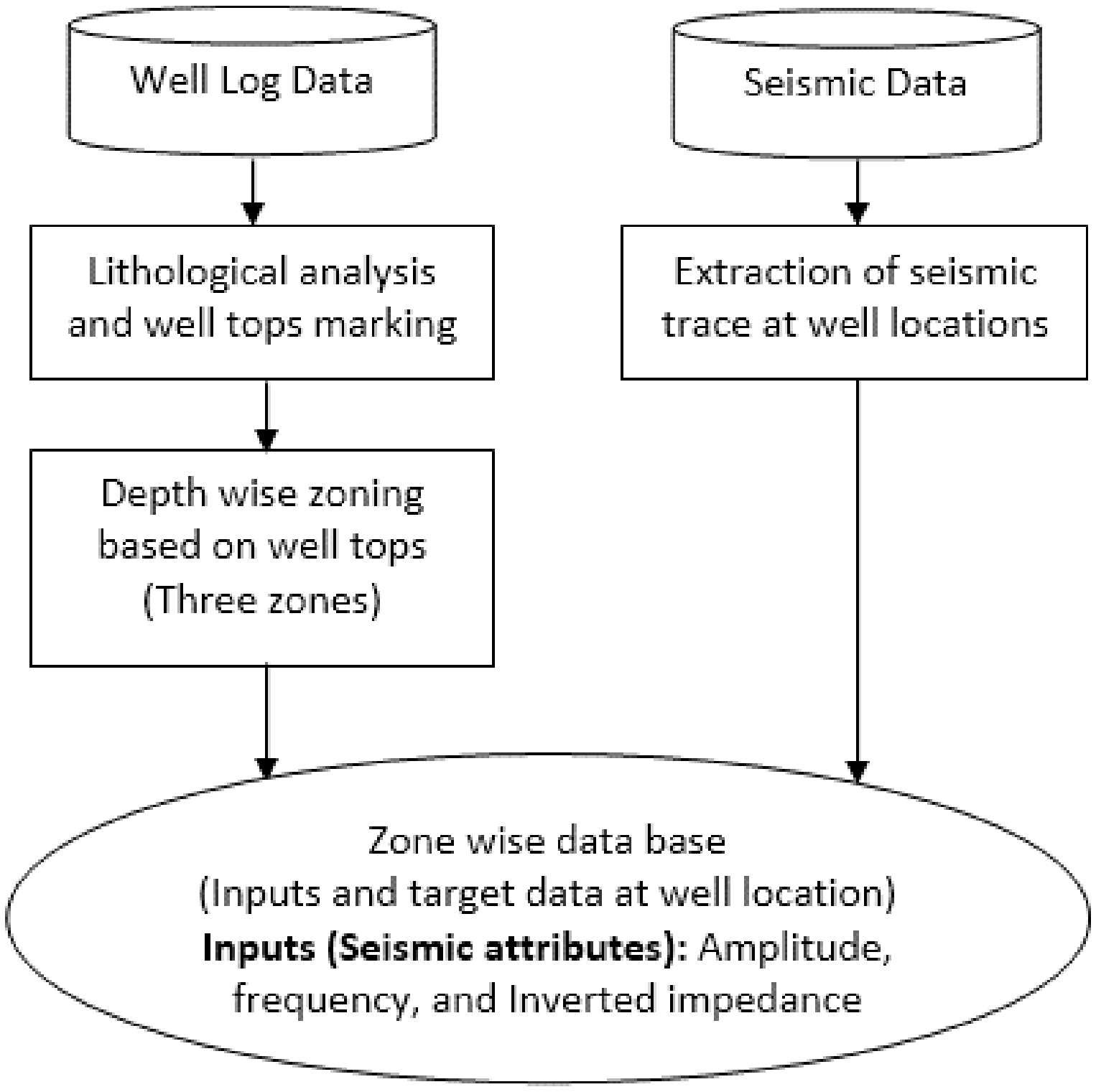

Fig. 3-10**:** Flowchart for preparation of data base.

---

Algorithm 3-1: Modular neural network approach

---

**Task :** Mapping between target property (sand fraction) and input predictors (Seismic attributes)

**Input**: Seismic attributes and sand fraction

a) Integration of seismic attributes and sand fraction signals

b) Zone/module wise division of data (*n*: total number of zones)

c) Decide blind well: k

d) Data partitioning – training and testing set

---

e) Data normalization based on min-max normalization OR z-score normalization

f) **for** $\underline{i}$ = 1 to $n$ ($i$: number of zones)

g) Selection of network structure (number of neuron, training algorithm)

h) Initialization of weights and biases, maximum epoch: $iter_{max}$, min error: $error_{min}$

i) **for** epoch=1: $iter_{max}$

j) Modify weights and biases following selected training algorithm

k) Calculate RMSE

l) **if** RMSE $\leq error_{min}$ || epoch=$iter_{max}$ break, **else** epoch= epoch+1, **end**

m) **end for**

n) Test the network using testing patterns of well k

o) **if** testing is satisfactory go to step p) **else** go to step g)

p) Freeze the network structure : $MANN_i$ ($i$ = 1 to $n$, here $n$=3)

q) Save the network structures and parameters for minimum error

r) **end for**

**Output**: Three sets of calibrated network parameters (weights and biases) w.r.t. three zones

Dataset from seven wells are used for training of the network, and then trained network is used to blindly model sand fraction for remaining one well. Separate training is carried out for each of the three modular networks keeping the learning algorithm and transfer functions same for all three networks. The optimal model is obtained by minimizing the RMSE between network output and target using selected state-of-art learning algorithm for each case. The testing of each model is carried out by using the zone-wise divided testing patterns corresponding to eighth well that is not included in the training set. Proposed MANN approach for the present study is described in Algorithm 3-1. Thus, three mapping functions are obtained using MANN approach [29] corresponding to three zones (Z-1, Z-2 and Z-3). These three trained networks are further used to obtain predicted sand fraction log for the whole study area. As indicated in Fig. 3-9 the predicted sand fraction logs from each modular network are concatenated to obtain the complete log profile. The obtained input-target relationships are used to estimate the lithological properties over the whole study area from seismic attributes.

### 3.3.2 The Proposed Framework

Seismic data are collected over a large study area, whereas well logs are available at specific well locations in the same region. Furthermore, the vertical resolution of seismic attributes is inferior compared to that of the well logs due to larger sampling interval. In general, the seismic data are helpful to model a reservoir; however, it is difficult to estimate the vertical distribution of reservoir properties with the help of seismic signals [11], [12]. Therefore, information of both - seismic and well logs is necessary to characterize a reservoir property with high-resolution in both vertical and horizontal directions. For example, sand/shale fraction, porosity, permeability and saturation are important petrophysical properties used in the interpretation of hydrocarbon reserves in details. Therefore, modeling of any such petrophysical characteristic has crucial importance in this research domain.

In the present study, sand fraction is estimated from three seismic signals (seismic amplitude, impedance and instantaneous frequency) using MANN concept. A framework, which includes pre-processing, modeling and validation, volumetric prediction, and post processing, to carry out sand fraction modeling is described.

The proposed workflow is implemented on a 64 bit MATLAB platform installed in the Intel(R) core (TM) i5 CPU @3.10 GHz computing system having 8.00 GB RAM. First, a combined dataset of seven wells is used to train three different neural networks according to depth wise zones (Z-1, Z-2, and Z-3) (refer to Fig. 3-9 and Fig. 3-10). Then, the trained networks are validated using the dataset of the remaining well. The three predicted log sections for the test well corresponding to each zone (Z-1, Z-2, and Z-3) are merged to obtain complete sand fraction log.

#### 3.3.2.1 Pre-processing

This step involves integration and normalization of target (sand fraction) and predictor variables (seismic amplitude, impedance and instantaneous frequency), followed by data partition into training and testing set.

*Integration of Seismic and Well Log Signals*

Integration of signals from different domains with the help of heuristic knowledge from human experts plays a major role in reservoir characterization [2]. Therefore, the first task in pre-processing is integration of seismic (which is in the time domain) and well log signals (which is in the depth domain) at each available well location. First, we extract the seismic attributes at eight available well locations. Then, data points in well log signals carrying missing values are excluded. It is followed by conversion of logs from the depth to the time domain using suitable velocity profile resulting from well-to-seismic tie. Then, the mismatch in sampling intervals of these two data sources (seismic and well logs) is addressed. Specifically, band-limited seismic attributes are sampled at an interval of two milliseconds, whereas the sampling interval of well logs is ~0.15 milliseconds for this particular dataset. Since, the sampling intervals of both the data are different, we apply Nyquist–Shannon sampling theorem [108], which states that a band-limited signal can be completely reconstructed from the samples, to reconstruct seismic attributes at each time instant corresponding to the well logs using cubic spline interpolation method [109]. Due to the removal of missing values from logs, the dataset is not uniform anymore. Finally, the dataset uniformly re-sampled at an interval of 0.10 milliseconds.

*Data Normalization*
Data normalization plays a crucial role for tuning the performance of machine learning algorithms. The predictors and target variables are normalized using the Z-score and min-max normalization, respectively.

*Data Partition*
A common approach in machine learning algorithms is to divide a dataset into training and testing sets for learning and validation, respectively. In this study, a combined dataset of seismic and well log signals corresponding to seven boreholes is used for training the networks whereas data of the remaining eighth well is used for testing the networks.

In this study, well tops guided zone wise prediction is carried out using the concept of MANN. Fig. 3-10 depicts a workflow for integration and division of the dataset into three separate zones (Z-1, Z-2, and Z-3) for further modeling of reservoir properties.

#### 3.3.2.2 Model Building and Validation

The learning starts after completion of pre-processing of the working dataset. Three networks corresponding to each of the three depth zones (Z-1, Z-2, and Z-3) are trained and tested. Each of the networks has three predictor variables corresponding to the presence of three input nodes in the network structure and a single output node representing target sand fraction. In this study, we opted for a single hidden layer for all cases. Selection of activation functions and training algorithm plays a crucial role in training of the network. In the present study, hyperbolic tangent sigmoid is used in the hidden layer [97], [99]; however, for the output layer log-sigmoid transfer function is used. In these type of iterative processes, the connecting weights are updated using the back propagation till the global minimum error is achieved. Conjugate gradient method is an advanced and effective method for error minimization [13]. Here, scaled-conjugate-gradient-back-propagation is selected over several other learning algorithms for its speed and simplicity [13] in training the networks. Number of neurons in the hidden layer and epochs are initialized with small values and gradually increased keeping the improvement of fitting between target and predicted sand fraction in consideration. However, the number of hidden layer neurons cannot be indefinitely increased; the possibility of overfitting has been avoided by keeping the maximum number of trainable parameters at least fifteen times lower than the number of available training patterns [13].

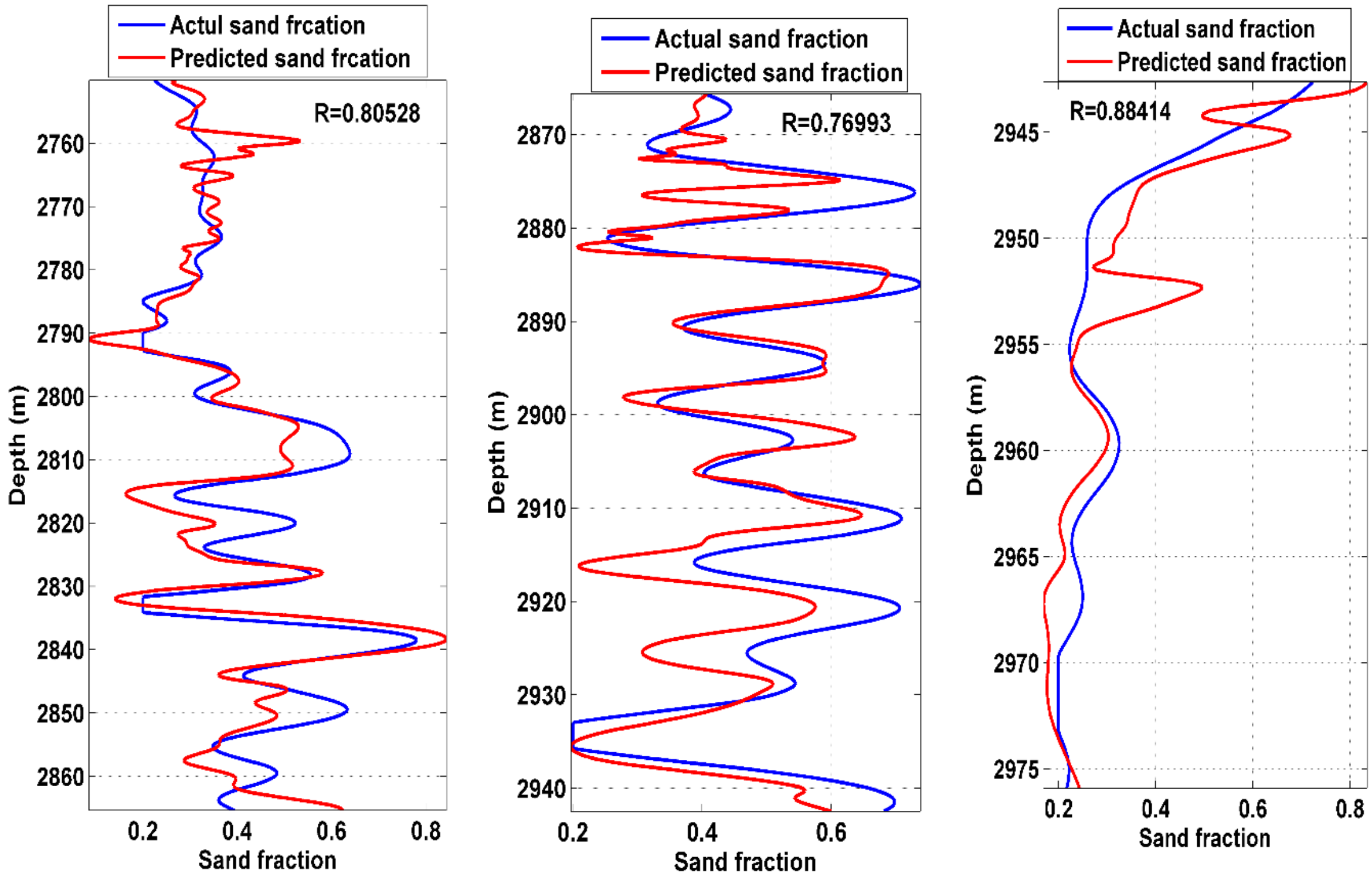

(a) (b) (c)

Fig. 3-11: Predicted and target sand fraction for three zones for well 6

Three separate networks are designed and trained for the three depth zones, and finally the trained networks are used for blind prediction. The performance of the trained networks is quantified in terms of four performance evaluators namely CC, RMSE, AEM and program execution time. It is important to carry out statistical analysis of the errors involved in the model. The calibrated networks, which performed well in blind prediction in terms of the four aforementioned performance evaluators, are saved and used in the next step, i.e., volumetric prediction.

Fig. 3-11 represents superimposed plots of target and network predicted sand fraction values for Zone 1, 2, and 3, respectively for well 6 only. Close observation of Fig. 3-11 reveals that the predicted logs follow the target ones with acceptable correlation coefficients (0.8058 for Z-1; 0.7699 for Z-2; and 0.8841 for Z-3). These high values of correlation coefficients indicate good prediction by the proposed framework. Similar results are obtained for other wells as well.

The correlation coefficients obtained by blind testing using three networks corresponds to three zones (Z-1, Z-2, and Z-3), and their average are compared with blind prediction coefficient using a single ANN for the overall depth range. Fig. 3-12–Fig. 3-15 present the results of performance comparison of the proposed workflow with an ANN in terms of performance evaluators for well 2, 4, and 6.

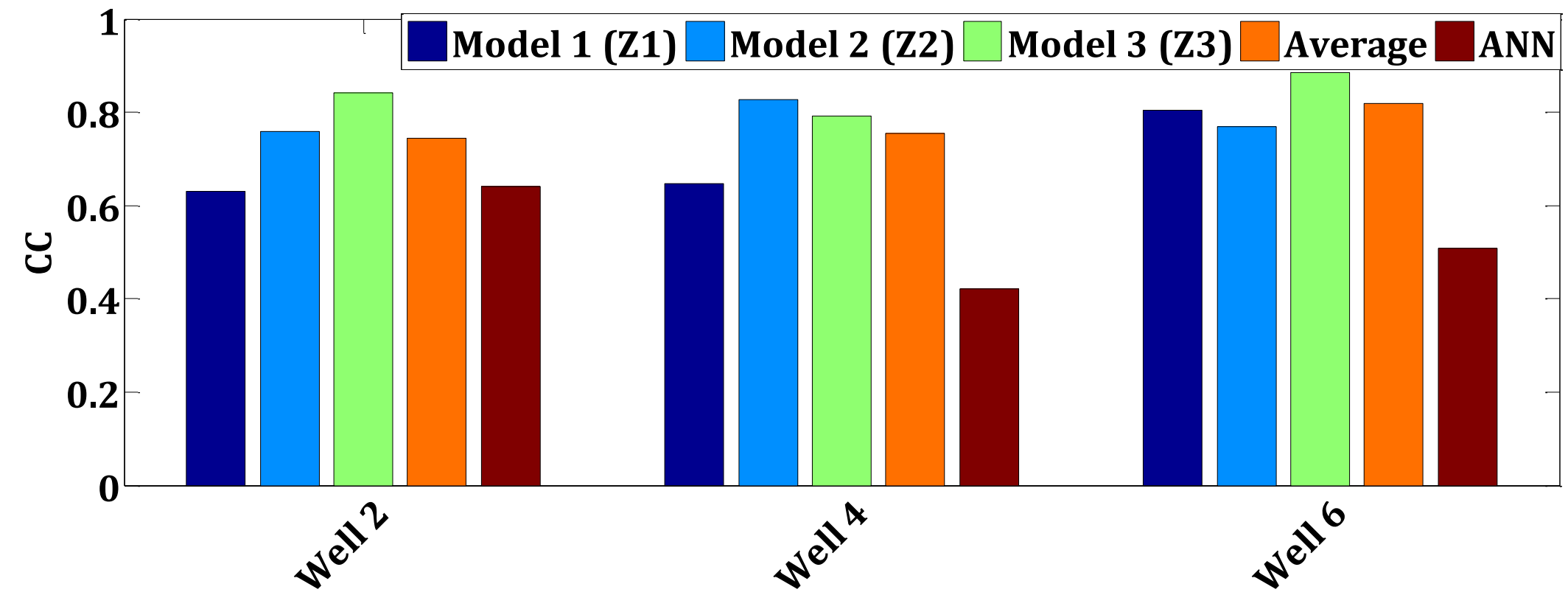

Fig. 3-12: Performance comparison of proposed workflow with ANN in terms of correlation coefficient (CC)

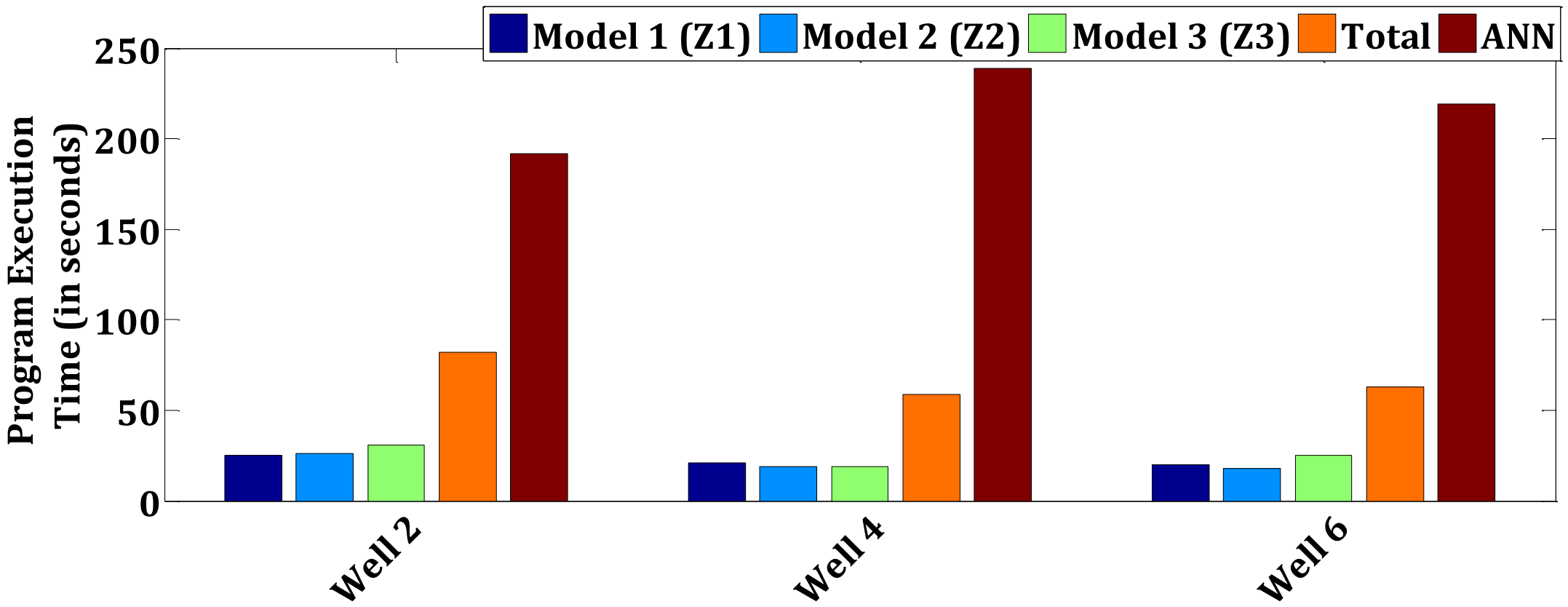

Fig. 3-13: Performance comparison of proposed workflow with ANN in terms of program execution time

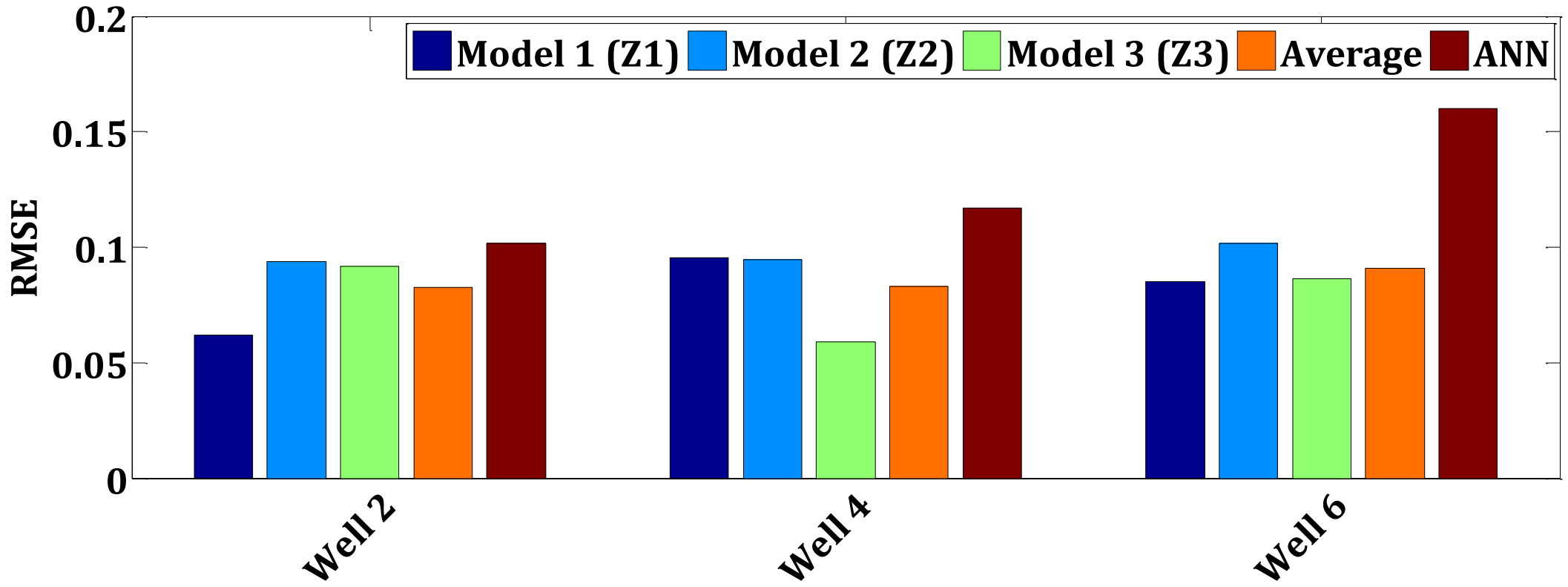

Fig. 3-14**:** Performance comparison of proposed workflow with ANN in terms of root mean square error (RMSE)

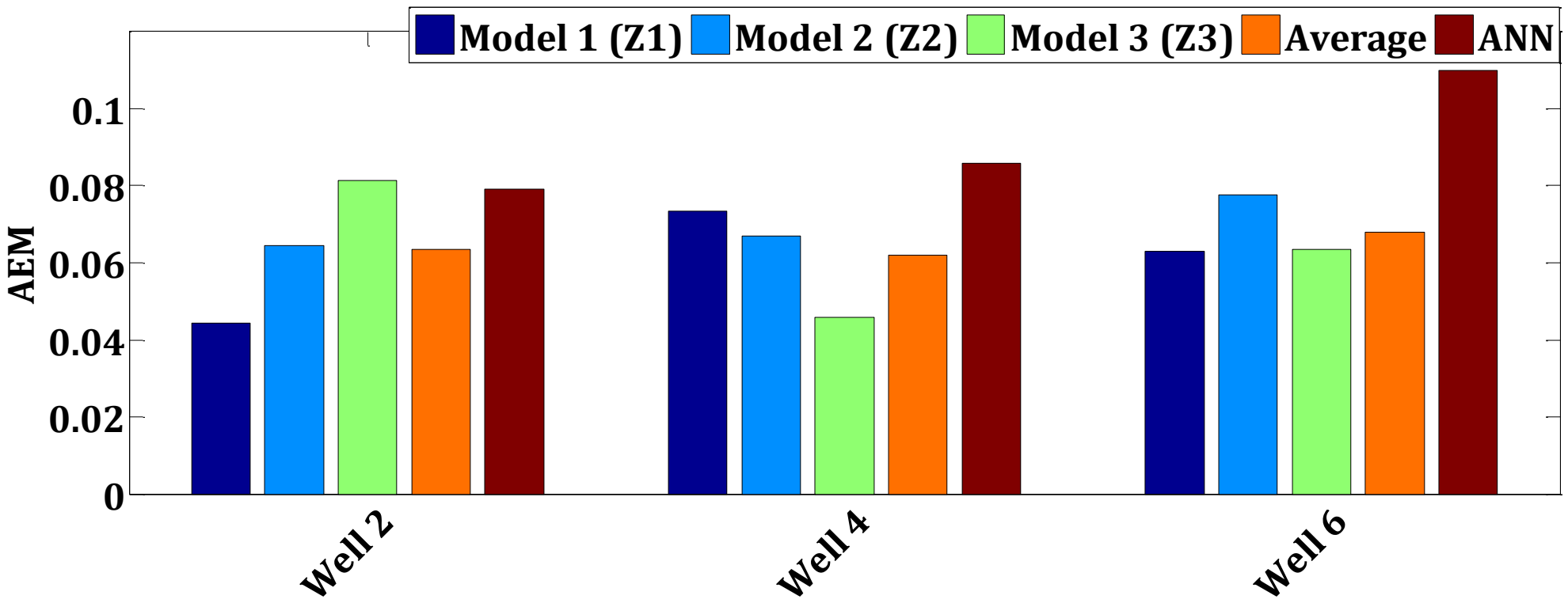

Fig. 3-15: Performance comparison of proposed workflow with ANN in terms of absolute error mean (AEM)

For example, in case of Well 4, first, the dataset is segregated into three sections following the well tops guided zonation. Then, three sets of training patterns are generated combining the samples belong to Well 1– 3, 5, – 8 for each of the segregated dataset in the previous stage. Three networks are initialized and trained using the training patterns corresponding to three zones (Z1, Z2, and Z3). Next, the calibrated networks are validated using testing patterns belong to Well 4 for each zone separately. In case of CC, RMSE, and AEM, average performance of the proposed framework is expressed by carrying out mean of the respective measures belong to three individual models. Fig. 3-12 and Fig. 3-14–Fig. 3-15 demonstrate CC, RMSE, and AEM respectively by the three individual models (Model 1, Model 2, and Model 3 for Z-1, Z-2, and Z3 respectively), their average performance, and the single ANN model used for the overall depth range. Contrarily, the total program execution time is resultant of summation of the three individual models. Fig. 3-13 depicts individual and total program execution times in seconds taken by the three models workflow along with that of the single ANN associated with whole depth range. As smaller networks deal with simpler structures and smaller dataset, acceptable accuracy is achieved with a reduced execution time in case of MANN approach.

### 3.3.2.3 Volumetric Prediction

This step is essential for visualization of reservoir characteristic at the boreholes and away from it after prediction from seismic attributes is carried out. As no direct relationship between seismic and well logs is evident in theory, which might be inherent, it is a challenging task to estimate lithological properties across the study area from seismic signals. Therefore, it would be beneficial for the geoscientists if a mapping between seismic and reservoir properties could be carried out by deriving a relationship between these two types of data from integrated dataset of seismic and lithological parameters at available well locations using MANN concept. The horizon or well tops information of the study area is available. Therefore, the dataset containing predictor attributes throughout the study area are segregated into three parts according to well top information. Then, for each zone, predicted sand fraction log is generated from seismic signals using tuned network parameters corresponding to a particular zone. Thus, a set of three logs is available for each particular trace point. These three logs can be concatenated accordingly to obtain the complete sand fraction log at a particular trace point. Hence, sand fraction logs are predicted for the study area from seismic input using tuned networks. Visualization at a specific in-line is demonstrated after predicting the sand fraction from three seismic attributes (seismic impedance, amplitude, and instantaneous frequency) over the area. In parallel, input attributes are also visualized across the in-line. It can be observed from the figures presented in the results section that predicted sand fraction requires post-processing step to improve the visualization quality.

Fig. 3-16 describes the variation of input seismic attributes and predicted sand fraction at an in-line corresponds to well 6. It can be observed from Fig. 3-16 (a)–(c) that the input attributes change smoothly throughout the study area. On the other hand, networks predicted sand fraction variation is not smooth.

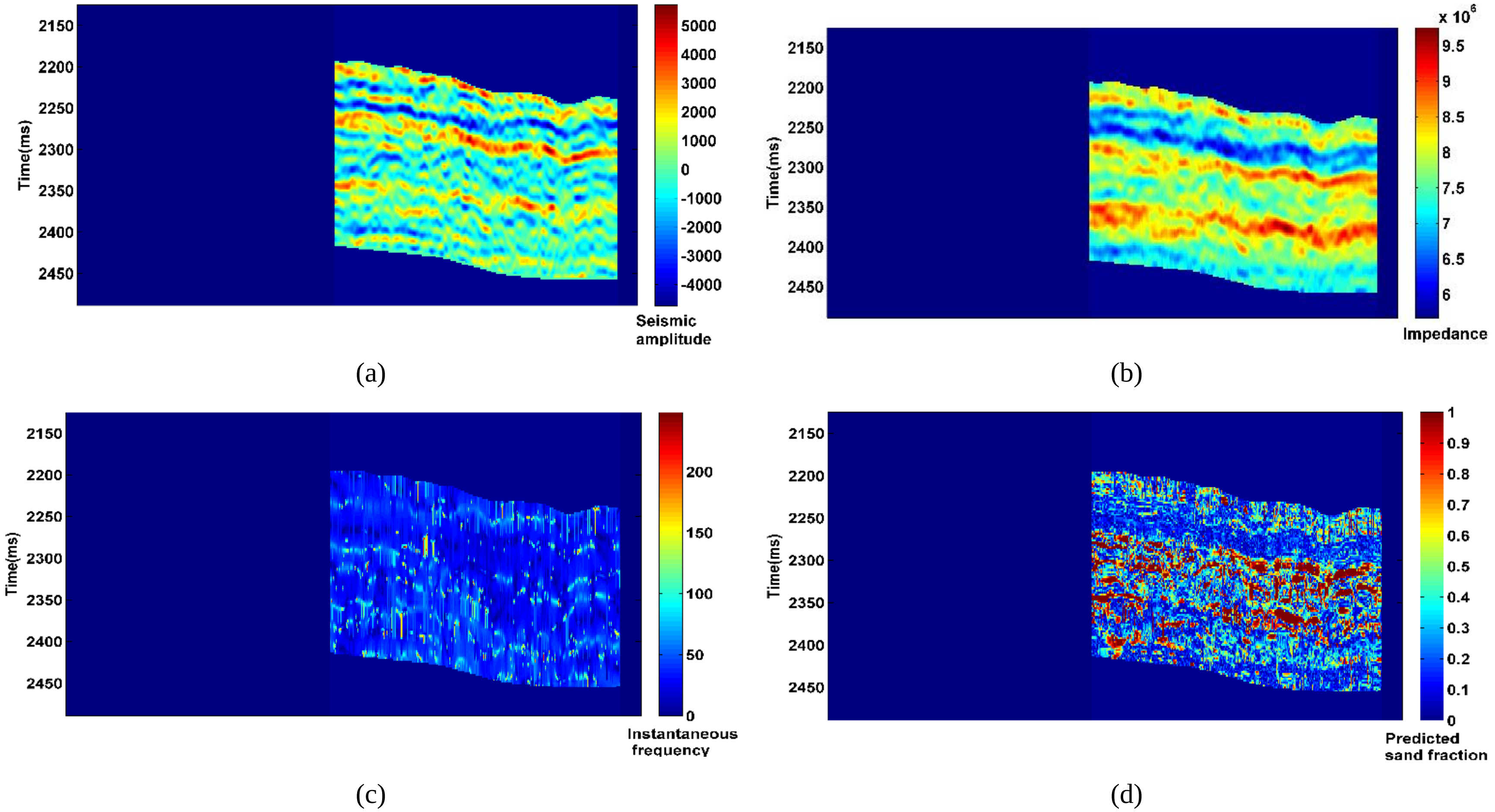

Fig. 3-16: Slice plots of input seismic attributes and predicted sand fraction correspond to inline number of well 6

#### 3.3.2.4 Post Processing

In this study, the predicted sand fraction is smoothened using moving average filter [73]. The necessity of the post-processing step is established by comparing the variation of seismic attributes and estimated sand fraction.

---

Algorithm 3-2: Moving Average Filter

---

**Task** : Reduce noise in predicted sand fraction volume

**Input** : Predicted sand volume matrix $X$ , window size $w$

a) Initialize: $w$

   $x_{j,k}$ - pixel value at $(j,k)$, $I_{j,k}^{w}$ be a window of size $w \times w$ centered at $(j,k)$

b) Compute $i_{j,k}^{mov,w}$–average of the pixel values in $I_{j,k}^{w}$

c) Replace $x_{j,k}$ by $i_{j,k}^{mov,w}$, thus obtain $X_{filt}$

d) If result is satisfactory, then stop, else go to step a).

**Output** : Filtered sand fraction matrix $X_{filt}$

---

Every matrix element in predicted sand fraction volume is considered as a pixel and smoothened using moving average filter respective to neighborhood of pixels within selected window size following Algorithm 3-2. In specific cases, where some of the neighborhood cell values are missing for a particular element, those missing values are replaced by NaN (not a number). For example, edge of the input matrix is filtered following above procedure. Here, the window size of the moving average filter used to smooth the predicted sand fraction is selected as 3×3 empirically.

This uneven variation in predicted sand fraction necessitates inclusion of a post-processing algorithm. We opt for a moving average filter based algorithm with a 3×3 window size. Implementation of the filtering technique on predicted sand fraction reduces noise in it. Comparing Fig. 3-16 (d) with Fig. 3-17, it can be observed that the variation of the latter is smoother than former. Thus, a realistic presentation of sand fraction variation over an area is obtained.

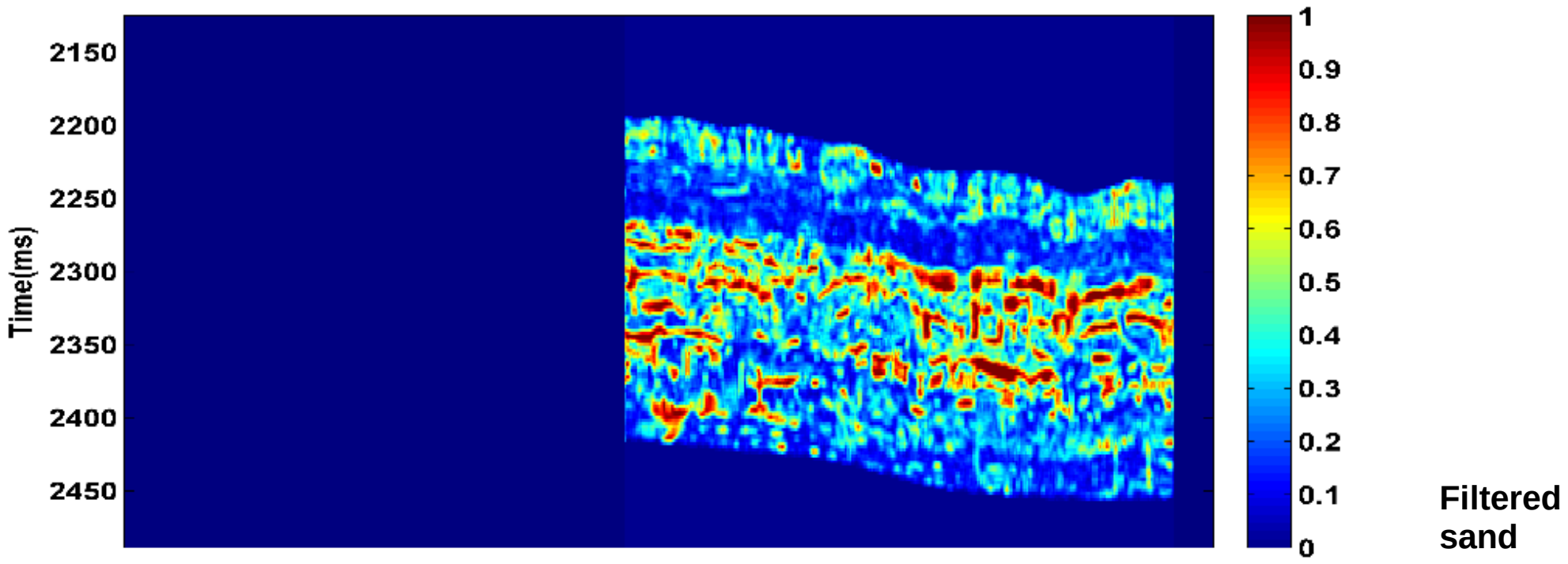

Fig. 3-17: Variation of smoothened sand fraction correspond to inline number of well 6

### 3.3.3 Discussion

The objective of the present study is to establish well tops guided prediction of a reservoir property using MANN concept over a single network while working on a large and complex dataset. Section 3.3 has presented the performance of the proposed workflow along with ANN by blind estimation of the sand fraction from the three seismic attributes (seismic impedance, amplitude, and instantaneous frequency). It is evident from the presented results that the proposed workflow has outperformed ANN in terms of higher correlation coefficient, reduced error measures and low program execution time by successfully calibrating a functional relationship between seismic and well log signals corresponding to each zone. Thus, MANN concept can be selected over ANN in case of complex large dataset. The post-processing on predicted sand fraction improves the visualization realistically.

The contributions of this study are consolidated as follows:

- Fusion of two concepts – similarity between logs belonging to same horizon and MANN
- Inclusion of seismic data as predictor variables
- The selection of number of modules based on well-top information
- Blind prediction
- The proposed workflow is established to produce better performance with reduced program execution time
- Enhanced visualization by post-processing

Next phase of research may be focused on estimation of 3D geo-cellular model for other characteristics involving seismic and well log signals at available well control points. Here, number of modules are decided based on well tops guided instead of trial-and-error methods.

## 3.4 Conclusion

The target property in this chapter i.e. sand fraction is distributed from zero to unity. The distribution of SF is not skewed at any particular point. Thus, the dataset is balanced from the perspective of the SF. The performance analysis carried out in Section 3.2 has established the regularization scheme proposed in Chapter 2. The three alternative approaches based on FT, WD, and EMD have cross-validated each other in terms of multiple performance indicators. Each of the three schemes has yielded better result compared to the case where the original SF was used as the target. Therefore, user may select either of the three schemes. The regularized SF is modelled from seismic attributes using ANN. The selection of the ANN structure and the initialization of parameters are crucial job to attain acceptable prediction performance. The performances are quantified in terms of multiple evaluators. With the increase in number of predictor variables, hidden layers, and neurons in each hidden layer, the structure of ANN will become more complex which in turn would increase the difficulty to train the network. The availability of enough number of training patterns is required for learning of the ANN. In case of large number of predictor attributes and smaller amount of training patterns, the dimensionality of the dataset need to be reduced in the pre-processing stage itself. Principal Component Analysis (PCA), forward sequential selection approach can be opted for dimensionality reduction. However, for this work, dimensionality reduction was not required owing to the presence of large training datasets.

In case of MANN, the problem is divided into multiple sub-problems. For each case, individual modules are trained using smaller datasets which in turn reduces the complexity in learning. Therefore, with the availability of well tops information, the complete dataset is divided into three zone wise datasets. Then, three individual ANN models are trained and tested separately.

Other petrophysical properties such as porosity, permeability, shale fraction, etc. can be modelled from seismic attributes using the frameworks proposed in this chapter.

# Chapter 4. Classification of Water Saturation

Evaluation of hydrocarbon reservoir requires classification of petrophysical properties from available dataset. However, characterization of reservoir attributes is difficult due to the nonlinear and heterogeneous nature of the subsurface physical properties. In this context, present study proposes a generalized one-class classification framework based on Support Vector Data Description (SVDD) to classify a reservoir characteristic– water saturation into two classes (Class high and Class low) from four logs namely gamma ray, neutron porosity, bulk density, and P-sonic using an imbalanced dataset. The comparison is carried out among the proposed framework and different supervised classification algorithms in terms of g-metric means and execution time. Experimental results show that the proposed framework has outperformed other classifiers in terms of these performance evaluators. Then, the proposed framework is modified and seismic attributes are used as predictor variables. The modified framework has predicted class labels of water saturation (Class low/Class high) from seismic information over the study area.

This chapter is designed as follows. First, the theory of SVDD is described in brief. Then, the framework to classify water saturation from well logs is presented. Finally, the modified framework to classify the water saturation from seismic attributes is described.

### 4.1 Support Vector Data Description (SVDD)

Large dataset can be characterized using data description techniques. Significant efforts have been made for the classification of real world datasets. SVDD, an extension of SVM, is widely used approach for the data classifications [65], [66].

In general, data are described by defining a closed boundary around the data. This closed boundary is defined by hypersphere, $F(R,a)$ where '$a$' represents centre and '$R$' is the radius. Volume of the hypersphere should be minimized for the data description [65]–[67], [110]. Outlier in the data can be characterized by defining slacks variables $\varepsilon_i \geq 0$. In this case, the minimization term of error function is given by

$$F(R,a) = R^2 + C\sum_i \varepsilon_i \left\| x_i - a \right\|^2 \leq R^2 + \varepsilon_i \qquad (4\text{-}1)$$

where,

$$\left\| x_i - a \right\|^2 \leq R^2 + \varepsilon_i, \ \text{for all } i \qquad (4\text{-}2)$$

Kernel function $K(x_i, x_j) = \phi(x_i).\phi(x_j)$ is used for smooth data description. Then the SVDD function can be represented as

$$L = \sum_i \alpha_i K(x_i, x_j) - \sum_{i,j} \alpha_i \alpha_i K(x_i, x_j) \ \text{for all } \alpha_i : 0 \leq \alpha_i \leq C \qquad (4\text{-}3)$$

Optimization of (4-3) gives the data description which can be obtained by several algorithms available in the literature, and Lagrange multipliers should satisfy the normalization constraint $\sum_i \alpha_i = 1$. The values of α is can be found out by minimizing $L$. We have used a Gaussian kernel

$$K\left(x_i, x_j\right) = e^{-q x_i - x_j} \quad (4\text{-}4)$$

to represent the dot product $\phi\left(x_i\right).\phi\left(x_j\right)$ as discussed in [65], [66], [111]. In order to calculate the radius we have to look for the support vectors. Firstly, $R^2(x)$ in terms of the kernel function for each of the point is found out. Then, we get

$$R^2\left(x\right) = K\left(x, x\right) - 2\sum_i \alpha_i K\left(x_i, x\right) + \sum_{i,j} \alpha_i \alpha_j K\left(x_i, x_j\right) \quad (4\text{-}5)$$

Now the support vectors are those data objects which lie on the surface of the hypersphere i.e., for which $C = \alpha_i$. The contours are formed by the data points along the cluster boundaries. For the purpose of our work, we take the radius of the circle $R$ to be the maximum of values $R(x)$ for the support vectors. Any data point lying beyond $R$ is considered to be an outlier. In one-class classification using SVDD, the minority class patterns are used as the target in the training phase to construct the hypersphere. Once the hypersphere is constructed, the classifier is evaluated by using majority class patterns as testing dataset. For imbalanced dataset, the improvement in one-class classifier performance compared to its two-class counterpart is apparent [112], [113].

## 4.2 Development of a Framework to Classify Water Saturation from Well Logs

The first important contribution of this chapter is to propose a generalized framework based on SVDD [65], [66] to characterize the water saturation from input well logs. Next, a comparative analysis is presented to demonstrate the effectiveness of the proposed classification method over other classifiers (discriminant[110], [114], naive Bayes [110], [115], support vector machine based classifier [116], [117]).

The rest of Section 4.2 is structured as follows: first, the data used in this study is described; after that the proposed classification framework is described. Then, a brief description of performance evaluators used in this work is given. After that, experimental results are reported. Finally, we conclude this chapter with a discussion and future scope.

### 4.2.1 Data Description

The well logs used in this work are acquired from four closely spaced boreholes located in an onshore hydrocarbon field of India. Henceforward, these aforementioned wells are to be referred as A, B, C, and D, respectively. The borehole data contains several logs such as gamma ray content (GR), bulk density (RHOB), P-sonic (DT), neutron porosity (NPHI), spontaneous potential (SP), acoustic impedance (AI) and different resistivity logs such as deep resistivity (RT), medium resistivity (RM) and shallow resistivity (RS) logs. Reservoir characteristics, e.g., sand fraction, porosity, water saturation, oil saturation etc. are derived from these log properties. Literature study reveals that GR, RHOB, DT, NPHI, SP among different logs are to be used as predictor variables to model or classify lithological properties. After selection of relevant features among available logs, we have used GR, RHOB, DT, and NPHI logs as input attributes to classify water saturation level. The rock properties of subsurface formations can be interpreted from these variables. The gamma radiation of different formations along the depth is represented by gamma ray log in American Petroleum Institute (API) unit. The density log is recorded in grams per cubic centimetre unit. It varies according to mineralogy and porosity values. Travel time of P-waves versus depth is recorded as P-sonic log in microsecond per feet. The fourth predictor variable i.e. neutron porosity log is attuned to read the true porosity and represented in per unit. In this work, the target variable is water saturation, which is an important characteristic in the petroleum industry representing the fraction of formation water present in the pore space.

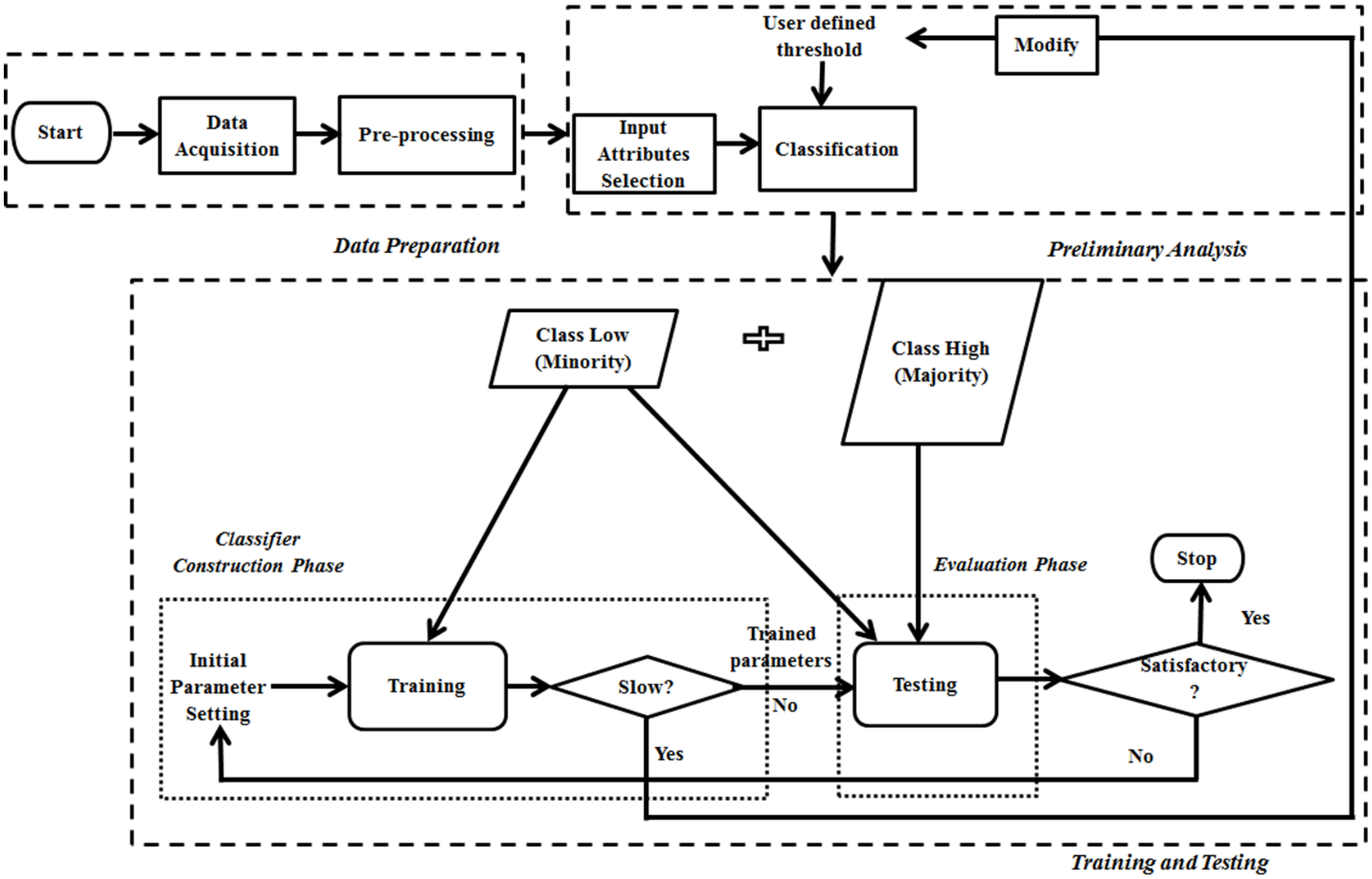

Fig. 4-1: Proposed framework for classification of imbalanced dataset

### 4.2.2 Proposed Classification Framework

In the recent years, SVDD and other kernel-based algorithms have been reported as popular techniques adapted for classification of imbalanced dataset in the field of hyperspectral image processing, outlier detection, document classification etc. In this work, an attempt has been made to construct an SVDD based framework to classify reservoir properties using an imbalanced geological dataset. The proposed generalized framework, which includes three steps namely- 1) data preparation, 2) preliminary analysis, and 3) training and testing, is represented in Fig. 4-1. These steps are briefly discussed in this section.

#### 4.2.2.1 Data Preparation

Well log data from four wells located in the western onshore hydrocarbon field of India are used in the present study. The procedure of data preparation is started with data acquisition as shown in Fig. 4-1. The log files contain a number of missing data values. These patterns are removed to make a data file of valid values only. Then we uniformly re-sample the data.

This stage is the starting point of the proposed framework. Well logs are selected and pre-processed. Fig. 4-2 represents plots of gamma ray, neutron porosity, bulk density, and resistivity logs along depth for well A. Similarly, Fig. 4-3 represents P-sonic, acoustic impedance, and water saturation logs along depth for the same well. Designing a classifier is required to classify water saturation log from available log variables. The selection of the input variables is carried out using Relief algorithm as discussed in the following section.

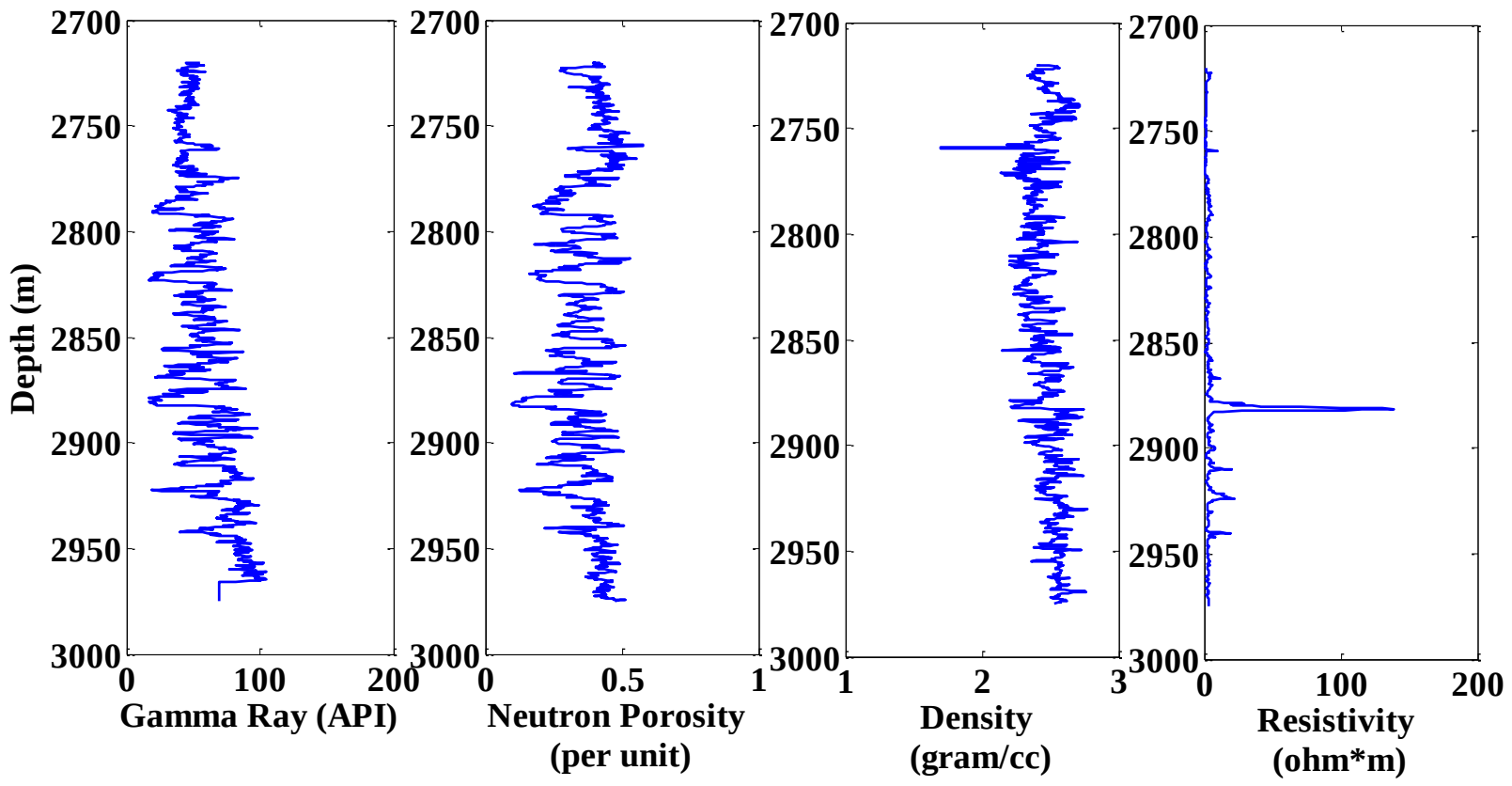

Fig. 4-2: Plots of gamma ray, neutron porosity, bulk density, and resistivity along depth for well A

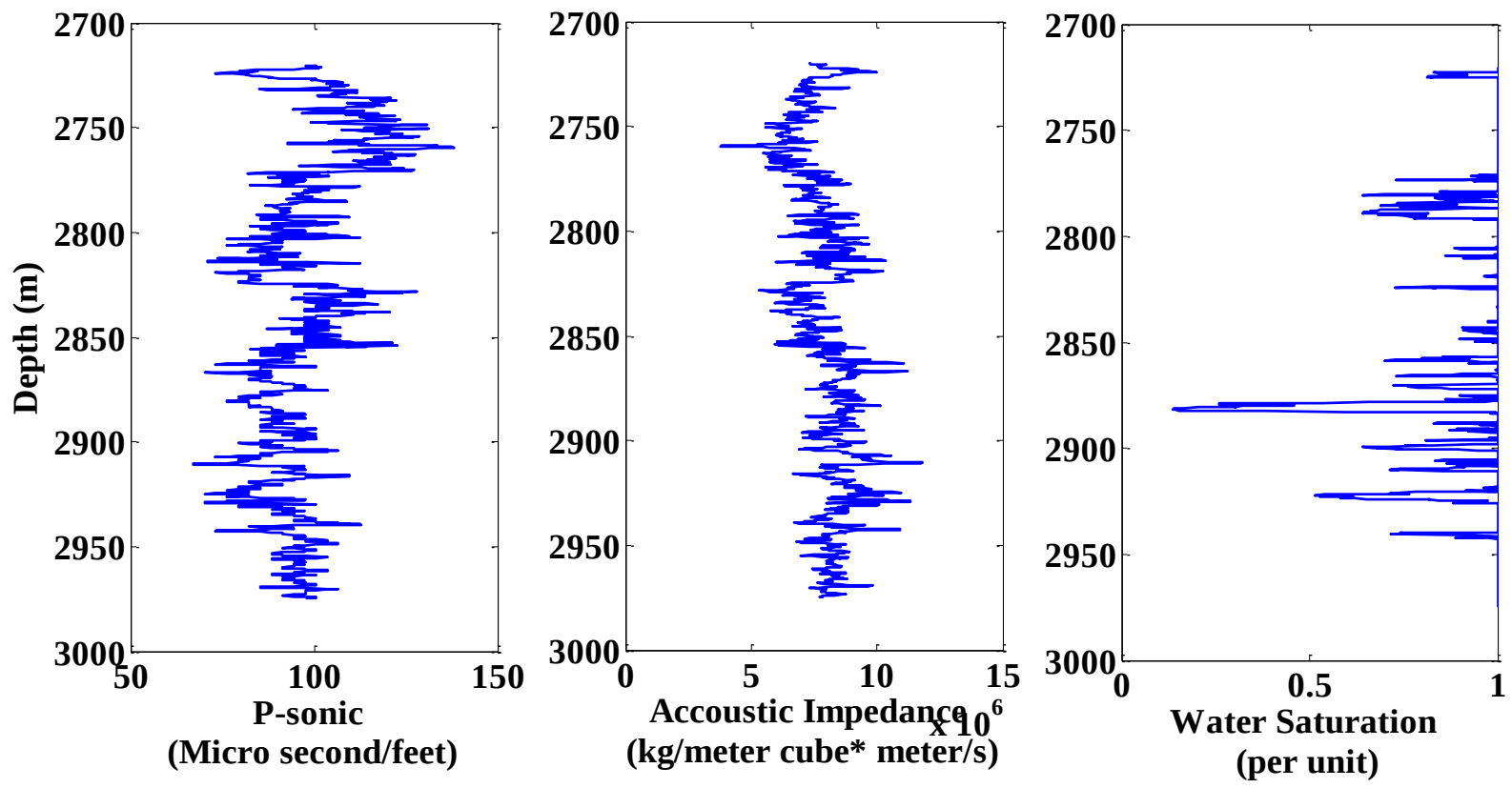

Fig. 4-3: Plots of P-sonic, acoustic impedance, and water saturation logs along depth for well A

#### 4.2.2.2 Preliminary Analysis

Feature selection plays a crucial role in tuning the performance of pattern classifiers. In the pre-processing stage, several number of “candidate features” are extracted from raw dataset. Then relevant features are selected using different algorithms i.e. mutual information, Relief algorithm, and its variants. Here, we use Relief algorithm [118], which identifies statistically relevant features and performs well in case of noisy dataset, to select input attributes before training the classifier. Designing a classifier with several inputs prolongs the training time along with unnecessary proliferation in the model complexity. Moreover, the generalization capability of a model enhances while using only relevant features as inputs.

Next, we classify the water saturation into two classes, namely- Class high and Class low using a user-defined threshold. Two factors guide the choice of threshold value. Firstly, saturation values belonging to the Class high must be as close to one as possible while in Class low it must be as close to zero as possible. This is done by observing the histogram of the saturation values. Secondly, the high computational complexity of the SVDD classifier has compelled us to set the threshold in a manner so as to have reasonable small number of patterns at least in one-class to have the classifier trained within reasonable time. This threshold value is modified depending on the training speed of the SVDD algorithm. After completion of the preliminary analysis, training and testing of SVDD based one-class classifier is started. Besides, selection of the threshold level is confirmed by expert geologists.

First, several attributes are extracted from the raw dataset. Then, four relevant attributes are selected from the six “candidate attributes” using Relief algorithm. The result of the Relief algorithm is represented in Fig. 4-4. It can be observed from the figure that GR, NPHI, RHOB, and DT logs are more relevant features related to water saturation in terms of predictor importance weight compared to RT and AI logs.

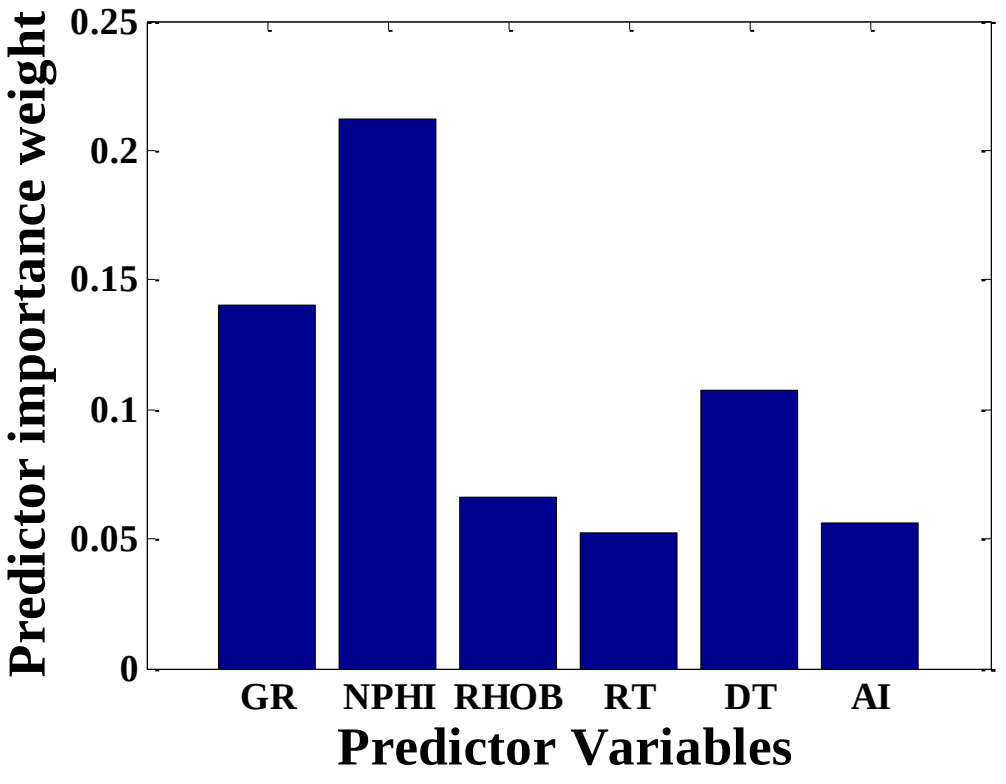

Fig. 4-4: Selection of relevant input attributes using Relief algorithm

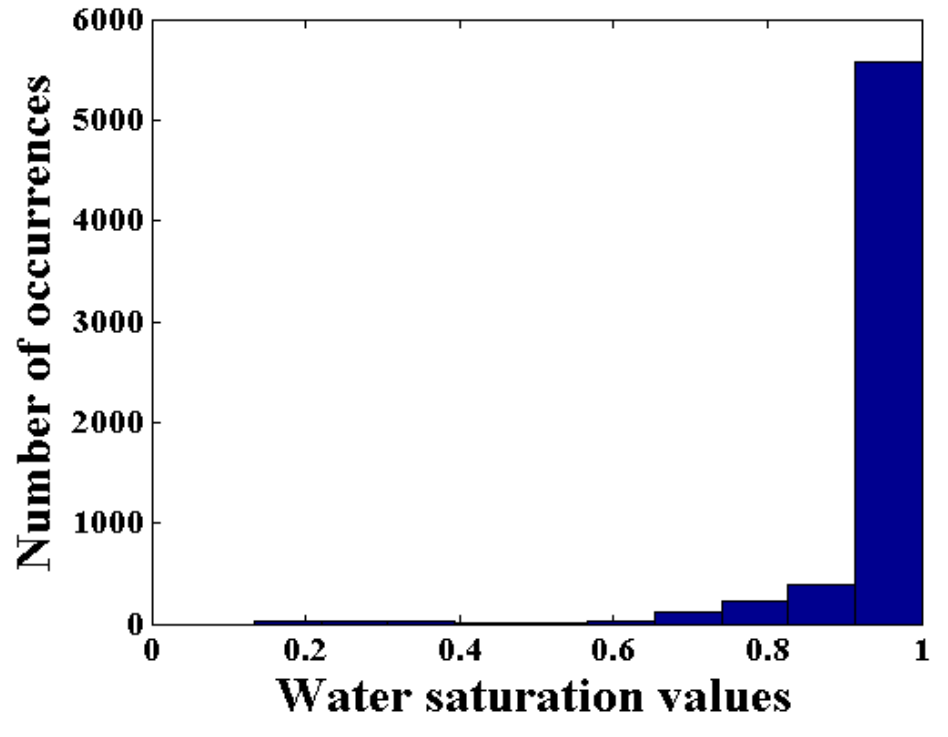

Fig. 4-5: Histogram plot for water saturation

After selection of the appropriate input attributes the next task is to classify water saturation into two classes using a user-defined threshold value. We consider two criteria as discussed in an earlier section for the selection of the threshold level to classify the water saturation values into two classes. For this particular problem, we choose 0.7 as the threshold value after verifying the constraints related to computational speed of SVDD algorithm and experienced geoscientists' view. Patterns with a saturation level greater than or equal to 0.7 are called Class high and the other patterns are called Class low. We have 3% of the whole data set in the Class low set. It can be observed from Fig. 4-5 that the distribution of water saturation values is skewed at one. Specifically, 97% of the total available patterns belong to Class high that is associated with higher values of water saturation. Therefore, Class low and Class high can be termed as minority and majority classes respectively.

#### 4.2.2.3 Training and Testing

The training and testing steps associated with the one-class classifier are shown in the lower part of Fig. 4-1. In this problem, the available patterns are significantly large in case of Class high compared to Class low. In other words, Class high and Class low can be invariably denoted as majority and minority classes. After training the SVDD using combined minority class patterns of remaining three wells, we test the performance of the classifier using majority class patterns of these three wells along with all the patterns (majority and minority) of the test well. The results reported in this article corresponds to the blind testing of the individual well when classifier learning is carried out using a kernel function and an initial $C$ value. For example, in case of blind prediction of well C, the SVDD hypersphere is constructed using patterns belong to minority class from combined dataset of remaining three wells using Gaussian kernel of width parameter of 2.0, and $C$ = 0.008 as initial parameter setting.

The input attributes (GR, RHOB, DT, and NPHI) of the training patterns are used to construct the SVDD hypersphere. Classification accuracy of SVDD is improved by adjusting few parameters: type of the kernel function and associated parameters, and radius of the hypersphere $C$. The kernel functions such as Gaussian, higher order polynomial (with order 2–10), radial basis function, exponential radial basis function, kernel parameters, are experimented with values of $C$ varying from 0 to 1. The classifier uses a Lagrangian function which is minimized using constrained optimization. It divides the patterns into two classes as true data that resides inside the hypersphere and outliers that reside outside the boundary of the hypersphere. The points which make the boundary of the hypersphere are called support vectors. In this work, we include these support vectors in the outlier class. The trained parameters are saved and applied to the majority class to test the classifier performance.

The performance of the proposed framework using one-class classifier based on SVDD is evaluated upon the accuracy of both positive and negative classes. Instead of employing confusion matrix, which is generally used to measure performance of classifier, here we use g-metric means [119]. This performance evaluator is often used in case of imbalanced dataset. G-metric means can be represented as

$$g=\sqrt{acc_P * acc_N} \qquad (4\text{-}6)$$

where, $acc_P$ and $acc_N$ represent sensitivity and specificity, respectively. Sensitivity indicates the accuracy on the positive instances i.e. (true positives/ (true positives + false negatives)) and similarly, specificity denotes the accuracy on the negative instances i.e. (true negatives/ (true negatives + false positives)).

Program execution time is also recorded to compare the performance of proposed framework with respect to other classifiers.

After completion of the training and testing stage, the classification performance achieved using this proposed framework is compared to other classifiers namely discriminant, naive Bayes, and SVM based classifier. SVM, naive base, and discriminant classifiers are optimized after initialization with appropriate parameter values using the same predictor variables. From the test output, the patterns classified as outliers and support vectors are considered to be majority class components; and data vectors are specified as minority class components. Then, comparison is carried out among these supervised classifiers depending on the blind testing result of each of the wells.

Table 4-1 and Table 4-2 represent comparison result of the proposed framework with other supervised classifiers.

Table 4-1: Performance comparison of classifiers in terms of g-metric mean

| **Well Name** | **G-Metric Mean** | | | |
|---|---|---|---|---|
| | ***SVM*** | ***Naive Bayes*** | ***Discriminant*** | ***Proposed Workflow (SVDD)*** |
| A | 0.75 | 0.75 | 0.70 | 0.78 |
| B | 0.61 | 0.50 | 0.59 | 0.65 |
| C | 0.71 | 0.81 | 0.80 | 0.83 |
| D | 0.74 | 0.80 | 0.68 | 0.90 |
| Average Performance | 0.70 | 0.71 | 0.69 | 0.79 |

Table 4-2**:** Performance comparison of classifiers in terms of program execution time

| **Well Name** | **Program Execution Time (In Seconds)** | | | |
|---|---|---|---|---|
| | ***SVM*** | ***Naive Bayes*** | ***Discriminant*** | ***Proposed Workflow (SVDD)*** |
| A | 50.0 | 44.1 | 32.3 | 30.2 |
| B | 40.2 | 34.0 | 43.1 | 40.5 |
| C | 43.1 | 30.1 | 45.1 | 19.3 |
| D | 43.1 | 54.7 | 40.3 | 26.4 |
| Average Performance | 44.1 | 40.7 | 40.2 | 29.1 |

It is evident from the Table 4-1 and Table 4-2 that the proposed classifier workflow outperformed other supervised classifiers in terms of g-metric means and program execution time.

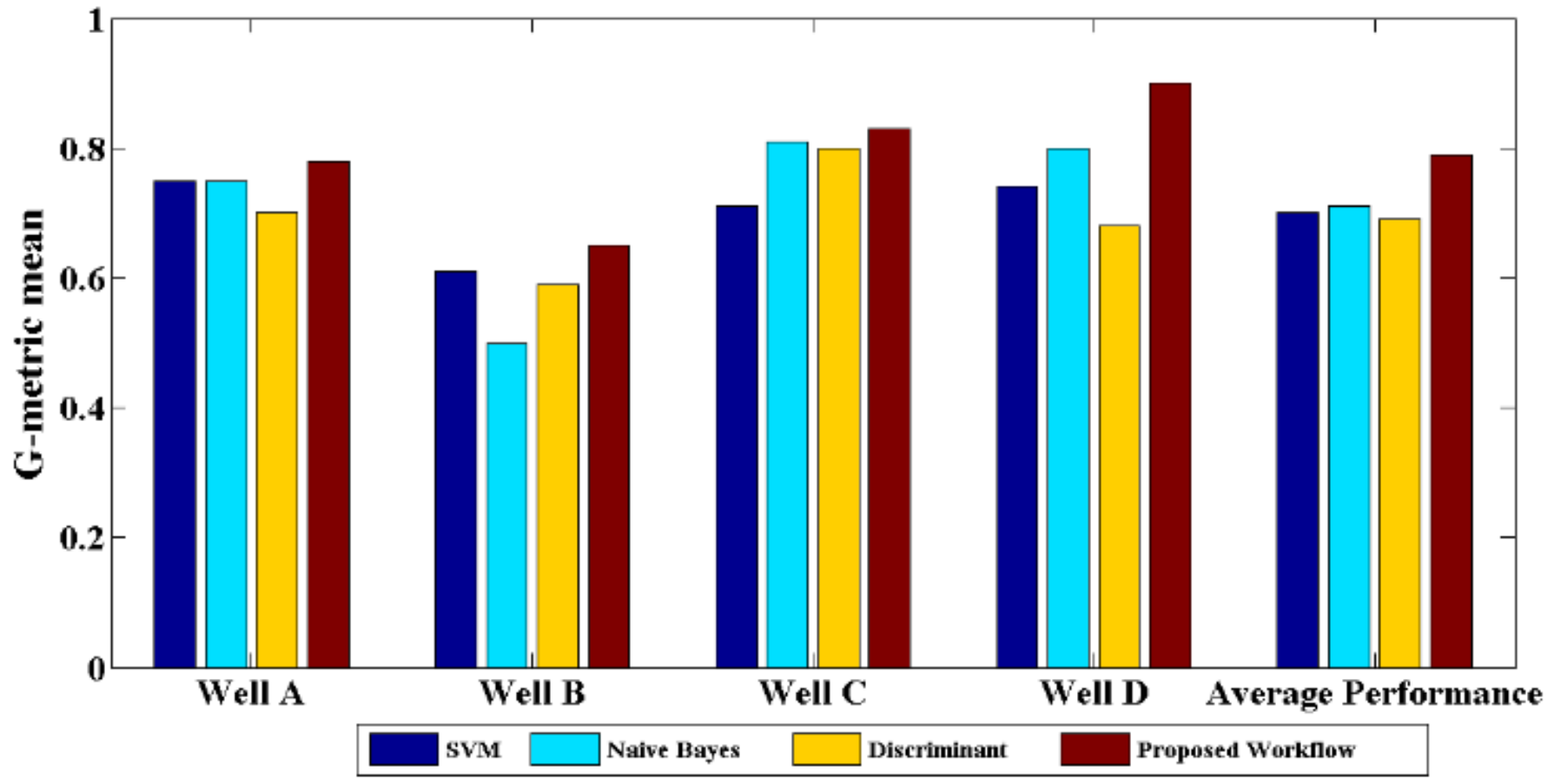

Fig. 4-6 : Bar plot describing performance of classifiers in terms of g-metric means

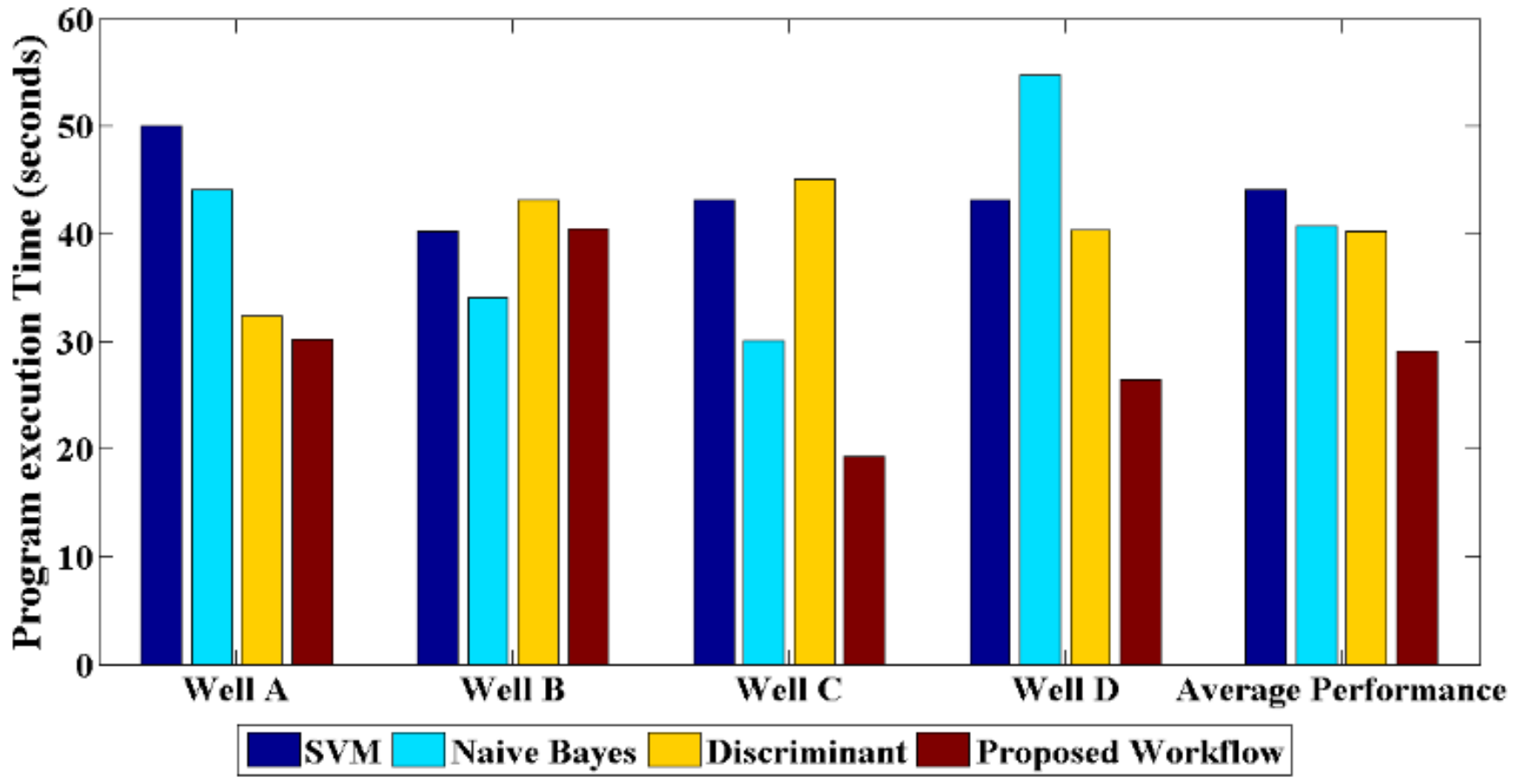

Fig. 4-7 : Bar plot describing performance of classifiers in terms of program execution time

Fig. 4-6 and Fig. 4-7 represent the result of performance comparison of the supervised classifiers in terms of g-metric means and program execution time respectively. Therefore, it can be inferred from the results that the proposed workflow based on SVDD can be used as a powerful tool to classify imbalanced dataset in reservoir characterization domain.

### 4.2.3 Discussion

In this work, a complete framework based on SVDD is proposed to classify water saturation from well logs using an imbalanced geological dataset. Comparative analysis reported in this section has shown that the proposed methodology has outperformed existing classifier algorithms in terms of performance evaluators (g-metric means and program execution time). This work can be extended with the inclusion of seismic attributes as inputs to the classifier based model. Integration of the seismic and limited number of available borehole data will help to produce 3D volume representing high and low water saturation values throughout a study area.

## 4.3 Development of a Framework to Classify Water Saturation from Seismic Attributes

Water saturation is an important property in reservoir engineering domain. Thus, satisfactory classification of water saturation from seismic attributes is beneficial for reservoir characterization. However, diverse and non-linear nature of the subsurface attributes makes the classification task difficult. Section 3.2 has proposed a generalized SVDD based novel classification framework to classify water saturation into two classes (Class high and Class low) from four well logs. In this section, the aforementioned framework is modified to use three seismic attributes such as seismic impedance, amplitude envelope, and seismic sweetness as predictor variables. Like previous section, g-metric means and program execution time are used to quantify the performance of the modified framework along with the established supervised classifiers. The documented results imply that the proposed framework is superior to the existing classifiers. The present study is envisioned to contribute in further reservoir modelling. The contributions of the present study are as follows:

- A complete classification framework integrating seismic and well log signals
- Blind prediction
- Comparison with other classifiers
- Water saturation level map over the area

### 4.3.1 Data Description

The dataset corresponding to the four wells (Well A, Well B, Well C, and Well D) as in Section 4.2 is used in this section. As an extension of the work carried out in Section 4.2, seismic attributes corresponding to the study area are included as predictor variables instead of well logs to achieve an area map of water saturation level. There are five seismic attributes acquired from the same study area such as seismic impedance, amplitude, instantaneous frequency, amplitude envelope and seismic sweetness. However, seismic impedance, amplitude envelope, and seismic sweetness are selected over amplitude and instantaneous frequency by Relief algorithm.

### 4.3.2 Proposed Classification Framework

A classification framework is designed to classify water saturation from seismic attributes using an imbalanced geological dataset in this section. There are four steps included in the workflow namely– data preparation, preliminary analysis, training and testing, volumetric classification and visualization of water saturation level map as demonstrated in Fig. 4-8. The steps in the proposed framework are designed by modifying the work done in Section 4.2 and briefly described in this section.

The research work carried out in this study are performed on a 64 bit MATLAB platform installed on a Intel(R) Core(TM) i5CPU @3.20 GHz workstation having 16 GB RAM. The following sections describe the experimental results achieved in every step of the proposed framework.

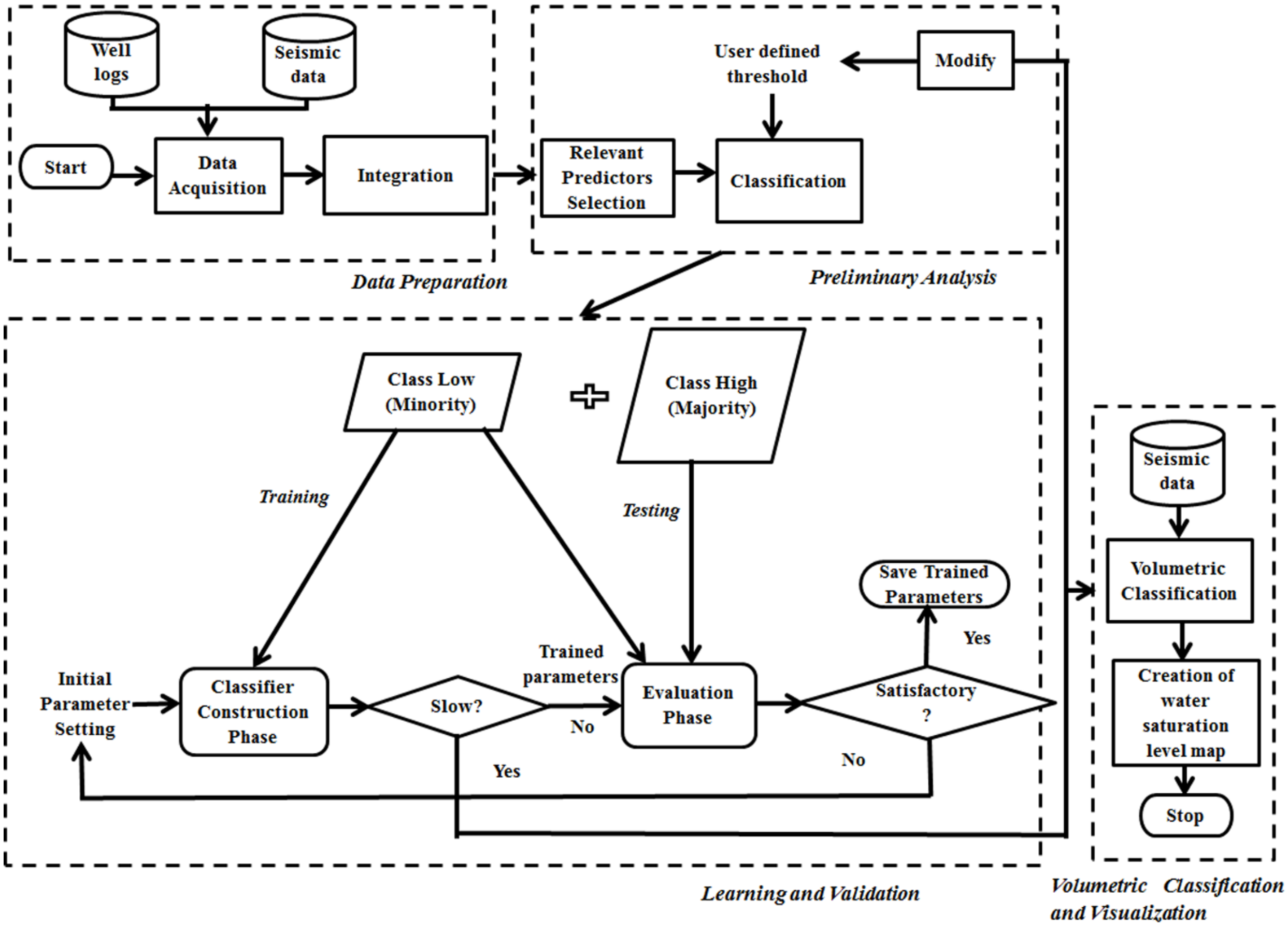

Fig. 4-8: Proposed classification framework

#### 4.3.2.1 Data Preparation

Seismic attributes along with water saturation log corresponding to four well locations are used in this study. As shown in the figure (Fig. 4-8), the procedure is started with data acquisition and integration of seismic and borehole data. First, the well logs are converted into the time domain from the depth domain using the time-depth relationships available at the four well locations. Then, seismic attributes at the four well locations are extracted from seismic volume. It is found that the sampling intervals of these dataset (seismic and well logs) are different. For example, the seismic patterns are sampled at an interval of two milliseconds, whereas the sampling interval of well logs is 0.15 milliseconds. Hence, we interpolate the band limited seismic signals at 0.15 milliseconds sampling interval corresponds to that of the well logs. Thus, the combined dataset of seismic attributes and water saturation is prepared to be used in the preliminary analysis stage.

Fig. 4-9 and Fig. 4-10 represent available five seismic attributes- (Fig. 4-9(a)) seismic impedance, (Fig. 4-9(b)) amplitude, (Fig. 4-9(c)) instantaneous frequency, (Fig. 4-10(a)) seismic amplitude envelope, (Fig. 4-10(b)) seismic sweetness and (Fig. 4-10(c)) water saturation along the Well A. The red dots on the seismic attributes represent original values at time interval of two milliseconds and the green curves represent reconstructed signals along the time interval of well log data. The blue curve in Fig. 4-10(c) represents water saturation along the Well A. It can be observed that water saturation distribution is biased towards maximum water saturation value (i.e. one).

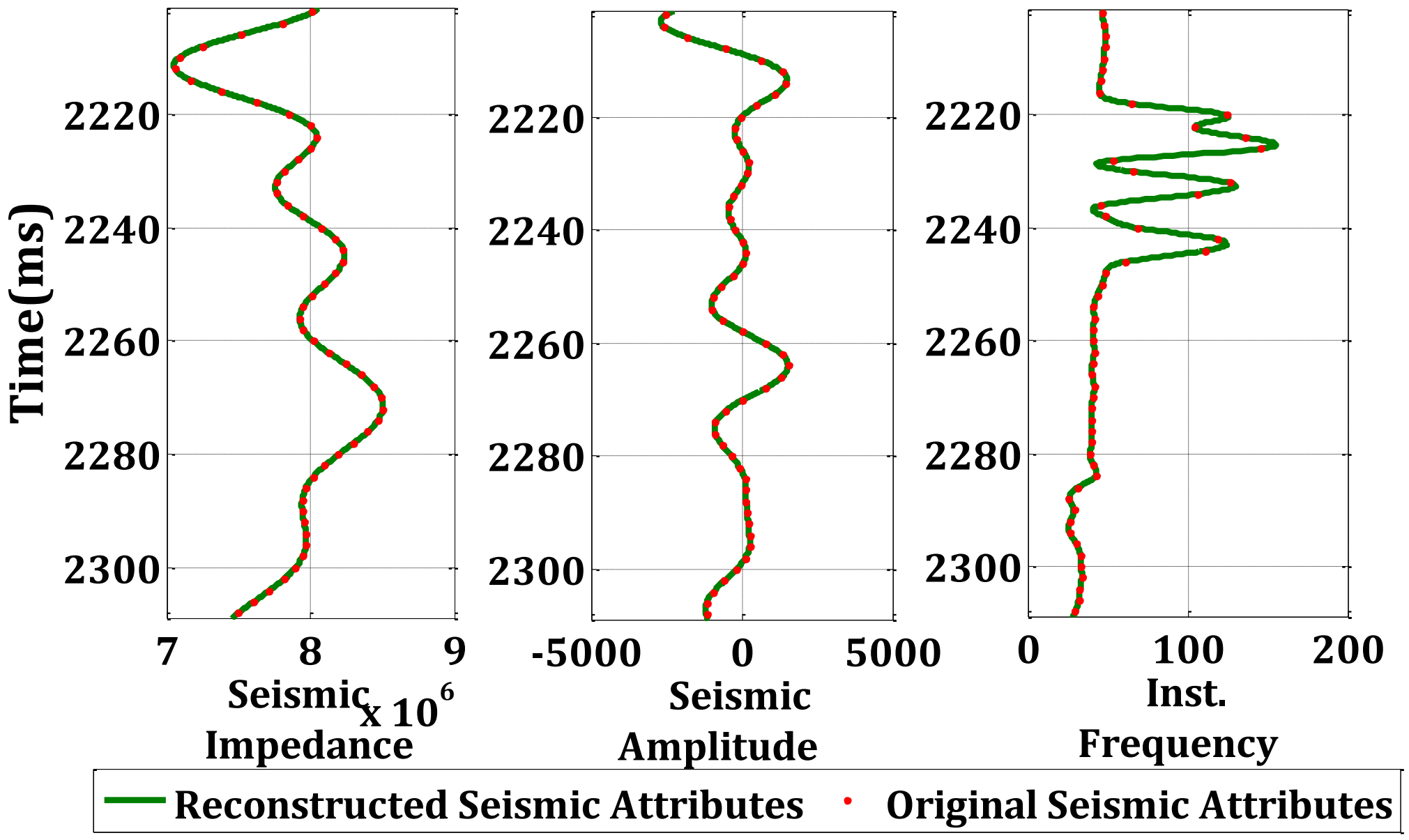

(a) (b) (c)

Fig. 4-9: Plots of (a) seismic impedance, (b) amplitude, and (c) instantaneous frequency along time (ms) for Well A

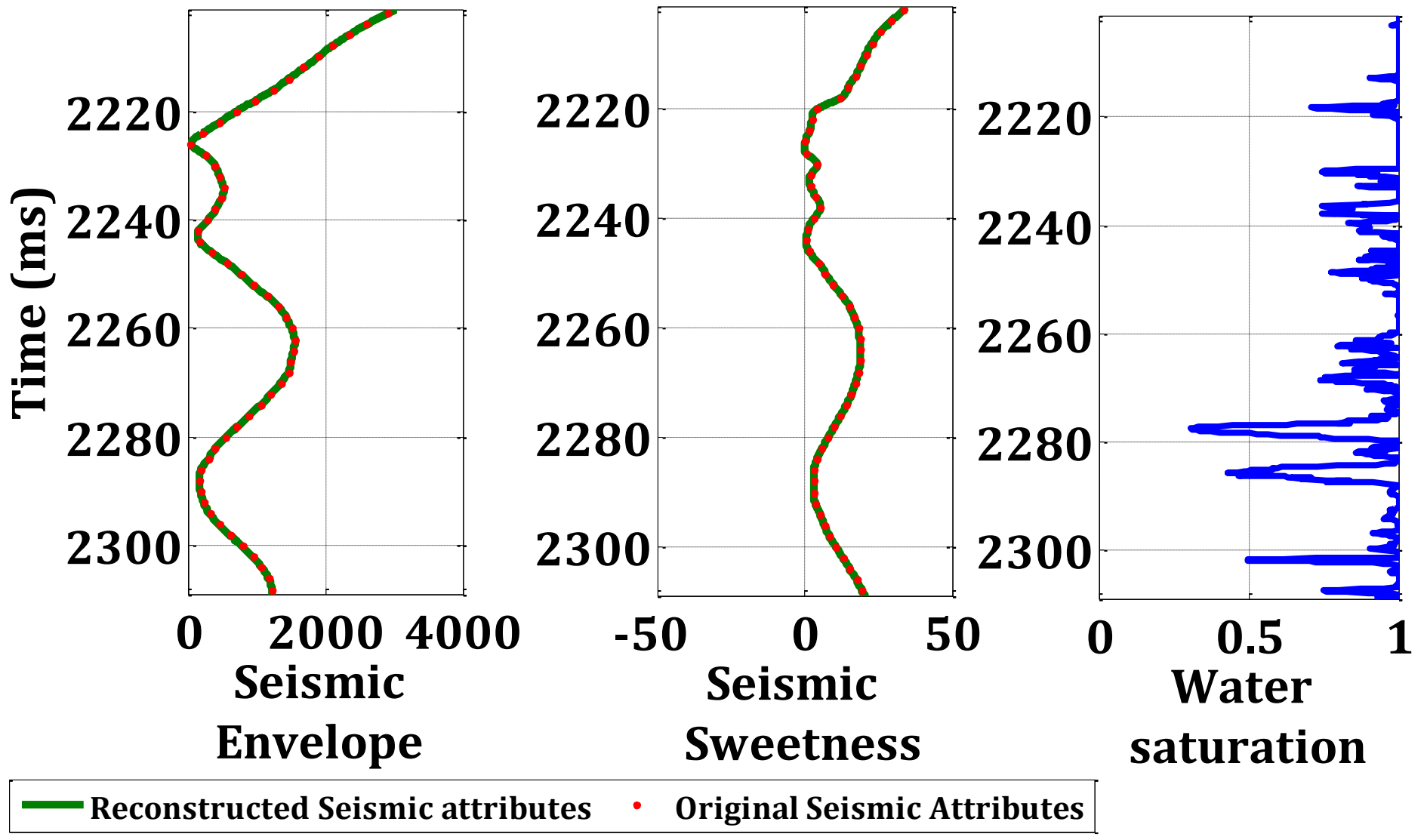

(a) (b) (c)

Fig. 4-10: Plots of (a) seismic amplitude envelope, (b) seismic sweetness, and (c) water saturation along time (ms) for Well A

#### 4.3.2.2 Preliminary Analysis

The performance of classifiers is dependent on the selection of relevant features. First, a number of "candidate features" are extracted from the raw dataset. Then, different algorithms i.e. mutual information, Relief algorithm [118], and its variants are used to identify relevant features among available features before starting to train the classifier. In this chapter, Relief algorithm selects statistically relevant features from a noisy dataset. Inclusion of unnecessary inputs in model elongates training time along with an increase in the model complexity. In contrary, application of relevant features as predictor variables enhances the generalization capability of a model [120]. The result of Relief algorithm is represented in Fig. 4-11. Fig. 4-11 reveals that seismic impedance, seismic amplitude envelope, and seismic sweetness are more relevant features with respect to water saturation in terms of predictor importance weight compared to amplitude and instantaneous frequency.

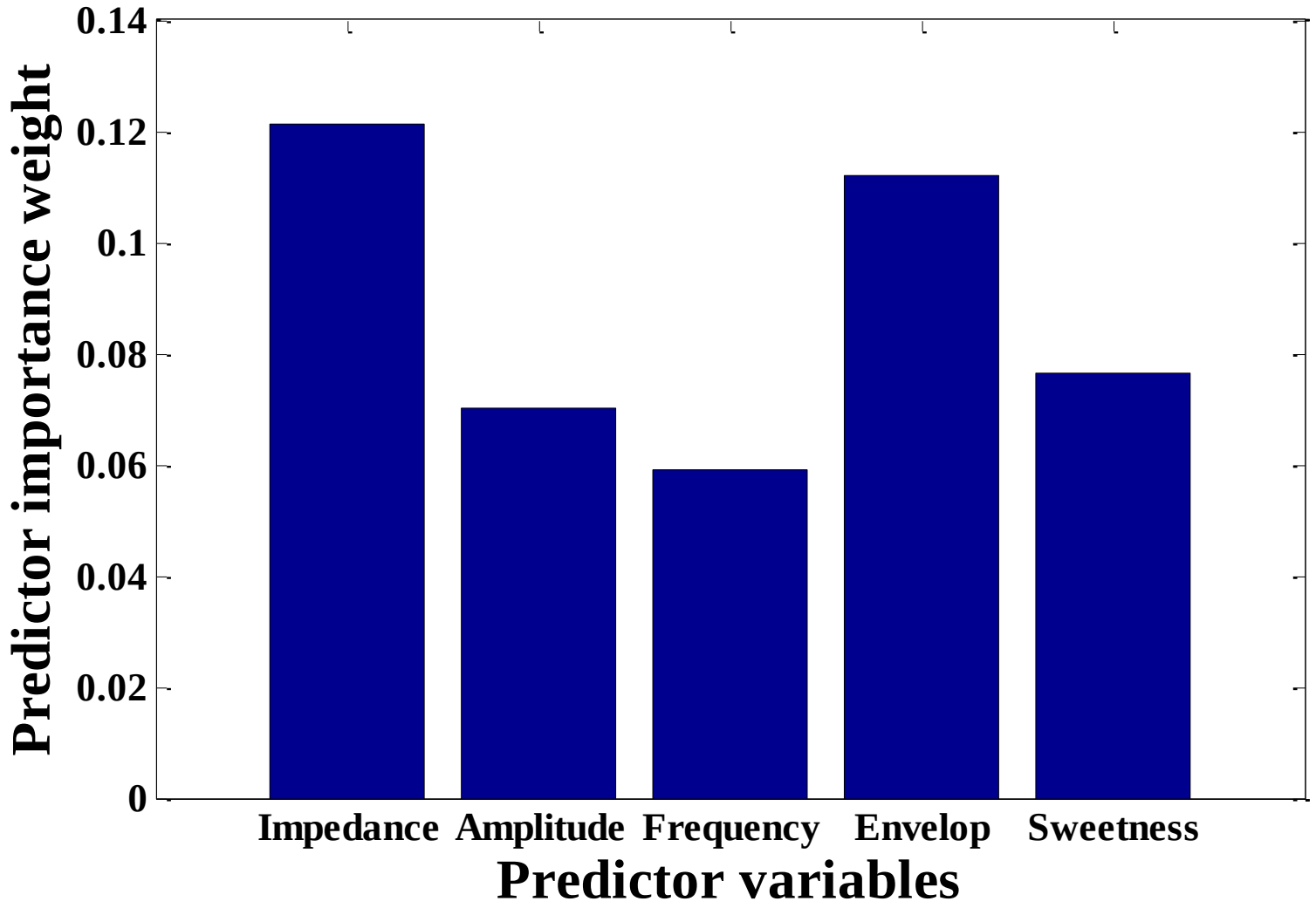

Fig. 4-11: Selection of relevant input attributes using Relief algorithm

Then, the water saturation is classified into two classes, namely- Class high and Class low using a user-defined threshold. The selection of the threshold level is governed by two constraints (as in Section 4.2). We selected the initial threshold level as 0.7 as in Section 4.2.

#### 4.3.2.3 Training and Testing

The lower part of Fig. 4-8 represents the training and testing steps associated with the classifier. For the working dataset, the number of available samples belongs to Class high is significantly large which in turn makes it majority class. Conversely, Class low is minority class due to the presence of small amount of samples belonging to this category in the working dataset. The division of training and testing pattern is carried out as in Section 4.2. The minority class (Class low) patterns belong to integrated dataset of three wells are used to train the classifier. The tuned classifier parameters are validated using the Class low patterns of test well and the combined majority class (Class high) samples of all the wells.

The input attributes (seismic impedance, amplitude envelope, and seismic sweetness) of training patterns are used to construct the SVDD hypersphere. Classification accuracy of SVDD is improved by adjusting multiple parameters such as the kernel function and associated parameters, and radius of the hypersphere $C$. We have experimented with different kernel functions such as Gaussian, higher order polynomial (with order of 2–10), radial basis function, and exponential radial basis function along with associated kernel parameters with $C$ values varying from 0 to 1. The task of the classifier is to minimize the Lagrangian function by constrained optimization as mentioned earlier in Section III. The data samples are categorized into three categories: true data (inside the hypersphere), outliers (outside the hypersphere), and support vectors (at the hypersphere periphery) by this optimization. As in [120], the support vectors are encompassed in the outlier category. The tuned parameters are tested using the majority class samples.

To establish the modified framework over existing classifier algorithms (e.g. ANN, and SVM based classifier), a comparison has been carried out. In all cases, the predictor attributes, and performance evaluators are same as the proposed framework. The division of training-testing samples and associated classification parameters are varied depending on respective classifiers. These classifiers are optimized with appropriate parameter values related to respective algorithms. The predictor variables are same (seismic impedance, seismic amplitude envelope, and seismic sweetness) as that of the proposed framework. The difference lies in the creation of training and testing data set. For these classifiers, the learning is carried out using the integrated dataset of three wells. The samples corresponding to the remaining fourth well are used to test the trained classifiers. Thus, majority and minority class components are collectively used to train the network instead of using only minority class patterns.

The performance of the modified framework (as in Fig. 4-8) is quantified using g-metric means [119], [120] and program execution time. G-metric means is associated with the accuracy of both positive and negative classes and often used in case of imbalanced dataset. Table 4-3 and Table 4-4 represent the comparison results of the proposed framework with other three classifiers in terms of g-metric mean and program execution time. It can be observed from Table 4-3 that the g-metric mean values in case of ANN based classifier are very poor. Then, the blind testing performance improves while using kernel-based algorithm SVM based classifier. Finally, our framework has yielded better performance compared to both– ANN and SVM based classifiers. As the number of patterns belongs to the minority class is insignificant compared to that of the majority class; hence, trained classifiers can detect the majority class testing patterns correctly. However, the minority class test patterns are also wrongfully classified in Class high (majority class). Hence, g-metric mean is poor. On the other hand, our framework is based on one-class classification. Therefore, it can detect minority class patterns in testing dataset yielding better g-metric means within reduced program execution time.

Table 4-3: G-metric mean comparison among the proposed framework and other classifiers

| **Well Name** | **Value of G-Metric Mean** | | |
|---|---|---|---|
| | ***ANN Based Classifier*** | ***SVM*** | ***Proposed Workflow (SVDD)*** |
| A | 0.28 | 0.48 | 0.72 |
| B | 0.26 | 0.65 | 0.74 |
| C | 0.34 | 0.55 | 0.69 |
| D | 0.20 | 0.62 | 0.65 |
| Average Performance | 0.27 | 0.57 | 0.7 |

Table 4-4 : Program execution time comparison among the proposed framework and other classifiers

| **Well Name** | **Value of program execution time (in seconds)** | | |
|---|---|---|---|
| | ***ANN Based Classifier*** | ***SVM*** | ***Proposed Workflow (SVDD)*** |
| A | 26.834 | 16.74 | 12.37 |
| B | 20.238 | 18.84 | 14.2 |
| C | 21.523 | 15.14 | 12.25 |
| D | 22.839 | 14.64 | 13.57 |
| Average Performance | 22.8585 | 16.34 | 13.09 |

The results in Table 4-3 and Table 4-4 are pictorially represented in Fig. 4-12 and Fig. 4-13 respectively. Fig. 4-12 and Fig. 4-13 reveal that the proposed framework has attained better performance compared to other classifiers with higher speed.

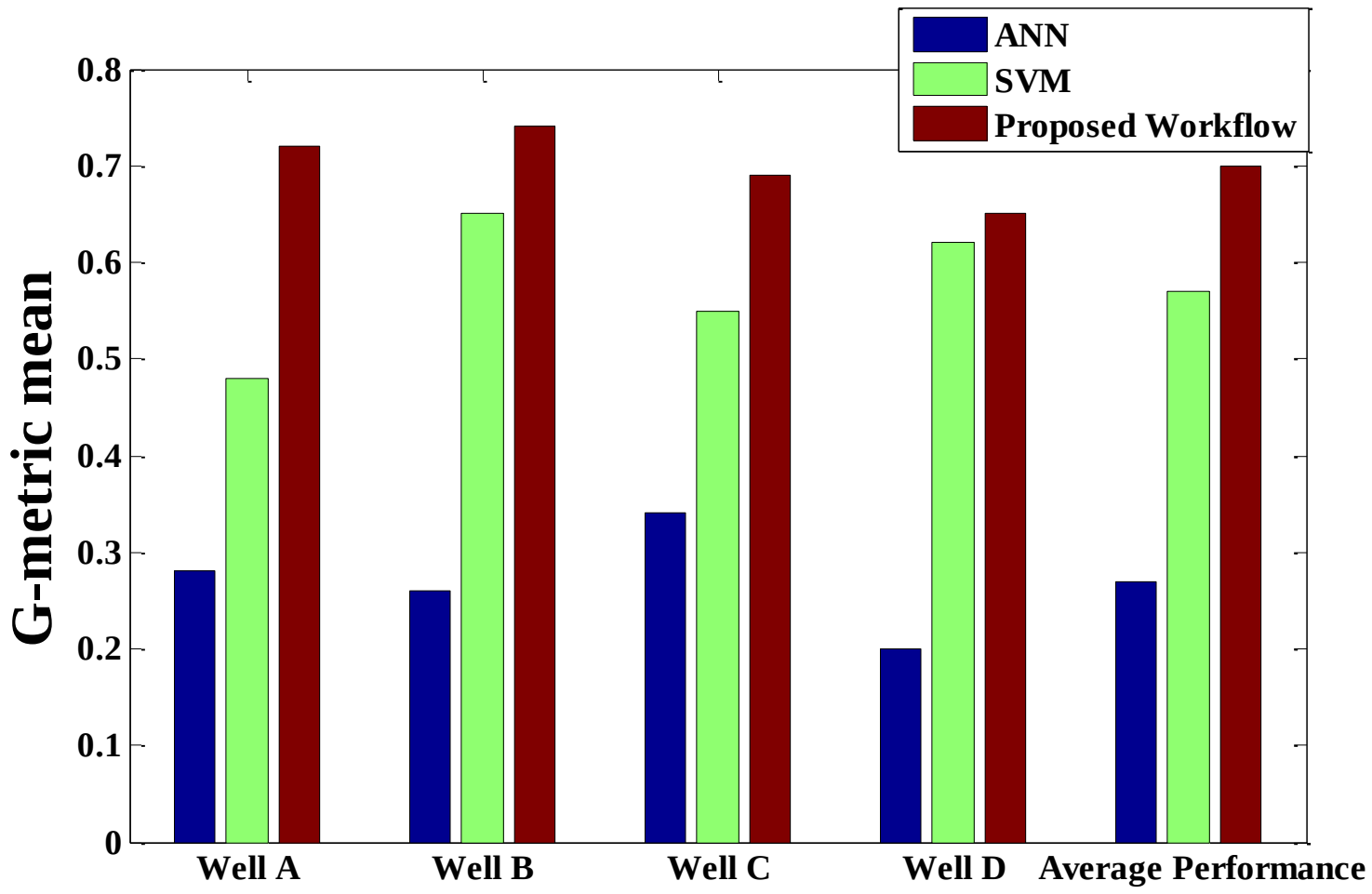

Fig. 4-12: Bar plot describing comparative performance analysis in terms of g-metric means

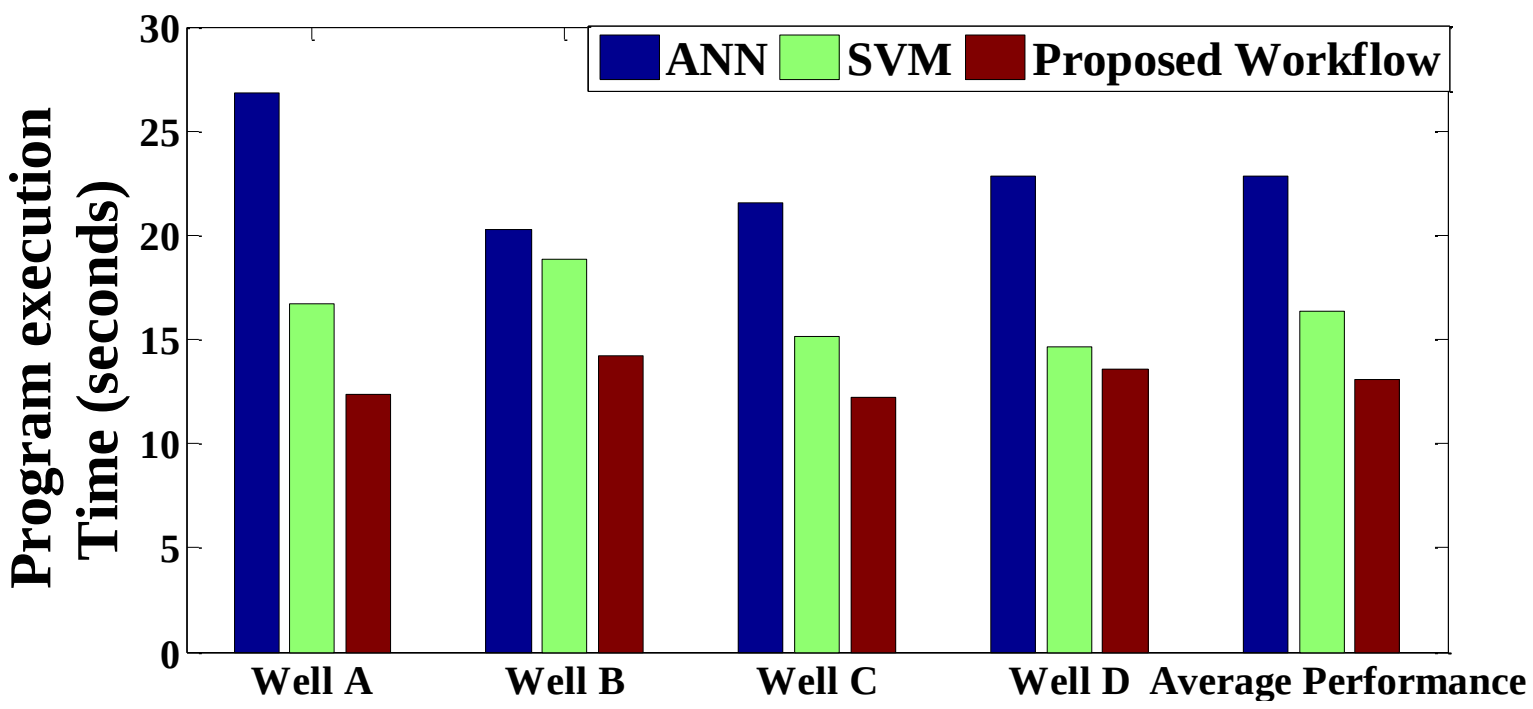

Fig. 4-13: Bar plot describing comparative performance analysis in terms of program execution time

#### 4.3.2.4 Volumetric Classification and Visualization

The trained parameters which yield acceptable results in the blind testing are saved. Then, the water saturation level in the study area can be estimated from seismic attributes. The saved SVDD parameters classify the water saturation level in Class high or Class low at any location in the study area using seismic attributes of the area. After the classification over the area, the variation of water saturation level is visualized at any selected part of the study area.

Fig. 4-14 represents the variation of seismic impedance, at a particular inline over the study area. The tuned classifier which was saved while blind prediction of well A is further used to classify water saturation over the study area from predictor seismic signals. Fig. 4-15 represents the distribution of water saturation level classified in two categories: Class high and Class low over the area at the same inline. Inside the study area, blue represents Class low and red colour represents Class high samples. It can be observed from Fig. 4-15 that the presence of Class high patterns is significant over that of the Class low samples throughout the area.

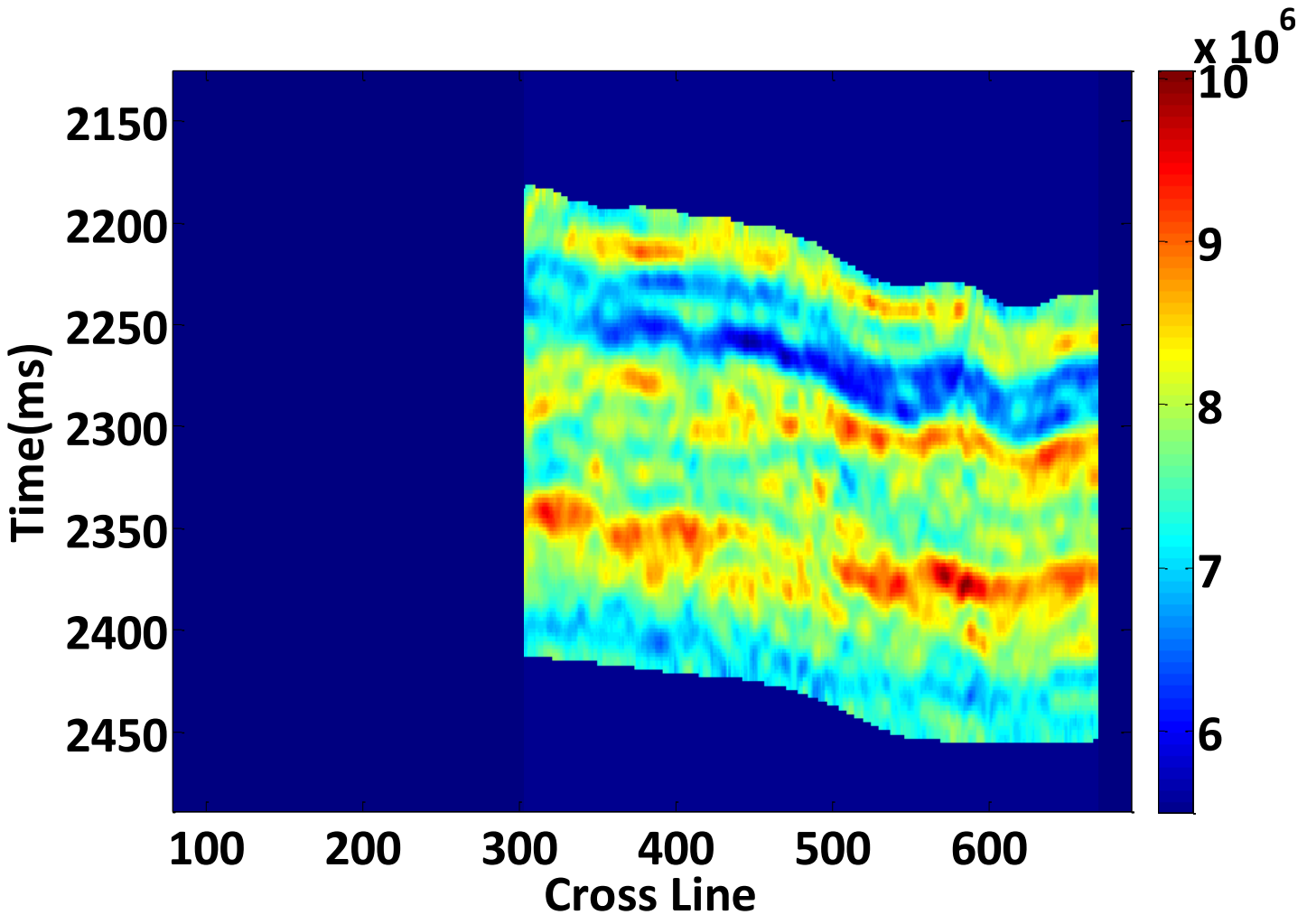

Fig. 4-14: Seismic impedance variation at a particular inline

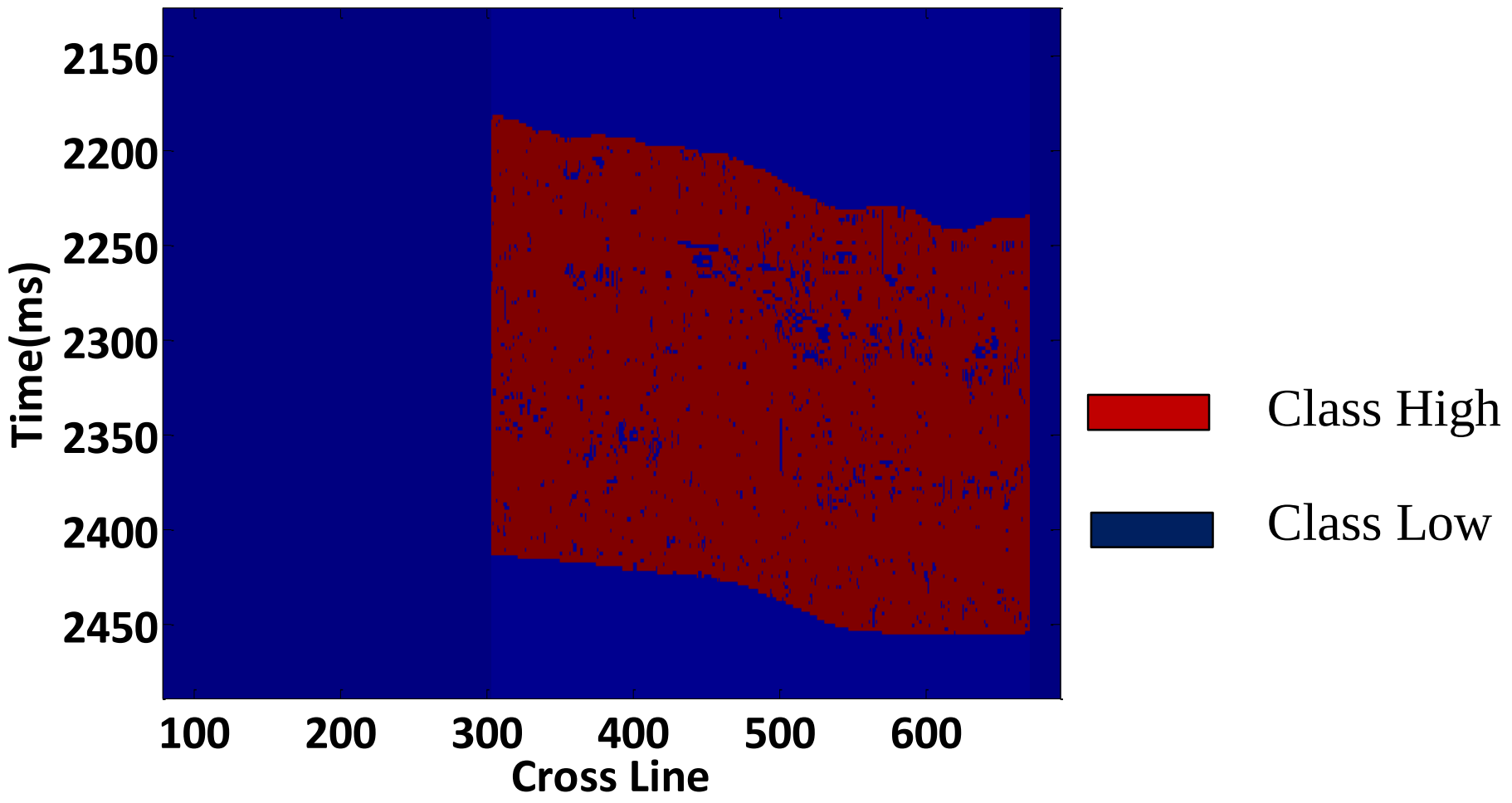

Fig. 4-15: Water saturation level variation at a particular inline

Section 4.3 has proposed a classification framework to classify water saturation levels from the seismic attributes using a small imbalanced dataset. In other words, the class labels (Class low/Class high) variation of water saturation can be predicted from the seismic attributes using the modified framework presented in Section 4.3. The area map representing high and low water saturation level is created using the proposed framework.

## 4.4 Conclusion

In this chapter, water saturation is classified from well logs and seismic attributes respectively using a dataset of four wells. Application of the SVDD to solve the class labels prediction problem using integrated dataset of seismic and borehole data in reservoir characterization field is the contribution of this chapter. Water saturation varies from zero to unity with a skew at unity. The class labels variation of other lithological properties having similar skewed nature can be predicted using the proposed frameworks from well logs and seismic attributes respectively. Although the frameworks have outperformed existing supervised classifiers in terms of performance evaluators, there is a scope of improvement in the selection of parameters associated with the SVDD algorithm. The SVDD parameters are selected empirically keeping the improvement in classification in view. In future, efforts can be made to automate the selection procedure using some evolutionary algorithms such as genetic algorithm, particle swarm optimization etc. The execution time taken by the SVDD algorithm is dependent on the number of training patterns available in the minor class. Thus the selection of the user defined classification threshold is carried out keeping this factor in mind along with experienced geophysicist's opinion. Apart from geological dataset, this framework can be implemented in binary class problems wherever number of training patterns pertaining to a particular class is very less compared to the other class.

# Chapter 5. Conclusion and Future Scope

In the present study, a novel pre-processing scheme is proposed to improve the prediction capability of machine learning algorithms by information filtering for prediction of a lithological property from seismic attributes in Chapter 2. As a result of this pre-processing scheme, the mutual dependency between predictor and seismic attributes are increased in the expense of the decrease in information content of the target property. The proposed scheme is implemented using seismic impedance, amplitude, and instantaneous frequency to model sand fraction using ANN. The issues associated with the data dimension and size of the training network and complexity associated with the selection of the ANN structure and parameters are discussed briefly. The selection of the network structures and initialization of network parameters are carried out empirically. In future research scope, selection of network structure and parameters can be automated using evolutionary algorithms. Then, in case of the working dataset with small number of training patterns and large dimension, inclusion of a dimensionality deduction algorithm in the pre-processing stage. However, for this study, the available training patterns are large enough compared to the data dimension. Here, post-processing schemes based on different spatial filters with selected window size are implemented to improve the visualization across the volume. Introduction of model based filtering based on variation of predictor attributes across the volume is a probable direction of research. Lithological properties having similar distributions as sand fraction i.e. varying between the minimum and maximum values can be modelled using the frameworks proposed in Chapter 3. For example, shale fraction, porosity, permeability, etc. can be predicted from seismic attributes using the generalized frameworks as in Chapter 3 depending on the availability of the well tops information.

Water saturation has an imbalanced distribution skewed at unity. Therefore, instead of predicting exact values of water saturation, class labels detection can serve the purpose of a reservoir engineer as the layers having low water saturation is of importance. In Chapter 4, the variation of the class labels (Class low/Class high) of water saturation is predicted using well logs and seismic data using a one-class classification framework based on SVDD. The performances of the proposed frameworks have been compared with other supervised algorithms. The selection of the SVDD parameters are crucial for obtaining good performance. It has been carried out empirically here. In future, the parameters can be automated using different evolutionary algorithms. The class labels of other characteristics having skewed distribution such as oil saturation can be modelled using the frameworks designed here.

On the whole this thesis discussed about creation of synthetic logs of lithological properties from seismic attributes of a study area after calibrating a functional relationships between the predictors and target using an integrated dataset of these two types of data at available well control points.

## 5.1 Dissemination out of this Work

Journal Papers

1. **S. Chaki**, A. Routray, and W. K. Mohanty, “A novel pre-processing scheme to improve the prediction of sand fraction from seismic attributes using neural networks,” *IEEE J. Sel. Topics Appl. Earth Observations and Remote Sens*., vol. 8, no. 4, pp. 1808-1820, 2015.
2. **S. Chaki**, A. K. Verma, A. Routray, W. K. Mohanty, and M. Jenamani, “Well tops guided prediction of reservoir properties using modular neural network concept: A case study from western onshore, India,” *J. Pet. Sci. Eng.*, vol. 123, pp. 155-163, Nov. 2014.
3. A. K. Verma, **S. Chaki**, A. Routray, W. K. Mohanty, M. Jenamani, “Quantification of sand fraction from seismic attributes using Neuro-Fuzzy approach,” *J. Appl. Geophysics*, vol. 111, pp. 141-155, Dec. 2014.

Conference Papers

1. **S. Chaki**, A. K. Verma, A. Routray, M. Jenamani, W. K. Mohanty, P. K. Chaudhuri, and S. K. Das, “Prediction of porosity and sand fraction from well log data using ANN and ANFIS : a comparative study,” in *10th Biennial Int. Conf. Expo. SPG*, Kochi, India, 2013.
2. **S. Chaki**, A. K. Verma, A. Routray, W. K. Mohanty, and M. Jenamani, “A one-class classification framework using SVDD : application to an imbalanced geological dataset,” in *Proc. IEEE Students’ Technology Symp. (TechSym)*, Kharagpur, India, 2014, pp. 76–81.
3. **S. Chaki**, A. K. Verma, A. Routray, W. K. Mohanty, and M. Jenamani, “A novel framework based on SVDD to classify water saturation from seismic attributes,” in *Fourth Int. Conf. Emerging Applicat. Inform. Technology (EAIT)*, Kolkata, India, 2014, pp. 64–69.

## 5.2 Future Scope

The following appears to be promising area for future research

- As initial parameters selection of machine learning algorithms crucial for achieving acceptable performance, it would be interesting to automate the initialization parameters by an appropriate metaheuristic algorithm
- Model based post–processing instead of spatial filtering on predicted lithological properties
- Uncertainty quantification associated with Reservoir characterization: Some interesting works have been reported in [121], [122] about application of uncertainty quantification in reservoir characterization. These publications can be used as a guiding point to explore uncertainty analysis related to modeling and data acquisition process of reservoir characterization.